\documentclass[aps, prd, twocolumn, lengthcheck, superscriptaddress, 
nofootinbib]{revtex4-1}

\usepackage{color}
\usepackage{hyperref}
\usepackage{amsfonts}
\usepackage{amsmath}
\usepackage{amssymb}
\usepackage{graphicx}
\usepackage{ulem}
\usepackage{epstopdf}  
\usepackage{float}
\usepackage{mathrsfs}
\usepackage{cases} 
\usepackage{bm}
\usepackage{latexsym} 
\usepackage{upgreek}
\usepackage{slashed}

\def\be{\begin{eqnarray}}
\def\ee{\end{eqnarray}}

\begin{document}
	
\newcommand*{\ssp}{\scriptscriptstyle}
\newcommand*{\gm}{\gamma}
\newcommand*{\nm}{\nonumber}

\title{Connecting dilaton thermal fluctuation with the Polyakov loop at finite temperature
}
\author{Bing-Kai Sheng}
\email{bingkai.sheng@ucas.ac.cn}
\affiliation{School of Fundamental Physics and Mathematical Sciences, Hangzhou Institute for Advanced Study, UCAS, Hangzhou, 310024, China}

\author{Yong-Liang Ma}
\email{ylma@nju.edu.cn}
\affiliation{School of Frontier Sciences, Nanjing University, Suzhou, 215163, China}
\date{\today}	

\begin{abstract}
 Understanding the character of the deconfinement phase transition is one of the fundamental challenges in particle physics. In this work, we derive a formula for the expectation value of the Polyakov loop---the order parameter of the deconfinement phase transition---in pure $\mathrm{SU(N_{\mathrm{c}})}$ gauge systems at finite temperatures starting from the  Coleman\textendash Weinberg-type effective potential encoding the trace anomaly of QCD. Our results are in good agreement with the Lattice QCD data and can effectively describe the large-$N_{\mathrm{c}}$ behaviors of the expectation value of the Polyakov loop. Notably, our findings predict the strongest first-order deconfinement phase transition as $N_{\mathrm{c}} \to +\infty$. Furthermore, to establish a relation between the dilaton field and the Polyakov loop, we also derive the scale transformation rule for temperature based on quantum statistical mechanics. The results of this work may shed a light on the connection between deconfinement phase transition and evolution of scale symmetry in the thermal system.
\end{abstract}	

\maketitle
	
\allowdisplaybreaks{}
\section{Introduction}
\label{sec:introduction}	  

It is well-known that quantum chromodynamics (QCD) with massless quarks possesses a scale symmetry, which is, however, broken at the quantum level by the trace anomaly~\cite{Coleman:1985rnk,book:91601717}. This phenomenon is characterized by the nonzero expectation value of the trace of the energy-momentum tensor, which is directly related to the nonzero gluonic condensate in the QCD vacuum, i.e., $\langle\Theta\rangle\equiv\langle T^{\mu}_{~\mu}\rangle\propto G^{2}$~\cite{Fujimoto:2022ohj,Li:2016uzn}. In nature, only hadrons without color charges are observed in low-energy experiments. Consequently, when studying hadron dynamics within the framework of low-energy effective field theories (EFTs) of QCD, the effects of trace anomaly are implemented through anomaly matching~\cite{Schechter:1980ak}. Typically, the trace anomaly in these EFTs is written in the form of a Coleman--Weinberg-type logarithmic potential in terms of a real scalar field $\chi$---termed by dilaton field~\cite{Schechter:1980ak,Meissner:1999pe,Goldberger:2007zk,Campbell:2011iw,Matsuzaki:2013eva}.
	
Although the trace anomaly is an inherent feature of the QCD vacuum, scale symmetry is expected to be (partially) restored in certain extreme environments~\cite{Harada:1999zj,Ma:2023ugl,Fujimoto:2022ohj,Boyd:1995zg,Boyd:1996bx,Cheng:2007jq,Borsanyi:2013bia,HotQCD:2014kol,Kurkela:2009gj,Annala:2019puf}, such as high temperatures and/or densities which exist in the early universe, ultra-relativistic heavy-ion collisions at RHIC and LHC, and cores of massive neutron stars. The restoration of scale symmetry under such high temperatures and/or high densities can be naively understood by considering that the quantum effects will be submerged by the medium at these extreme conditions, a phenomenon attributed to the asymptotic freedom of QCD.

As an {\it ab initio} calculation, Lattice QCD (LQCD) has yielded abundant results of the thermodynamic quantities associated with the trace anomaly at high temperatures, including pressure, energy density, and entropy density. In early studies~\cite{Boyd:1995zg,Boyd:1996bx}, the trace anomaly of a pure $\mathrm{SU(3)}$ gauge system in a thermal medium was calculated and the results indicate that at high temperatures the trace anomaly tends to vanish and the pressure $P$ approaches the Stefan-Boltzmann law $P\propto T^{4}$. Subsequently, the trace anomaly was calculated in the framework of full QCD including dynamical quarks~\cite{Cheng:2007jq,Borsanyi:2013bia,HotQCD:2014kol}. It was found that the trace anomaly are melted at very high temperatures, with the ratio $P/T^{4}$ converging to a constant. Although LQCD has significantly deepened our understanding of strong interactions from the first principle, challenges arise when considering the finite chemical potentials. The finite quark chemical potential spoils the Monte Carlo simulations of the path integral on a large discretized Euclidian space-time lattice, leading to the so-called fermion sign problem of LQCD~\cite{Dexheimer:2020zzs,Troyer:2004ge} (for more details, see e.g., Refs.~\cite{MUSES:2023hyz,book:94872761} and references therein). Consequently, in the context of dense nuclear matter and dense quark matter, Lattice QCD simulations are limited, and the credible tools are effective models/theories.

In the literature, many studies have suggested the existence of a confinement-deconfinement phase transition at extremely high temperatures and/or densities~\cite{book:94872761,Kaczmarek:2002mc,Ratti:2005jh,Abuki:2009dt,Andreev:2009zk,Burnier:2009bk,Noronha:2010hb,Colangelo:2010pe,Stoffers:2010sp,Megias:2010ku,Megias:2010gt,Borsanyi:2010bp,Fukushima:2010bq,Li:2011hp,Fukushima:2011jc,Fukushima:2012qa,Ruggieri:2012ny,Petreczky:2012rq,Adams:2012th,Alba:2013haa,Bellwied:2013cta,Alba:2014lda,Andersen:2014xxa,Kharzeev:2015kna,Fukushima:2017csk,Andersen:2021lnk,Bluhm:2024uhj}. Although the confinement phenomenon has not yet been analytically derived from the QCD, the description of the confinement-deconfinement phase transition for strongly coupled matter has been effectively facilitated through the definition of the Polyakov loop and its expectation value~\cite{Fukushima:2003fw,book:94872761}. The Polyakov loop is defined as the trace in color space of the Wilson line along the temporal direction of Euclidian space-time~\cite{book:94872761} 
\be
\label{Polyakov loop}
\Phi(\bm{x})=\frac{1}{N_{\mathrm{c}}}\mathrm{tr_{c}}\left\{\mathcal{P}\exp\bigg[-ig_{\mathrm{s}}\int_{0}^{\beta}\mathrm{d}x_{4}A_{4}^{a}(\bm{x},x_{4})T^{a}\bigg]\right\},
\ee
with color number $N_{\mathrm{c}}$, strong coupling constant $g_{\mathrm{s}}$, the generators $T^{a}$ of $\mathrm{SU(N_{\mathrm{c}})}$ group and $\beta=1/T$. Gauge field $A^{a}_{4}(\bm{x},x_{4}),a=1,2,\cdots,N_{\mathrm{c}}^{2}-1$  is defined in four-dimensional Euclidean space-time. Here, $\mathrm{tr_{c}}$ and $\mathcal{P}$ denote the trace in color space and the path-ordering respectively. It should be noted that the Polyakov loop~(\ref{Polyakov loop}) has a $\mathbb{Z}_{N_{\mathrm{c}}}$ symmetry---or center symmetry---due to the non-trivial periodic boundary condition of gluon field at finite temperatures, i.e., $A^{a}_{\mu}(\bm{x},x_{4}+\beta)=A^{a}_{\mu}(\bm{x},x_{4})$~\cite{book:94872761}. 
Therefore, the $\mathbb{Z}_{N_{\mathrm{c}}}$ symmetry must be taken into account when one constructs models involving the Polyakov loop~\cite{Fukushima:2003fw,Ratti:2005jh,Dexheimer:2009hi,Sakai:2010rp,Sasaki:2011wu,Ishii:2013kaa,Ishii:2015ira,Mattos:2021tmz}.

The order parameter of the deconfinement phase transition is the expectation value of the Polyakov loop, i.e., $\langle\!\langle\hat{\Phi}\rangle\!\rangle$. In the heavy quark limit $\langle\!\langle\hat{\Phi}\rangle\!\rangle \propto e^{-(F-F_{0})/T}$ where $F_{0}$ represents the free-energy of gluons and $F-F_{0}$ is the least work required to excite a quark in the thermal medium composed of gluons~\cite{book:94872761}. Consequently, if the system is in the confinement phase, it takes infinity least work to extract a single quark from the thermal medium, resulting in $\langle\!\langle\hat{\Phi}\rangle\!\rangle=0$. Conversely, when the expectation value of the Polyakov loop is not zero but finite, $F-F_{0}$ becomes finite as well, allowing for the existence of states with a single quark, indicating that the the system is in the deconfinement phase. 

In addition to the $N_{\mathrm{c}}=3$ Yang-Mills theory with three generations of quarks in the Standard Model, several recent works~\cite{Panero:2009tv,Huang:2020crf,Kang:2021epo} explored the restoration of scale symmetry in pure gauge systems with $N_{\mathrm{c}}>3$ and  investigted the related thermodynamic quantities such as pressures, energy densities, entropy densities, and the trace of the energy-momentum tensor. In Ref.~\cite{Panero:2009tv}, the thermal properties of pure gauge systems in large-$N_{\mathrm{c}}$ limit were investigated based on LQCD simulations. Furthermore, Ref.~\cite{Huang:2020crf} provided a detailed study of the dynamics associated with the dark confinement-deconfinement phase transition and its implications for gravitational-wave spectra. Overall, the evolution of the scale symmetry, or the physics of dilaton, and the confinement-deconfinement phase transition are of great interest across various fields, including cosmology, astrophysics, particle physics, and nuclear physics (see, e.g., Refs.~\cite{Gasperini:2007ar,Sasaki:2011ff,Crewther:2013vea,Ma:2020tsj,Fujimoto:2020tjc,Fujimoto:2022ohj,Zhang:2024sju} and references therein).
	
Inspired by the fact that both the dilaton potential, which encodes the trace anomalies of QCD, and
the Polyakov loop potential, which models the deconfinement phase transition, can be expressed in the
logarithmic forms, we attempted to establish a connection between the trace anomaly of QCD and the deconfinement phase transition in Ref.~\cite{Sheng:2023rnd} considering that both are related to the configuration of gluons. We obtained an effective potential of the Polyakov loop from the Coleman--Weinberg-type dilaton potential and evaluated the parameters in the relation between the dilaton and the Polyakov loop using the LQCD data of $\langle\!\langle\hat{\Phi}\rangle\!\rangle$ in pure $\mathrm{SU(3)}$ gauge sector~\cite{Kaczmarek:2002mc,Ratti:2005jh} and these parameters are functions of temperature. While results are qualitatively in agreement with those of LQCD, the critical temperature of the deconfinement phase transition from our ansatz, $T_{\mathrm{c}}\approx 356\text{MeV}$, is much higher than the LQCD result, $T_{\mathrm{c},\text{LQCD}}\approx 270\text{MeV}$. We attributed this discrepancy to the mean-field approximation (MFA) employed in our analysis and expect that the thermal fluctuation of the dilaton field may compensate for this gap, particularly at the vicinity of the critical temperature. 

In this work, we will focus on the thermal fluctuation of dilaton field and construct a more reliable relationship between the dilaton field $\chi$ and the Polyakov loop $\Phi$. We will first introduce two types of thermal fluctuations: one directly corresponds to the random thermal motion of dilaton particles near the critical temperature, while the other characterizes the strongly coupled gluons in the deconfinement phase and hence depends on the Polyakov loop. Considering that the scale transformation properties of these quantities are crucial for establishing the relationship, we will formulate the scale transformation properties for thermodynamic quantities, especially temperature $T$, within the framework of quantum statistical mechanics. We will illustrate the scale invariance of the Polyakov loop, a key element of our approach. Notably, in a scale-symmetric system, once we know the scale transformations for pressure $P$, temperature $T$ and chemical potential $\mu$ we can straightforwardly write down the analytical form of the equation of state (EoS) instead of doing tedious calculations from a given microscopic model or theory. Finally, we will present the expression of the dilaton fluctuation as a function of $\Phi$, $\Phi^{*}$ and $T$. This expression takes the form of a power series in terms of the Polyakov loop with the coefficients proportional to $T$. 

As the most important conclusion of this work, by minimizing the temperature-modified Coleman--Weinberg-type potential, we will detail the derivation of expectation value  $\langle\!\langle\hat{\Phi}\rangle\!\rangle$ at both the next-leading order (NLO) and the next-next-leading order (NNLO) of the power series. We will also compare our results numerically with the LQCD data for pure $\mathrm{SU(3)}$, $\mathrm{SU(4)}$ and $\mathrm{SU(5)}$ gauge systems~\cite{Kaczmarek:2002mc,Ratti:2005jh,Mykkanen:2012ri}. Moreover, we will qualitatively compare our findings with results from 4-8PLM Polyakov potentials~\cite{Kang:2021epo} for several large-$N_{\mathrm{c}}$ scenarios, specifically, $N_{\mathrm{c}}=6,7,\cdots,11$, where the parameters are fitted using the equations of state obtained from LQCD simulations. It is found that our results are quantitatively in good agreement with the LQCD data for $N_{\mathrm{c}}=3,4,5$, including the deconfinement phase transition temperature for $N_{\mathrm{c}}=3$, i.e., $T_{\mathrm{c}}(N_{\mathrm{c}}=3)\approx270\text{MeV}$, and the qualitative results in large-$N_{\mathrm{c}}$ cases. In particular, our formulae predict the strongest first-order deconfinement phase transition for the limit $N_{\mathrm{c}}\to +\infty$. 

We would like to emphasize that an analytical expression of $\langle\!\langle\hat{\Phi}\rangle\!\rangle$ as a function of temperature was derived many years ago within the framework of gauge-string duality for the pure $\mathrm{SU(3)}$ gauge system in Ref.~\cite{Andreev:2009zk}. The author calibrated the parameters using LQCD data and found that the results align quite well with LQCD simulations. In this work, we derive the analytical expression for $\langle\!\langle\hat{\Phi}\rangle\!\rangle$ using field theory, focusing on the relationship between the trace anomaly and the deconfinement phase transition. The good consistency of our results with that of LQCD and the effective description of the large-$N_{\mathrm{c}}$ behavior of $\langle\!\langle\hat{\Phi}\rangle\!\rangle$ indicate that we may insights into the confinement-deconfinement phase transition by investigating the trace anomaly and the potential relations between these two crucial phenomena. These findings may deepen our understanding of the evolution of the universe, the mechanism of the electroweak symmetry breaking, and the phase diagram of QCD.
	
The paper is organized as follows: In Sec.~\ref{sec:concepts}, we will briefly review the basic concepts of the expectation value and the fluctuation of the dilaton field in thermal field theory. Sec.~\ref{sec:ScalTran} formulates the scale transformation rules of thermodynamic quantities and presents in general the expression of the dilaton fluctuation as a power series with respect to the Polyakov loop. In Sec.~\ref{sec:ExpDilaton}, we will derive the formulae of $\langle\!\langle\hat{\Phi}\rangle\!\rangle$ as functions of temperature, up to NLO and NNLO of the power series, by minimizing the temperature-modified Coleman--Weinberg-type potential. In Sec.~\ref{sec:three}, we calibrate the parameters in the formulae and compare our results with the LQCD data. Using the fixed constants, we then discuss the large-$N_{\mathrm{c}}$ behavior of the Polyakov loop. Finally, we present a summary and conclusions in Sec.~\ref{sec:Sum}. As supplementary material of the formulation of the scale transformation properties of thermodynamic quantities, we discuss the Liouville equation in classical and quantum field theories and the quantum scale transformation in Heisenberg picture, respectively, in App.~\ref{app:Liouville} and App.~\ref{app:ScaleHeisenberg}. In addition, we discuss the scale-covariance of the four laws of thermodynamics in App.~\ref{app:4laws} to demonstrate the self-consistency of our approach.

\section{Basic concepts of the dilaton field in thermal field theory}
\label{sec:concepts}

\subsection{Concepts of the expectation value and the fluctuation of the dilaton field}

In general, the dilaton field $\chi$ can be decomposed into two parts: the expectation value $\langle\!\langle\hat{\chi}\rangle\!\rangle$ and the fluctuation $\delta_{\chi}(\bm{x},t;T)$, namely
\be
\label{expecchi plus fluc}
\chi(\bm{x},t;T) & = & \langle\!\langle\hat{\chi}\rangle\!\rangle + \delta_{\chi}(\bm{x},t;T).
\ee
Since at finite temperatures, the system is in a mixed state described by the density operator $\hat{\rho}=e^{-\hat{H}/T}/Z$ and the configuration of $\chi$ is affected by the heat reservoir, the temperature variable $T$ is explicitly written in $\chi(\bm{x},t;T)$ and $\delta_{\chi}(\bm{x},t;T)$. As for the expectation value $\langle\!\langle\hat{\chi}\rangle\!\rangle$, it is defined as an ensemble average
\be
\langle\!\langle\hat{\chi}\rangle\!\rangle = \mathrm{Tr}[\hat{\rho}\hat{\chi}(\bm{x},t;T)].
\ee 

Consider the eigenvectors of the energy-momentum vector $\hat{P}^{\mu}=(\hat{H},\hat{\bm{P}})$ in the thermal system, denoted by $\{|n;T\rangle\}$, where $T$ represents temperature and $n$ signifies other quantum numbers. The trace can be calculated as follows
\be
\label{expecchi in Pmu repre}
\langle\!\langle\hat{\chi}\rangle\!\rangle & = &
\frac{1}{Z}\sum_{n}\langle n;T|e^{-\hat{H}/T}\hat{\chi}(\bm{x},t;T)|n;T\rangle \nm \\
& = & \frac{1}{Z}\sum_{n}e^{-E_{n}/T}\langle n;T|\hat{\chi}(\bm{x},t;T)|n;T\rangle,
\ee
where $E_n$ is the eigenvalue of the Hamiltonian of state $|n;T\rangle$ and $Z = {\rm Tr} \left(e^{-\hat{H}/T}\right)$ is the normalization factor. By using the translation operator of space-time coordinates for the dilaton field
\be
\label{transl of chi}
\hat{\chi}(\bm{x},t;T)=e^{i(\hat{H}t-\hat{\bm{P}}\cdot\bm{x})}\hat{\chi}(\bm{0},0;T)e^{-i(\hat{H}t-\hat{\bm{P}}\cdot\bm{x})}~,
\ee
one can conclude that, $\langle\!\langle\hat{\chi}\rangle\!\rangle$ is spacetime-independent but simply a function of temperature.

We next consider the limit of zero temperature for $\langle\!\langle\hat{\chi}\rangle\!\rangle$. Isolating the vacuum state $|n=\Omega;T\rangle$ of the system which has energy $E_\Omega$, we can rewrite (\ref{expecchi in Pmu repre}) as
\be
\label{expecchi in Pmu repre 2}
\langle\!\langle\hat{\chi}\rangle\!\rangle & = & \left(\langle\hat{\chi}\rangle+\sum_{n\neq\Omega}e^{-(E_{n}-E_{\Omega})/T}\langle n;T|\hat{\chi}(\bm{0},0;T)|n;T\rangle\right) \nm \\
& &{} \times\frac{e^{-E_{\Omega}/T}}{Z},
\ee
where $\langle\hat{\chi}\rangle$ is the vacuum expectation value (VEV) of the dilaton 
\be
\langle\hat{\chi}\rangle = \langle\Omega;T|\hat{\chi}(\bm{0},0;T)|\Omega;T\rangle.
\ee
Here, it should be noted that the vacuum state $|\Omega;T\rangle$ is conceptually distinct from that in quantum field theories defined at zero temperature. In fact, the vacuum state at zero temperature is $|\Omega\rangle=\lim\limits_{T\to0^{+}}|\Omega;T\rangle$ where $0^{+}$ means the right limit approaching zero since the absolute zero temperature cannot be carried out in accordance with the third law of thermodynamics.

From the second terms in the parenthesis of Eq.~(\ref{expecchi in Pmu repre 2}), it is clear what zero temperature signifies. When the energy differences between the excited states and the ground state $E_{n}-E_{\Omega}>0,n\neq\Omega$ are much larger than the temperature $T$, we have $\langle\!\langle\hat{\chi}\rangle\!\rangle\approx e^{-E_{\Omega}/T}\langle\hat{\chi}\rangle/Z$ and $Z\approx e^{-E_{\Omega}/T}$. Consequently, the ensemble average of the dilaton field $\langle\!\langle\hat{\chi}\rangle\!\rangle$ becomes equal to the VEV $\langle\hat{\chi}\rangle$. In this case, the properties of the system are dominated by the physics at zero temperature, allowing us to neglect the effects of the heat reservoir on this system. Incidentally, when $E_{n}-E_{\Omega}\gg T,n\neq\Omega$, the system can be considered to be in the pure state $|\Omega;T\rangle$ due to
\be
\label{dens op low T}
\hat{\rho} & = & \frac{e^{-\hat{H}/T}}{Z} = \frac{1}{Z}\sum_{n}e^{-E_{n}/T}|n;T\rangle\langle n;T| \nm \\
& \approx & |\Omega;T\rangle\langle\Omega;T|.
\ee

For the fluctuation $\delta_{\chi}(\bm{x},t;T)$, it describes the excitations of the dilaton field in spacetime and becomes an operator in the canonical quantization formalism, i.e.,
\be 
\hat{\delta}_{\chi}(\bm{x},t;T)=\hat{\chi}(\bm{x},t;T)-\langle\!\langle\hat{\chi}\rangle\!\rangle\hat{I},
\ee
where $\hat{I}$ is the identity operator of Hilbert space. From the definition of the expectation value one can easily conclude $\langle\!\langle\hat{\delta}_{\chi}\rangle\!\rangle=0$.

In addition, one can write down the eigen-equation of the dilaton operator
\be
\label{chi eigen eq}
\hat{\chi}(\bm{x},t;T)|\chi,t;T\rangle=\chi(\bm{x};T)|\chi,t;T\rangle,
\ee
where the eigenvalue $\chi(\bm{x};T)$ at point $\bm{x}$ is an arbitrary real number since the dilaton field is a real scalar field. The eigenvector $|\chi,t;T\rangle$ is the direct product of $|\chi(\bm{x};T),t;T\rangle$, namely 
\be
|\chi,t;T\rangle=\prod\limits_{\bm{x}\in\mathbb{R}^{3}}|\chi(\bm{x};T),t;T\rangle,
\ee 
which satisfies the orthonormal and complete relations
\be
\label{ortho chi}
& & \langle\chi,t;T|\chi',t;T\rangle=\delta_{\chi\chi'}(T) \nm\\
& & \qquad \qquad \qquad \;\;\; \equiv\prod_{\bm{x}\in\mathbb{R}^{3}}\delta[\chi(\bm{x};T)-\chi'(\bm{x};T)],\\
\label{compl chi}
& & \int[\mathrm{D}\chi(T)]|\chi,t;T\rangle\langle\chi,t;T|=\hat{I},
\ee
where $\mathrm{D}\chi(T)\equiv\prod\limits_{\bm{x}\in\mathbb{R}^{3}}[\mathrm{d}\chi(\bm{x};T)]$. Obviously, $|\chi,t;T\rangle$ is also the eigenvector of $\hat{\delta}_{\chi}(\bm{x},t;T)$ with eigenvalue $\delta_{\chi}(\bm{x};T)=\chi(\bm{x};T)-\langle\!\langle\hat{\chi}\rangle\!\rangle$.

\subsection{Thermodynamics of the pure dilaton theory}

With the above discussions, we can investigate the thermodynamics of the system composed of dilatons. Consider Lagrangian
\be
\label{dilaton Lagragian}
\mathcal{L}_{\chi} & = & \frac{1}{2}\partial_{\mu}\chi\partial^{\mu}\chi-\widetilde{\mathcal{V}}(\chi),
\ee 
where the Coleman\textendash Weinberg-type potential $\widetilde{V}(\chi)$ is
\be
\label{CW potential T}
\widetilde{\mathcal{V}}(\chi) & = & \frac{\tilde{m}_{\sigma}^{2}\tilde{f}_{\sigma}^{2}}{4}\left(\frac{\chi}{\tilde{f}_{\sigma}}\right)^{4}\left[\mathrm{ln}\left(\frac{\chi}{\tilde{f}_{\sigma}}\right)-\frac{1}{4}\right].
\ee 
Throughout this work, we use an over wave-line to indicate the temperature dependence of the parameters. And $\tilde{f}_{\sigma}$ is exactly equal to $\langle\!\langle\hat{\chi}\rangle\!\rangle$ with the boundary condition $\tilde{f}_{\sigma}(T=0)=f_{\sigma}\approx 150\text{MeV}$~\cite{Sheng:2023rnd}.

From Lagrangian~(\ref{dilaton Lagragian}), one obtains the Hamiltonian density as
\be
\label{dilaton Hamiltonian}
\mathcal{H}_{\chi}=\frac{1}{2}\pi^{2}+\frac{1}{2}(\bm{\nabla}\chi)^{2}+\widetilde{\mathcal{V}}(\chi)
\ee
with canonical momentum density $\pi=\partial_{0}\chi$. In principle, we can evaluate the partition function $Z=\mathrm{Tr}\exp(-\int\mathrm{d}^{3}x\hat{\mathcal{H}}_{\chi}/T)$ and obtain all the thermodynamic quantities of the system formed by dilaton particles. Here, instead, our primary objective is to establish a possible relationship between the dilaton field and the Polyakov loop. To achieve this, it is crucial to identify the approximations not only simplify calculation but also capture the dominated physics involved.

When studying the deconfinement phase transition at high temperatures, one commonly adopted ansatz is to treat the gluon field as a static and homogeneous background field~\cite{Fukushima:2003fw,Ratti:2005jh}. In this work, we apply this ansatz and regard the Polyakov loop (\ref{Polyakov loop}) as a homogeneous background field as well. Regarding the dilaton field, since it relates to the gluonic configuration through anomaly matching~\cite{Schechter:1980ak,Ma:2019ery}
\be
\label{ano mat}
\theta^{\mu}_{~\mu}=\frac{\beta}{2g_{s}}G^{a}_{\mu\nu}G^{a,\mu\nu}={} -\frac{m_{\sigma}^{2}}{4f_{\sigma}^{2}}\chi^{4},
\ee
it is also static and homogeneous. Therefore, the dynamic of the dilaton field is actually neglected in the following. 

With respect to the fact that the thermal fluctuation of the dilaton field will play a crucial role when the temperature approaches the critical value $T_{\mathrm{c}}$ and the gluonic degree of freedom (DoF) starts to arise, we include a thermal fluctuation of the dilaton field $\delta_{\chi,\mathrm{th}}(\Phi,\Phi^{*};T)$ related to the Polyakov loop. 
Then, the configuration of the dilaton field is expressed as
\be
\label{expecchi plus thfluc}
\chi(T)=\tilde{f}_{\sigma}+\delta_{\chi,\mathrm{th}}(\Phi,\Phi^{*};T).
\ee
Furthermore, we formally decompose the thermal fluctuation into two parts
 \be
 \label{thermal fluc}
 \delta_{\chi,\mathrm{th}}(\Phi,\Phi^{*};T)=\delta^{(1)}_{\chi,\mathrm{th}}(T)+\delta^{(2)}_{\chi,\mathrm{th}}(\Phi,\Phi^{*};T).
 \ee
In this decomposition, $\delta^{(1)}_{\chi,\mathrm{th}}$ denotes the thermal fluctuation of dilaton particle in the vicinity of the critical temperature $T_{\mathrm{c}}$ which is independent of the Polyakov loop. In contrast, $\delta^{(2)}_{\chi,\mathrm{th}}$ represents the thermal fluctuation correlated with the Polyakov loop. The physical meaning of Eqs. (\ref{expecchi plus thfluc}) and (\ref{thermal fluc}) is clear: at zero temperature, $\delta^{(1)}_{\chi,\mathrm{th}}=\delta^{(2)}_{\chi,\mathrm{th}}=0$ and $\chi(T=0)=\tilde{f}_{\sigma}(T=0)=f_{\sigma}$. The system is in a stable state corresponding to the minimal value of the potential $\widetilde{\mathcal{V}}(\chi)$, specifically $\widetilde{\mathcal{V}}(\chi=f_{\sigma})=-m_{\sigma}^{2}f_{\sigma}^{2}/16$, indicating that the scale symmetry is spontaneously broken. When the system is heated, the random thermal motion of dilaton is dramatically excited and, as the temperature approaches the critical temperature $T_{\mathrm{c}}$, the gluonic DoF emerges. Therefore, the thermal fluctuation $\delta^{(1)}_{\chi,\mathrm{th}}$ contributes substantially at $T\sim T_{\mathrm{c}}$ and the Polyakov loop in $\delta^{(2)}_{\chi,\mathrm{th}}$ obtains a finite value after the deconfinement phase transition at rather high temperature.

Now, we can evaluate the thermodynamic potential based on the above discussions. The Hamiltonian reduces to the Coleman--Weinberg-type potential, i.e., $H_{\chi}=\int\mathrm{d}^{3}x\mathcal{H}_{\chi}=V\widetilde{\mathcal{V}}(\chi)$ with $V$ being the volume of the system. Then the partition function can be calculated by utilizing the eigenvectors of the configuration of $\chi$ as
\be
\label{parti func chi}
Z_{\chi} & = & \mathrm{Tr}e^{-\hat{H}_{\chi}/T}=\int[\mathrm{D}\chi(T)]\langle\chi;T|e^{-V\widetilde{\mathcal{V}}(\hat{\chi})/T}|\chi;T\rangle \nm \\
& = & \int[\mathrm{D}\chi(T)]\langle\chi;T|\chi;T\rangle e^{-V\widetilde{\mathcal{V}}(\chi)/T} .
\ee
Note that since the dilaton field is static and homogeneous, the time symbol $t$ in $|\chi,t;T\rangle$ is omitted. By using Eq.~(\ref{ortho chi}), we have 
\be
\label{parti func chi}
Z_{\chi} & = & Z_{\chi,\mathrm{div}}\overline{Z}_{\chi},
\ee
where
\be
Z_{\chi,\mathrm{div}} & = & \left[\prod_{\bm{x}\in\mathbb{R}^{3}}\delta(0)\right], \nm\\
\overline{Z}_{\chi} & = & \int[\mathrm{D}\chi(T)] e^{-V\widetilde{\mathcal{V}}(\chi)/T},
\ee
with $Z_{\chi,\mathrm{div}}$ being a divergent constant factor. Then, the thermodynamic potential $\Omega_\chi$ becomes
\be
\label{Omega chi}
\Omega_{\chi} & = &{} -\frac{T\ln Z_{\chi}}{V} = 
\overline{\Omega}_{\chi}+\Omega_{\chi,\mathrm{div}} ,
\ee
with 
\be
\overline{\Omega}_{\chi} & = &{} \frac{-T\ln\overline{Z}_{\chi}}{V},\nm\\
\Omega_{\chi,\mathrm{div}} & = &{} \frac{-T\ln Z_{\chi,\mathrm{div}}}{V}.
\ee
Consequently, one can express the entropy density as
\be
\label{entropy chi}
s_{\chi}=-\frac{\partial\Omega_{\chi}}{\partial T}=-\frac{\partial\overline{\Omega}_{\chi}}{\partial T}-\frac{\partial\Omega_{\chi,\mathrm{div}}}{\partial T}~.
\ee
The second term in the right hand side gives a temperature-independent and divergent contribution to the entropy density. This divergent term itself is unphysical since a temperature-independent value of the entropy cannot enter into the entropy change of a thermodynamic process. For this reason, we subtract $\Omega_{\chi,\mathrm{div}}$ in Eq.~(\ref{Omega chi}).

Finally, we calculate the functional integral $\overline{Z}_{\chi}$. Since it is difficult to complete the integral straightforwardly, we apply the stationary phase approximation (SPA)~\cite{Hell:2009by} which yields 
 \be\label{SPA parti func}
 \overline{Z}_{\chi}\approx\mathcal{N}e^{-V\widetilde{\mathcal{V}}(\overline{\chi})/T},
 \ee
 where $\mathcal{N}$ is a constant and $\overline{\chi}$ is determined by the stationary point of the exponential factor. Then, the thermodynamic potential $\Omega_{\chi}$ becomes
 \be
 \label{therm potential SPA}
 \Omega_{\chi} & = & \overline{\Omega}_{\chi}=\widetilde{\mathcal{V}}(\overline{\chi})+\frac{-T}{V}\ln\mathcal{N}.
 \ee
The second term of $\overline{\Omega}_{\chi}$ can also be subtracted since it contributes to a constant entropy density. The value of $\overline{\chi}$ is given by
 \be\label{chi bar}
 \overline{\chi}=\tilde{f}_{\sigma}+\delta^{(1)}_{\chi,\mathrm{th}}(T)+\delta^{(2)}_{\chi,\mathrm{th}}(\langle\!\langle\hat{\Phi}\rangle\!\rangle,\langle\!\langle\hat{\Phi}\rangle\!\rangle^{*};T),
 \ee
 where the expectation value of the Polyakov loop is the solution of
 \be\label{eom Phi}
 \left.\frac{\partial\Omega_{\chi}}{\partial\Phi}\right|_{\Phi=\langle\!\langle\hat{\Phi}\rangle\!\rangle}=0~, \quad \left.\frac{\partial\Omega_{\chi}}{\partial\Phi^{*}}\right|_{\Phi^{*}=\langle\!\langle\hat{\Phi}\rangle\!\rangle^{*}}=0~.
 \ee
 
Now, the key issue that we need to address is the construction of an explicit expression of $\delta^{(1)}_{\chi,\mathrm{th}}$ and $\delta^{(2)}_{\chi,\mathrm{th}}$ with respect to $T$, $\Phi$ and $\Phi^{*}$. In previous work~\cite{Sheng:2023rnd}, we utilized the MFA and proposed the relation between the dilaton field and the Polyakov loop of the form~\footnote{In Ref.~\cite{Sheng:2023rnd}, we used the symbols $\langle\chi\rangle$ and $\langle\Phi\rangle$ to represent the expectation values of the dilaton field and the Polyakov loop, respectively. Here, we scrupulously represent them as $\langle\!\langle\hat{\chi}\rangle\!\rangle$ and $\langle\!\langle\hat{\Phi}\rangle\!\rangle$}
\begin{eqnarray}
\frac{\langle\!\langle\hat{\chi}\rangle\!\rangle}{f_{\sigma}} & = & 1-a\langle\!\langle\hat{\Phi}\rangle\!\rangle^{*}\langle\!\langle\hat{\Phi}\rangle\!\rangle - \frac{b}{2}\left(\langle\!\langle\hat{\Phi}\rangle\!\rangle^{*3}+\langle\!\langle\hat{\Phi}\rangle\!\rangle^{3}\right) \nonumber \\
& &{} - c\left(\langle\!\langle\hat{\Phi}\rangle\!\rangle^{*}\langle\!\langle\hat{\Phi}\rangle\!\rangle\right)^{2}~,
\label{eq:mfa}
\end{eqnarray}
where $f_{\sigma}$ was fixed at zero temperature.
Exactly speaking, this ansatz based on MFA has a theoretical flaw when we consider the scale transformation of Eq.~(\ref{eq:mfa}). Once we go beyond the MFA, we have to address whether Eq.~(\ref{eq:mfa}) has scale covariance? Actually, we only know the scale transformation property of dilaton field at this moment, therefore, if we wonder how the Polyakov loop transforms under scale transformation we have to investigate how the temperature $T$---a quantity that the polyakov loop explicitly depends on---transforms under scale transformation.

\section{Scale transformations of temperature and chemical potentials}
\label{sec:ScalTran}

In this section, we discuss the scale transformation properties of temperature $T$ and chemical potentials $\mu_{i}, i=1,2,\cdots,K$ corresponding to conserved charges $\hat{Q}_{i}, i=1,2,\cdots,K$ which commute with each other. For this purpose, we first discuss how the Liouville equation transforms under scale transformation.

\subsection{Scale covariance of the Liouville equation}

We consider the fundamental equation of quantum statistical mechanics---the Liouville equation
\be\label{Liouville eq}
\frac{\mathrm{d}\hat{\rho}(t)}{\mathrm{d}t}=0
\ee
where $\hat{\rho}(t)$ is the density operator. Note that throughout this paper, for the sake of clarity, the time $t$ of all conserved dynamical variables, such as the density operator $\hat{\rho}=\hat{\rho}(t)$ and Hamiltonian $\hat{H}=\hat{H}(t)$, is kept. When the density operator has its classical counterpart, namely, it is a functional of fields $\hat{\phi}^{r}(\bm{x},t)$ and densities of canonical momenta $\hat{\pi}_{r}(\bm{x},t)$, i.e., $\hat{\rho}(t)=\rho[\hat{\phi}^{r}_{(t)},\hat{\pi}_{r,(t)};t]$, the Liouville equation is rewritten as
\be
\label{Liouville eq(ensemble)}
\frac{\mathrm{d}\hat{\rho}(t)}{\mathrm{d}t} & = & \frac{\partial \rho[\hat{\phi}^{r}_{(t)},\hat{\pi}_{r,(t)};t]}{\partial t} \nm\\
& &{} +\frac{1}{i}\Big[\rho[\hat{\phi}^{r}_{(t)},\hat{\pi}_{r,(t)};t],H[\hat{\phi}^{r}_{(t)},\hat{\pi}_{r,(t)}]\Big] \nm \\
& = & 0~.
\ee
Here, the index $r$ denotes the intrinsic DoFs of the field $\phi^{r}(\bm{x},t)$ such as spin, isospin, and so on. It is important to note that for complex fields, we regard their real parts and imaginary parts as independent variables, consequently, the index $r$ contains both the real and imaginary parts. As a result, both $\phi^{r}(\bm{x},t)$ and $\pi_{r}(\bm{x},t)$ are considered real variables. For more details of the Liouville equation in the context of field theory and conventions and notations, see Appendix~\ref{app:Liouville}.

When a thermodynamic system has scale symmetry, the Liouville equation describing the dynamics of the ensemble for this system must be covariant under scale transformation $x'^{\mu}=\lambda x^{\mu}=e^{a}x^{\mu},a\in\mathbb{R}$. We next argue that, due to this covariance,  the density operator is a scale-invariant object, namely $\hat{\rho}'(t')=\hat{\rho}(t)$ or equivalently $\rho[\hat{\phi}'^{r}_{(t')},\hat{\pi}'_{r,(t')};t']=\rho[\hat{\phi}^{r}_{(t)},\hat{\pi}_{r,(t)};t]$.

It should be emphasized that, in general, the density operator involves some parameters $\beta_{j}, j=1,2,\cdots,L$ in addition to time $t$. A typical example is the density operator of a grand canonical ensemble, in which the parameters are temperature $T$ and chemical potentials $\mu_{i}$. For the sake of clarity, we restore these suppressed parameters in $\hat{\rho}(t)$ such that $\hat{\rho}(t) = \hat{\rho}(t,\beta_{1},\cdots,\beta_{L})=\rho[\hat{\phi}^{r}_{(t)},\hat{\pi}_{r,(t)};t,\beta_{1},\cdots,\beta_{L}]$. Since these parameters may have certain scale dimensions, the scale invariance of the density operator is rewritten as $\hat{\rho}'(t',\beta'_{1},\cdots,\beta'_{L})=\hat{\rho}(t,\beta_{1},\cdots,\beta_{L})$, or equivalently $\rho[\hat{\phi}'^{r}_{(t')},\hat{\pi}'_{r,(t')};t',\beta'_{1},\cdots,\beta'_{L}]=\rho[\hat{\phi}^{r}_{(t)},\hat{\pi}_{r,(t)};t,\beta_{1},\cdots,\beta_{L}]$. The Liouville equation of the form (\ref{Liouville eq(ensemble)}) can be rewritten as
\be
\label{Liouville eq(ensemble) t beta}
\frac{\partial \hat{\rho}(t,\beta_{1},\cdots,\beta_{L})}{\partial t}+\frac{1}{i}\Big[\hat{\rho}(t,\beta_{1},\cdots,\beta_{L}),\hat{H}(t)\Big]=0.
\ee

With the above convention and notation, we can write down the following scale transformation of the density operator
\be\label{scale trans rho}
\hat{\rho}'(t',\beta'_{1},\cdots,\beta'_{L})&=\hat{D}(t',a)\hat{\rho}(t',\beta'_{1},\cdots,\beta'_{L})[\hat{D}(t',a)]^{\dagger} \nm \\
&\equiv\mathscr{D}_{\rho}\big(\hat{\rho}(t,\beta_{1},\cdots,\beta_{L})\big),
\ee
where, as shown in Appendix \ref{app:ScaleHeisenberg}, the scale transformation operator $\hat{D}(t,a)$ takes the form $\hat{D}(t,a)=\exp[-ia\hat{Q}_{{\ssp\mathrm{D}}}(t)]$ with $\hat{Q}_{{\ssp\mathrm{D}}}(t)$ being the conserved charge of scale symmetry. It should be noted that $\hat{D}(t,a)$ has an explicit time dependence due to the explicit time dependence of $\hat{Q}_{{\ssp\mathrm{D}}}(t)$, and $\mathscr{D}_{\rho}$ denotes the scale transformation of the density operator, which is the same as the distribution probability $\rho(t,\beta_{1},\cdots,\beta_{L})$ of a phase point in the phase space of classical theory, i.e., $\rho'(t',\beta'_{1},\cdots,\beta'_{L})=\mathscr{D}_{\rho}\big(\rho(t,\beta_{1},\cdots,\beta_{L})\big)$. 

Next, we need to confirm that the above transformed density operator (\ref{scale trans rho}) is indeed a solution of Eq.~(\ref{Liouville eq(ensemble) t beta}), or in other words, the Liouville equation is scale-covariant under the scale transformation (\ref{scale trans rho}) of the density operator. We first replace time $t$ and the parameters $(\beta_{1},\cdots,\beta_{L})$ in Eq.~(\ref{Liouville eq(ensemble) t beta}) with the scale-transformed ones and obtain
\be
\label{Liouville eq(ensemble) t' beta'}
\frac{\partial\hat{\rho}(t',\beta'_{1},\cdots,\beta'_{L})}{\partial t'}+\frac{1}{i}\Big[\hat{\rho}(t',\beta'_{1},\cdots,\beta'_{L}),\hat{H}(t')\Big]=0,
\ee
substituting Eq.~(\ref{scale trans rho}) into Eq.~(\ref{Liouville eq(ensemble) t' beta'}), we have
\be
\label{scale trans Liouville eq(ensemble)3}
& & \left\{\frac{\partial\hat{D}(t',a)}{\partial t'}[\hat{D}(t',a)]^{\dagger}\right\}^{\dagger}\hat{\rho}'(t',\beta'_{1},\cdots,\beta'_{L}) \nm\\
& & {} +\frac{\partial\hat{\rho}'(t',\beta'_{1},\cdots,\beta'_{L})}{\partial t'} \nm\\
& & {} +\hat{\rho}'(t',\beta'_{1},\cdots,\beta'_{L})\frac{\partial \hat{D}(t',a)}{\partial t'}[\hat{D}(t',a)]^{\dagger} \nm \\
& & {} +\frac{1}{i}\Big[\hat{\rho}'(t',\beta'_{1},\cdots,\beta'_{L}),\hat{D}(t',a)\hat{H}(t')[\hat{D}(t',a)]^{\dagger}\Big]=0.\nm\\
\ee	
For a system with scale symmetry, $\hat{Q}_{{\ssp\mathrm{D}}}(t)$ is a conserved charge operator, i.e., $\mathrm{d}\hat{Q}_{{\ssp\mathrm{D}}}(t)/\mathrm{d}t=0$, that is,  
\be
\label{dD over dt}
\frac{\mathrm{d}\hat{D}(t',a)}{\mathrm{d}t'}=\frac{1}{i}\Big[\hat{D}(t',a),\hat{H}(t')\Big]+\frac{\partial \hat{D}(t',a)}{\partial t'}=0.
\ee 	 
Then, we can obtain
\be
\label{pD over pt Ddagger}
\frac{\partial\hat{D}(t',a)}{\partial t'}[\hat{D}(t',a)]^{\dagger} & = & i\Big[\hat{D}(t',a),\hat{H}(t')\Big][\hat{D}(t',a)]^{\dagger} \nm \\
& = & i\bigg(\hat{H}'(t')-\hat{H}(t')\bigg).
\ee
Substituting Eq.~(\ref{pD over pt Ddagger}) into Eq.~(\ref{scale trans Liouville eq(ensemble)3}), we finally obtain
\be
\label{scale trans Liouville eq(ensemble) final}
\frac{\partial\hat{\rho}'(t',\beta'_{1},\cdots,\beta'_{L})}{\partial t'}+\dfrac{1}{i}\Big[\hat{\rho}'(t',\beta'_{1},\cdots,\beta'_{L}),\hat{H}(t')\Big]=0.
\nm\\
\ee

Comparing Eq.~(\ref{scale trans Liouville eq(ensemble) final}) and Eq.~(\ref{Liouville eq(ensemble) t' beta'}), we find that after transformation~(\ref{scale trans rho}), the time evolution of $\hat{\rho}'(t',\beta'_{1},\cdots,\beta'_{L})$ is totally the same as that of $\hat{\rho}(t',\beta'_{1},\cdots,\beta'_{L})$ since the evolution rules of both $\hat{\rho}'(t',\beta'_{1},\cdots,\beta'_{L})$ and $\hat{\rho}(t',\beta'_{1},\cdots,\beta'_{L})$ are dominated by the same Hamiltonian $\hat{H}(t')$. Consequently, we say that for a system with scale symmetry, the Liouville equation (\ref{Liouville eq(ensemble) t' beta'}) or (\ref{Liouville eq(ensemble) t beta}) is scale-covariant.

The next we should know mapping $\mathscr{D}_{\rho}$ in Eq.~(\ref{scale trans rho}). This issue can be addressed by the scale covariance of the Liouville equation. The mapping $\mathscr{D}_{\rho}$ actually provides the relation at the same spacetime point between $\hat{\rho}(t,\beta_{1},\cdots,\beta_{L})$ and $\hat{\rho}'(t',\beta'_{1},\cdots,\beta'_{L})$ and both of them actually describe the same mixed state of a system at that spacetime point. 
For this reason, if a system has scale symmetry, the covariance of the Liouville equation yields
\be
\label{Liouville eq(ensemble) totally primed}
\frac{\partial\hat{\rho}'(t',\beta'_{1},\cdots,\beta'_{L})}{\partial t'}+\frac{1}{i}\Big[\hat{\rho}'(t',\beta'_{1},\cdots,\beta'_{L}),\hat{H}'(t')\Big]=0.
\nm\\
\ee
Because of $x^{\prime \mu}=\lambda x^{\mu}$ and $\hat{H}'(t')=\lambda^{-1}\hat{H}(t)$, we have
\be
\label{Liouville eq(ensemble) totally primed 3}
\frac{\partial\hat{\rho}'(t',\beta'_{1},\cdots,\beta'_{L})}{\partial t}+\frac{1}{i}\Big[\hat{\rho}'(t',\beta'_{1},\cdots,\beta'_{L}),\hat{H}(t)\Big]=0. \nm\\
\ee
Comparing Eq.~(\ref{Liouville eq(ensemble) totally primed 3}) and Eq.~(\ref{Liouville eq(ensemble) t beta}), we obtain
\be
\label{rho prime eq C rho}
\hat{\rho}'(t',\beta'_{1},\cdots,\beta'_{L})=C\hat{\rho}(t,\beta_{1},\cdots,\beta_{L}),
\ee
where $C$ is a constant which must be unity due to $\mathrm{Tr}[\hat{\rho}'(t',\beta'_{1},\cdots,\beta'_{L})]=\mathrm{Tr}[\hat{\rho}(t,\beta_{1},\cdots,\beta_{L})]=1$. Consequently, $\mathscr{D}_{\rho}$ is an identity mapping.

\subsection{Scale transformations of temperature, chemical potentials and the Polyakov loop}

Now, we are ready to derive the scale transformation properties of temperature and chemical potentials. We recall the density operator of a grand canonical ensemble
\be
& & \hat{\rho}(t,T,V,\mu_{1},\cdots,\mu_{K}) \nonumber\\
& &  =  \frac{1}{Z}\exp\left\{-\left[\hat{H}(t)-\sum_{i=1}^{K}\mu_{i}\hat{Q}_{i}(t)\right]/T\right\},
\ee
with the grand canonical partition function 
\be
\displaystyle Z & \equiv & Z(t,T,V,\mu_{1},\cdots,\mu_{K}) \nonumber\\
& = & \mathrm{Tr}\exp\left\{-\left[\hat{H}(t)-\sum_{i=1}^{K}\mu_{i}\hat{Q}_{i}(t)\right]/T\right\}.
\ee
It should be noted that for a system at thermal equilibrium, the density operator of a grand canonical ensemble is not explicitly time dependent, i.e. $\partial\hat{\rho}(t,T,V,\mu_{1},\cdots,\mu_{K})/\partial t=0$ and is a solution of the Liouville equation (\ref{Liouville eq(ensemble) t beta}). Here, we retain the time variable $t$ in the density operator to ensure the uniformity of the conventions and notation.

Using the scale transformation Eq.~(\ref{scale trans rho}), we have
\be
\label{scale trans grand-rho}
& &\hat{\rho}(t,T,V,\mu_{1},\cdots,\mu_{K}) \nm \\
& &\qquad{} =\hat{D}(t',a)\hat{\rho}(t',T',V',\mu'_{1},\cdots,\mu'_{K})[\hat{D}(t',a)]^{\dagger}.
\nm\\
\ee
That is
\begin{widetext}
\be
\label{scale trans grand-rho 1}
\frac{\exp\bigg\{-\Big[\hat{H}(t)-\sum\limits_{i=1}^{K}\mu_{i}\hat{Q}_{i}(t)\Big]/T\bigg\}}{Z(t,T,V,\mu_{1},\cdots,\mu_{K})} & = & \hat{D}(t',a)\frac{\exp\bigg\{-\Big[\hat{H}(t')-\sum\limits_{i=1}^{K}\mu'_{i}\hat{Q}_{i}(t')\Big]/T'\bigg\}}{Z(t',T',V',\mu'_{1},\cdots,\mu'_{K})}[\hat{D}(t',a)]^{\dagger} \nm \\
& = & \frac{\exp\bigg\{-\Big[\hat{H}'(t')-\sum\limits_{i=1}^{K}\mu'_{i}\hat{Q}'_{i}(t')\Big]/T'\bigg\}}{Z(t',T',V',\mu'_{1},\cdots,\mu'_{K})}.
\ee
And, the partition function $Z(t',T',V',\mu'_{1},\cdots,\mu'_{K})$ can be written explicitly as
\be
\label{PF prime variables}
Z(t',T',V',\mu'_{1},\cdots,\mu'_{K}) & = & \mathrm{Tr}\Bigg(\exp\bigg\{-\Big[\hat{H}(t')-\sum\limits_{i=1}^{K}\mu'_{i}\hat{Q}_{i}(t')\Big]/T'\bigg\}\Bigg) \nm \\
& = & \mathrm{Tr}\Bigg([\hat{D}(t',a)]^{\dagger}\hat{D}(t',a)\exp\bigg\{-\Big[\hat{H}(t')-\sum\limits_{i=1}^{K}\mu'_{i}\hat{Q}_{i}(t')\Big]/T'\bigg\}\Bigg) \nm \\
& = & \mathrm{Tr}\Bigg(\hat{D}(t',a)\exp\bigg\{-\Big[\hat{H}(t')-\sum\limits_{i=1}^{K}\mu'_{i}\hat{Q}_{i}(t')\Big]/T'\bigg\}[\hat{D}(t',a)]^{\dagger}\Bigg) \nm \\
& = & \mathrm{Tr}\Bigg(\exp\bigg\{-\Big[\hat{H}'(t')-\sum\limits_{i=1}^{K}\mu'_{i}\hat{Q}'_{i}(t')\Big]/T'\bigg\}\Bigg) \nm \\
& = & Z'(t',T',V',\mu'_{1},\cdots,\mu'_{K}).
\ee
Substituting Eq.~(\ref{PF prime variables}) into Eq.~(\ref{scale trans grand-rho 1}) and using the scale transformations of the Hamiltonian and the conserved charges illustrated in Appendix~\ref{app:ScaleHeisenberg} $\hat{H}'(t')=\lambda^{-1}\hat{H}(t)$ and $\hat{Q}'_{i}(t')=\hat{Q}_{i}(t)$, we have
\be
\label{Z prime over Z}
\frac{Z'(t',T',V',\mu'_{1},\cdots,\mu'_{K})}{Z(t,T,V,\mu_{1},\cdots,\mu_{K})}\hat{I} 
& = & \exp\bigg\{\Big[\frac{1}{T}-\dfrac{\lambda^{-1}}{T'}\Big]\hat{H}(t)+\sum\limits_{i=1}^{K}\Big[\frac{\mu'_{i}}{T'}-\dfrac{\mu_{i}}{T}\Big]\hat{Q}_{i}(t)\bigg\},
\ee	
\end{widetext}	
considering $\Big[\hat{H}(t),\hat{Q}_{i}(t)\Big]=0$ and $\Big[\hat{Q}_{i}(t),\hat{Q}_{j}(t)\Big]=0$, for $i,j=1,\cdots,K$. 

For the Hamiltonian and the conserved charges of a scale symmetric system, the right hand side of Eq.~(\ref{Z prime over Z}) is proportional to the identity operator of Hilbert space.
This yields relations
\be
\label{scale trans T mu}
T'=\lambda^{-1}T, \quad \mu'_{i}=\lambda^{-1}\mu_{i},~i=1,\cdots,K~,
\ee
which are the scale transformation properties  of temperature and chemical potentials. Meanwhile, we find that the partition function is a scale invariant object, i.e., $Z'(t',T',V',\mu'_{1},\cdots,\mu'_{K})=Z(t,T,V,\mu_{1},\cdots,\mu_{K})$. 

Based on the above argument, we conclude that the Polyakov loop is scale invariant. This conclusion can be arrived at by considering the scale transformed Polyakov loop
\be
\label{Polyakov loop prime}
\Phi'(\bm{x}') & = & \frac{1}{N_{\mathrm{c}}}\mathrm{tr_{c}}\Bigg\{\mathcal{P}\exp\bigg[-ig_{\mathrm{s}}\int_{0}^{\beta'}\mathrm{d}x'_{4}A'^{a}_{4}(\bm{x}',x'_{4})T^{a}\bigg]\Bigg\}.
\nm\\
\ee
Utilizing the scale transformation properties $x'_{4}=\lambda x_{4}$, $A'^{a}_{4}(\bm{x}',x'_{4})=\lambda^{-1}A^{a}_{4}(\bm{x},x_{4})$ and $\beta'=1/T'=\lambda/T=\lambda\beta$, one can easily obtain
\be\label{Polyakov loop scale invariant}
\Phi'(\bm{x}')  
& = & \Phi(\bm{x}),
\ee
which means that the Polyakov loop is scale invariant. This conclusion is the key point of connecting the dilatonic thermal fluctuations with the Polyakov loop. 

Additionally, one can derive the scale transformation properties of other thermodynamic quantities such as pressure, entropy, etc. and argue that the four laws of thermodynamics are scale covariant. This indicates that the discussion in the work is self-consistent. We leave the details of the argument in Appendix~\ref{app:4laws}.

\subsection{Construction of EoS}

Based on the above discussion, we now show that for any scale-symmetric system, the equation of state can be written down straightforwardly instead of calculating the partition function based on a certain model or theory. In other words, we can find the basic building blocks of a scale-symmetric EoS and construct it directly. This process is similar to that of constructing a microscopic Lagrangian with certain symmetries as long as we know the symmetry transformation rules of fields.

Generally speaking, EoSs are functions of temperature and chemical potentials, i.e. $P=f_{P}(T,\mu_{1},\cdots,\mu_{K})$. For any scale symmetric thermodynamic system, the EoS itself should be covariant under scale transformations of temperature, chemical potentials, and pressure, namely
\be
\label{scale trans EOS}
P^\prime =f_{P}(T',\mu'_{1},\cdots,\mu'_{K}) \to \lambda^{-4}P.
\ee 
Then, the function $f_{P}$ must have the following property
\be
& & f_{P}\left(\lambda^{-1}T,\lambda^{-1}\mu_{1},\cdots,\lambda^{-1}\mu_{K}\right) \nm\\
& & \qquad\qquad{} \to 
\lambda^{-4}f_{P}(T,\mu_{1},\cdots,\mu_{K}).
\label{fP property}
\ee
By using this constraint, we can write down the scale symmetric terms of the EoS for a thermodynamic system. For simplicity, we consider a system that has only one component of matter, or equivalently, has only one kind of conserved charge $\hat{Q}$ as a particle number variable, i.e. $P=f_{P}(T,\mu)$. Therefore, the EoS can be generally expressed as
\be\label{scale sym terms P}
P=\mathcal{C}_{1}T^{4}+\mathcal{C}_{2}\mu^{4}+\mathcal{C}_{3}T^{2}\mu^{2}+\mathcal{C}_{4}T^{3}\mu+\mathcal{C}_{5}T\mu^{3}+\cdots, 
\nm\\
\ee
where $\mathcal{C}_{1},\mathcal{C}_{2},\cdots$ are coefficients determined by the specified microscopic model and $\cdots$ stands for the other scale-covariant terms.

A typical example is the Stefan-Boltzmann law which states that when massless particles dominate the DoFs in a thermodynamic system, the pressure scales as $P\propto T^{4}$ at high temperatures or $P\propto\mu_{\mathrm{B}}^{4}$ at high baryon densities~\cite{Fujimoto:2022ohj}. Obviously, $P\propto T^{4}$ and $P\propto \mu_{\mathrm{B}}^{4}$ are simply two special cases of Eq.~(\ref{scale sym terms P}) at, respectively, high temperature and chemical potential limit. 

Another example is the pressure of a non-interacting fermion gas with a small fermion mass at high temperature and low chemical potential. This is expressed as~\cite{Kapusta_Gale_2023}
\be\label{non int FG P}
P=\dfrac{7\pi^{2}}{180}T^{4}+\dfrac{(2\mu^{2}-m^{2})T^{2}}{12}+\cdots.
\ee
If one were taken the massless limit $m\to 0$, it is transparent that the term $T^{2}\mu^{2}$ emerges in addition to the term $T^{4}$.

Consider the Euler theorem for a homogeneous function $f_{m}^{(\alpha_{1},\cdots,\alpha_{n})}(x_{1},\cdots,x_{n})$ with property
\be
\label{homogeneous func}
& & f_{m}^{(\alpha_{1},\cdots,\alpha_{n})}(\lambda^{\alpha_{1}}x_{1},\cdots,\lambda^{\alpha_{n}}x_{n}) \nm\\
& & \qquad\qquad{} = \lambda^{m} f_{m}^{(\alpha_{1},\cdots,\alpha_{n})}(x_{1},\cdots,x_{n}),
\ee 
it satisfies the differential equation of the form
\be
\label{PDE of homogeneous func}
& & \sum_{i=1}^{n}\!\!\alpha_{i}x_{i}\frac{\partial f_{m}^{(\alpha_{1},\cdots,\alpha_{n})}(x_{1},\cdots,x_{n})}{\partial x_{i}}\nm\\
& & \qquad\qquad {} = m f_{m}^{(\alpha_{1},\cdots,\alpha_{n})}(x_{1},\cdots,x_{n}).
\ee
Then, from Eq.~(\ref{fP property}) one has
\be
\label{PDE of scale symm P}
-\left(T\frac{\partial P}{\partial T}+\sum_{i=1}^{K}\mu_{i}\frac{\partial P}{\partial \mu_{i}}\right)={}-4P,
\ee
which can be rewritten as
\be
\label{PDE of scale symm LogP}
T\frac{\partial\left(\frac{1}{4}\ln P\right)}{\partial T}+\sum_{i=1}^{K}\mu_{i}\frac{\partial\left(\frac{1}{4}\ln P\right)}{\partial \mu_{i}}=1.
\ee
Introducing the homogeneous functions $f_{P,(m)}(T,\mu_{1},\cdots,\mu_{K}),m=1,2,\cdots$ with property $f_{P,(m)}(\lambda T,\lambda \mu_{1},\cdots,\lambda \mu_{K})=\lambda^{m}f_{P,(m)}(T,\mu_{1},\cdots,\mu_{K})$, considering Eq.~(\ref{PDE of homogeneous func}), we have equations
\be
\label{PDF of fPm}
T\dfrac{\partial\left(\dfrac{1}{m}\ln f_{P,(m)}\right)}{\partial T}+\sum_{i=1}^{K}\mu_{i}\dfrac{\partial\left(\dfrac{1}{m}\ln f_{P,(m)}\right)}{\partial \mu_{i}}=1.
\ee
Comparing Eq.~(\ref{PDE of scale symm LogP}) with Eq.~(\ref{PDF of fPm}), one can obtain the solution
\be\label{solution scale symm P}
\dfrac{1}{4}\ln P=\sum_{m}\dfrac{\mathfrak{c}_{m}}{m}\ln f_{P,(m)}+\dfrac{1}{4}\ln P_{0},
\ee
where $\mathfrak{c}_{m},m=1,2,\cdots$ and $P_{0}$ are constants and $\displaystyle \sum_{m}\mathfrak{c}_{m}=1$. Namely, we find a solution of Eq.~(\ref{PDE of scale symm P}) with form
\be\label{solution scale symm P final}
P=P_{0}\exp\bigg[4\sum_{m}\dfrac{\mathfrak{c}_{m}}{m}\ln f_{P,(m)}(T,\mu_{1},\cdots,\mu_{K})\bigg].
\ee

On the other hand, Eq.~(\ref{PDE of scale symm P}) originated from the scale covariance of the EoS can indeed give the traceless of energy-momentum tensor $\epsilon-3P=0$. Considering the thermodynamic equation
\be
& & \mathrm{d}P=s\mathrm{d}T+\sum_{i=1}^{K}n_{i}\mathrm{d}\mu_{i},\label{differential P}
\ee
with entropy density $s$ and particle number densities $n_{i}$, Eq.~(\ref{PDE of scale symm P}) yields
\be
\label{PDE of scale symm P rew}
-4P ={} -Ts-\sum_{i=1}^{K}\mu_{i}n_{i}.
\ee
In combination with the thermodynamic equation
\be
& & \epsilon+P=Ts+\sum_{i=1}^{K}\mu_{i}n_{i},\label{energy density plus P}
\ee
we finally have $\epsilon-3P=0$. 

Before concluding this section, we would like to emphasize that for any scale-symmetric system, the partition function itself must satisfy a special differential equation. Let us recall Eq.~(\ref{PF prime variables}) and the scale invariance of the partition function, i.e. $Z'(t',T',V',\mu'_{1},\cdots,\mu'_{K})=Z(t,T,V,\mu_{1},\cdots,\mu_{K})$. We are immediately conscious of the property
\be
\label{prop of scale symm Z}
& &Z(\lambda t,\lambda^{-1}T,\lambda^{3}V,\lambda^{-1}\mu_{1},\cdots,\lambda^{-1}\mu_{K}) \nm \\
& & \qquad\qquad\qquad\qquad {} = Z(t,T,V,\mu_{1},\cdots,\mu_{K}),
\ee
which yields the following differential equation of the partition function due to Eq.~(\ref{PDE of homogeneous func}):
\be
\label{PDE of scale symm Z}
t\frac{\partial Z}{\partial t}-T\frac{\partial Z}{\partial T}+3V\frac{\partial Z}{\partial V}-\sum_{i=1}^{K}\mu_{i}\frac{\partial Z}{\partial \mu_{i}}=0.
\ee
As for equilibrium states, $\partial Z/\partial t=0$ and utilizing the relation between the pressure and the partition function $Z=\exp(PV/T)$, we obtain the following result based on the above equation:
\be
\label{PDE of P from PDE of Z}
TZ\left(-\frac{PV}{T^{2}}+\frac{V}{T}\frac{\partial P}{\partial T}\right)-3VZ\frac{P}{T}+\sum_{i=1}^{K}\mu_{i}Z\frac{V}{T}\frac{\partial P}{\partial \mu_{i}}=0,
\nm\\
\ee
which can be simplified to
\be
\label{PDE of P from PDE of Z sim}
\frac{ZV}{T}\left(4P-T\frac{\partial P}{\partial T}-\sum_{i=1}^{K}\mu_{i}\frac{\partial P}{\partial \mu_{i}}\right)=0.
\ee
And the nonzero value of $ZV/T$ gives rise to Eq.~(\ref{PDE of scale symm P}) meaning that the logic is self-consistent.

\section{Expressions of dilaton fluctuations}

\label{sec:ExpDilaton}

In this section, we shall construct the possible expressions of $\delta^{(1)}_{\chi,\mathrm{th}}(T)$ and $\delta^{(2)}_{\chi,\mathrm{th}}(\Phi,\Phi^{*};T)$ based on the scale transformation of temperature Eq.~(\ref{scale trans T mu}). 

The Polyakov loop has a non-trivial $\mathbb{Z}_{N_{\mathrm{c}}}$ transformation behavior under the gauge transformation of gluon field, that is~\cite{book:94872761}
\be
\label{center sym trans}
\Phi'(\bm{x})=\exp\left[\dfrac{2\pi i k}{N_{\mathrm{c}}}\right]\Phi(\bm{x})~, \quad k=1,2,\cdots,N_{\mathrm{c}}~,
\ee
which arises from the non-trivial periodic boundary condition of gluon field at finite temperatures, i.e. $A^{a}_{\mu}(\bm{x},x_{4}+\beta)=A^{a}_{\mu}(\bm{x},x_{4}),a=1,2,\cdots,N_{\mathrm{c}}^{2}-1$~\cite{book:94872761}. Therefore, $\delta^{(2)}_{\chi,\mathrm{th}}(\Phi,\Phi^{*};T)$ must be a $\mathbb{Z}_{N_{\mathrm{c}}}$-symmetric function. Then, we can expand it as
\be
\label{delta2 expa}
\frac{\delta^{(2)}_{\chi,\mathrm{th}}}{\tilde{f}_{\sigma}} & = & \sum_{k=0}^{+\infty}\sum_{k'=0}^{+\infty}\frac{c_{kk'}(T)}{2}\Big[\varphi^{k}(\varphi^{*})^{k'}+(\varphi^{*})^{k}\varphi^{k'}\Big].
\ee 
Here, for simplicity, we denote the $\mathbb{Z}_{N_{\mathrm{c}}}$ invariant $\Phi^{N_{\mathrm{c}}}$ as $\varphi\equiv\Phi^{N_{\mathrm{c}}}$. 

In general, in $\delta^{(2)}_{\chi,\mathrm{th}}(\Phi,\Phi^{*};T)$, there are numerous combinations of the Polyakov loop that are $\mathbb{Z}_{N_{\mathrm{c}}}$-invariant and Hermitian, such as $\Phi^{*}\Phi,(\Phi^{*}\Phi)^{2},\Phi^{N_{\mathrm{c}}}+(\Phi^{N_{\mathrm{c}}})^{*},\cdots$. Some of them, like $\Phi^{*}\Phi$ and $(\Phi^{*}\Phi)^{2}$, are invariant under the $\mathrm{U(1)}$ transformation applied to $\Phi$, i.e. $\Phi'(\bm{x})=e^{i\theta}\Phi(\bm{x})$. Considering that the deconfinement phase transition is associated with the spontaneous breaking of the $\mathbb{Z}_{N_{\mathrm{c}}}$ symmetry at high temperatures, and recognizing that $\mathbb{Z}_{N_{\mathrm{c}}}$ is a subgroup of $\mathrm{U(1)}$, we consider that these $\mathrm{U(1)}$-invariant combinations, which exhibit a larger symmetry than $\mathbb{Z}_{N_{\mathrm{c}}}$, may not be the primary objects that can capture the essence of the center symmetry, specifically the $\mathbb{Z}_{N_{\mathrm{c}}}$ symmetry at high temperatures. Consequently, we only choose $\Phi^{N_{\mathrm{c}}}$ and $(\Phi^{N_{\mathrm{c}}})^{*}$ as the basic building blocks of $\delta^{(2)}_{\chi,\mathrm{th}}$. The coefficients in Eq.~(\ref{delta2 expa}) have the following property:
\be\label{property of ckk'}
c_{kk'}=
\begin{cases}
	c_{k'k},~&k\neq k' \\
	0,~&k=k'\neq 0	
\end{cases}	
.
\ee

We come back to the scale transformation $x'^{\mu}=\lambda x^{\mu},\lambda>0$. Due to the scale invariance of the Polyakov loop~(\ref{Polyakov loop scale invariant}), we have $\varphi'(\bm{x}')=\varphi(\bm{x})$. Since the thermal fluctuations of the dilaton field transform in the same way as $\chi'(\bm{x}',t';T')=\lambda^{-1}\chi(\bm{x},t;T)$, $\delta^{(1)}_{\chi,\mathrm{th}}(T')=\lambda^{-1}\delta^{(1)}_{\chi,\mathrm{th}}(T)$ and
\be
\label{scale trans delta2}
\delta^{(2)}_{\chi,\mathrm{th}}(\Phi'(\bm{x}'),[\Phi'(\bm{x}')]^{*};T') = \frac{1}{\lambda}\delta^{(2)}_{\chi,\mathrm{th}}(\Phi(\bm{x}),[\Phi(\bm{x})]^{*};T).
\nm \\
\ee
Considering the scale transformation of temperature~(\ref{scale trans T mu}), we obtain
\be
& & \delta^{(1)}_{\chi,\mathrm{th}}(\lambda^{-1}T)=\lambda^{-1}\delta^{(1)}_{\chi,\mathrm{th}}(T)~, \label{delta1 eq} \\ 
 & & c_{kk'}(\lambda^{-1}T)=\lambda^{-1}c_{kk'}(T)~. \label{ckk eq}
\ee
One can immediately be aware of $\delta^{(1)}_{\chi,\mathrm{th}}(T)\propto T$ and $c_{kk'}(T)\propto T$. More gingerly, $\delta^{(1)}_{\chi,\mathrm{th}}(T)$ and $c_{kk'}(T)$ are homogeneous functions of $T$. According to the Euler theorem for homogeneous functions, the differential equations of $\delta^{(1)}_{\chi,\mathrm{th}}(T)$ and $c_{kk'}(T)$ are
\be
& & T\dfrac{\mathrm{d}\delta^{(1)}_{\chi,\mathrm{th}}}{\mathrm{d}T}=\delta^{(1)}_{\chi,\mathrm{th}}, \label{ODE delta1} \\
& & T\dfrac{\mathrm{d}c_{kk'}}{\mathrm{d}T}=c_{kk'}, \label{ODE ckk}
\ee
which yield the solutions
\be
& & \delta^{(1)}_{\chi,\mathrm{th}}(T)=\bar{\delta}T, \label{sol ODE delta1} \\
& & c_{kk'}(T)=\bar{c}_{kk'}T, \label{sol ODE ckk}
\ee	
with $\bar{\delta}$ and $\bar{c}_{kk'}$ being constants.

\subsection{The next to leading order of $\delta_{\chi,\mathrm{th}}^{(2)}/\tilde{f}_{\sigma}$ with respect to $\Phi^{N_{\mathrm{c}}}$}
\label{subsec: NLO}

Let us consider the next to leading order (NLO) expansion of $\delta^{(2)}_{\chi,\mathrm{th}}(\Phi,\Phi^{*};T)/\tilde{f}_{\sigma}$, $\delta^{(2)}_{\chi,\mathrm{NLO}}/\tilde{f}_{\sigma}$. From Eq.~(\ref{delta2 expa}) we have
\be
\label{re expand delta chi Nc}
\frac{\delta^{(2)}_{\chi,\mathrm{NLO}}}{\tilde{f}_{\sigma}} 
& = & T\bar{c}_{00}\left[1+\dfrac{\bar{c}_{01}}{\bar{c}_{00}}\left(\varphi^{*}+\varphi\right)\right].
\ee
Since in the pure-gauge sector, $\varphi^{*}=\varphi$ due to $\Phi^{*}=\Phi$~\cite{Ratti:2005jh}, we can  rewrite Eq.~(\ref{re expand delta chi Nc}) as follows
\be
\label{re expand delta chi Nc pG}
\frac{\delta^{(2)}_{\chi,\mathrm{NLO}}}{\tilde{f}_{\sigma}}=\dfrac{T}{T_{0,N_{\mathrm{c}}}}\left(1-a_{N_{\mathrm{c}}}\varphi\right).
\ee
with conventions $T_{0,N_{\mathrm{c}}}\equiv1/\bar{c}_{00},~a_{N_{\mathrm{c}}}\equiv-2\bar{c}_{01}/\bar{c}_{00}$. It should be noted that the constants $T_{0,N_{\mathrm{c}}}$ and $a_{N_{\mathrm{c}}}$, similarly $\bar{c}_{kk'}$, are generally functions of $N_{\mathrm{c}}$ although are independent of temperature.

We proceed to address the minimum problem of the thermodynamic potential~(\ref{therm potential SPA}). The first-order derivative reads
\be
\label{dOmega dPhi NLO}
\frac{\partial \Omega_{\chi}}{\partial \Phi} 
& = &{} -\frac{N_{\mathrm{c}}a_{N_{\mathrm{c}}}\tilde{m}^{2}_{\sigma}\tilde{f}^{2}_{\sigma}T}{T_{0,N_{\mathrm{c}}}}\Phi^{N_{\mathrm{c}}-1}\left(\dfrac{\chi}{\tilde{f}_{\sigma}}\right)^{3}\ln\left(\dfrac{\chi}{\tilde{f}_{\sigma}}\right).
\nm\\
\ee
The saddle point equation at $\Phi=\langle\!\langle\hat{\Phi}\rangle\!\rangle$ yields
\be
\label{eq expec Phi explicit}
{} -\frac{N_{\mathrm{c}}a_{N_{\mathrm{c}}}\tilde{m}^{2}_{\sigma}\tilde{f}^{2}_{\sigma}T}{T_{0,N_{\mathrm{c}}}}\langle\!\langle\hat{\Phi}\rangle\!\rangle^{N_{\mathrm{c}}-1}\left(\frac{\overline{\chi}}{\tilde{f}_{\sigma}}\right)^{3}\ln\left(\frac{\overline{\chi}}{\tilde{f}_{\sigma}}\right)=0.
\ee
Then, we obtain the following three algebraic equations:
\be
& & \langle\!\langle\hat{\Phi}\rangle\!\rangle_{(1)}=0, \nm\\
& & \frac{\overline{\chi}_{(2)}}{\tilde{f}_{\sigma}}=1+\frac{\delta^{(1)}_{\chi,\mathrm{th}}(T)}{\tilde{f}_{\sigma}}+\frac{T}{T_{0,N_{\mathrm{c}}}}\left(1-a_{N_{\mathrm{c}}}\varphi_{(2)}\right)=1, \nm\\
& & \frac{\overline{\chi}_{(3)}}{\tilde{f}_{\sigma}}=1+\frac{\delta^{(1)}_{\chi,\mathrm{th}}(T)}{\tilde{f}_{\sigma}}+\frac{T}{T_{0,N_{\mathrm{c}}}}\left(1-a_{N_{\mathrm{c}}}\varphi_{(3)}\right)=0, \nm\\
\label{sol eq expec Phi explicit}
\ee
where $\varphi_{(2)}\equiv(\langle\!\langle\hat{\Phi}\rangle\!\rangle_{(2)})^{N_{\mathrm{c}}}$ and $\varphi_{(3)}\equiv(\langle\!\langle\hat{\Phi}\rangle\!\rangle_{(3)})^{N_{\mathrm{c}}}$. 

The first solution $\langle\!\langle\Phi\rangle\!\rangle_{(1)}=0$ is trivial. As for the second and the third solutions, i.e. $\langle\!\langle\hat{\Phi}\rangle\!\rangle_{(2)}$ and $\langle\!\langle\hat{\Phi}\rangle\!\rangle_{(3)}$, we will not consider $\langle\!\langle\hat{\Phi}\rangle\!\rangle_{(3)}$ due to
\be
\label{Omega at Phi2,3}
\Omega_{\chi}(\langle\!\langle\hat{\Phi}\rangle\!\rangle_{(3)})=0>\Omega_{\chi}(\langle\!\langle\hat{\Phi}\rangle\!\rangle_{(2)})=-\frac{\tilde{m}_{\sigma}^{2}\tilde{f}_{\sigma}^{2}}{16}.
\ee
Therefore, we obtain $\langle\!\langle\hat{\Phi}\rangle\!\rangle_{(2)}$ from Eq.~(\ref{sol eq expec Phi explicit}) as
\be
\label{varphi2 NLO}
\varphi_{(2)}=\frac{1}{a_{N_{\mathrm{c}}}}\left(1+\frac{\delta_{\chi,\mathrm{th}}^{(1)}(T)}{\tilde{f}_{\sigma}}\cdot\frac{T_{0,N_{\mathrm{c}}}}{T}\right),
\ee
namely
\be\label{Phi2 NLO}
\langle\!\langle\hat{\Phi}\rangle\!\rangle_{(2)}=\left[\dfrac{1}{a_{N_{\mathrm{c}}}}\left(1+\dfrac{\delta_{\chi,\mathrm{th}}^{(1)}(T)}{\tilde{f}_{\sigma}}\cdot\dfrac{T_{0,N_{\mathrm{c}}}}{T}\right)\right]^{\frac{1}{N_{\mathrm{c}}}}.
\ee

In the pure $\mathrm{SU(N_{\mathrm{c}})}$ gauge sector, the deconfinement phase transition is first order when $N_{\mathrm{c}}\geq 3$, and its strength increases with larger $N_{\mathrm{c}}$~\cite{Panero:2009tv,Lucini:2002ku,Lucini:2003zr,Lucini:2005vg,Holland:2003kg,Holland:2003mc,Pepe:2004rc}. Currently, we consider the first-order deconfinement phase transition as a known input since we do not yet have a method to derive it from the first principle of QCD. We will study the temperature dependence of $\delta^{(1)}_{\chi,\mathrm{th}}/\tilde{f}_{\sigma}$ below the critical temperature ($T<T_{\mathrm{c},N_{\mathrm{c}}}$) and above the critical temperature ($T>T_{\mathrm{c},N_{\mathrm{c}}}$). Note that, up to now, we do not know the critical temperature $T_{c,N_{\mathrm{c}}}$ for $N_{\mathrm{c}}>3$.

When the system is in the confinement phase ($T<T_{\mathrm{c},N_{\mathrm{c}}}$), the order parameter $\langle\!\langle\hat{\Phi}\rangle\!\rangle_{(2)}=0$ and we have $\delta^{(1)}_{\chi,\mathrm{th}}/\tilde{f}_{\sigma}=-T/T_{0,N_{\mathrm{c}}}$.

Another point we should check is whether $\langle\!\langle\hat{\Phi}\rangle\!\rangle_{(2)}=0$ is the minimum point of $\Omega_{\chi}$. This cannot be accomplished by using the second-order derivative $\partial^{2}\Omega_{\chi}/\partial\Phi^{2}$ since it vanishes at $\langle\!\langle\hat{\Phi}\rangle\!\rangle_{(2)}=0$. For this reason, we consider an infinitesimal shift around $\Phi=\langle\!\langle\hat{\Phi}\rangle\!\rangle_{(2)}=0$
\be\label{shift of Phi}
\Phi=\langle\!\langle\hat{\Phi}\rangle\!\rangle_{(2)}+\delta\Phi=\delta\Phi>0.
\ee
In terms of Eq.~(\ref{re expand delta chi Nc pG}), we have
\be
\label{shift of chi}
\left.\frac{\chi(T)}{\tilde{f}_{\sigma}}\right|_{\Phi} 
& = & 1+\frac{\delta^{(1)}_{\chi,\mathrm{th}}(T)}{\tilde{f}_{\sigma}}+\frac{\delta^{(2)}_{\chi,\mathrm{th}}(\delta\Phi,\delta\Phi;T)}{\tilde{f}_{\sigma}} \nm \\
& = & 1-\frac{a_{N_{\mathrm{c}}}T}{T_{0,N_{\mathrm{c}}}}(\delta\Phi)^{N_{\mathrm{c}}}.
\ee
If we choose $a_{N_{\mathrm{c}}}/T_{0,N_{\mathrm{c}}}>0$, $\chi(T)/\tilde{f}_{\sigma}$ at the point $\Phi=\langle\!\langle\hat{\Phi}\rangle\!\rangle_{(2)}+\delta\Phi=\delta\Phi$ is smaller than unity and hence
\be
\label{dOmegadPhi shift}
\left.\frac{\partial \Omega_{\chi}}{\partial \Phi}\right|_{\Phi=\delta\Phi} & = & {} -\frac{3a_{N_{\mathrm{c}}}T\tilde{m}_{\sigma}^{2}\tilde{f}_{\sigma}^{2}}{T_{0,N_{\mathrm{c}}}}\left[\frac{\chi(T)}{\tilde{f}_{\sigma}}\right]^{3}\ln\left[\frac{\chi(T)}{\tilde{f}_{\sigma}}\right](\delta\Phi)^{2} \nm \\
& > & 0.
\ee
Then, we can conclude that the thermodynamic potential $\Omega_{\chi}$ is a monotonic increasing function near the point $\Phi=\langle\!\langle\hat{\Phi}\rangle\!\rangle_{(2)}=0$, namely, this point gives the minimal value of $\Omega_{\chi}$.

At high temperatures $T>T_{\mathrm{c},N_{\mathrm{c}}}$, the color DoFs, i.e., gluons, escape from the glueballs and interact with each other. In this case, scale symmetry is not fully restored by the heat reservoir.  Only when $T\gg T_{\mathrm{c},N_{\mathrm{c}}}$, the interacting gluons decouple due to the asymptotic freedom and the scale symmetry restore completely, which is reflected in the EoS $P\propto T^{4}$. With this picture, at $T>T_{\mathrm{c},N_{\mathrm{c}}}$ we assume that only the thermal fluctuation related to the Polyakov loop, i.e. $\delta^{(2)}_{\chi,\mathrm{th}}(\langle\!\langle\hat{\Phi}\rangle\!\rangle_{(2)},\langle\!\langle\hat{\Phi}\rangle\!\rangle_{(2)};T)$ plays a crucial role in the configuration $\overline{\chi}_{(2)}$ of Eq.~(\ref{sol eq expec Phi explicit}). 

For the sake of clarity, we rewrite the expression of $\overline{\chi}_{(2)}$ more explicitly as
\be
\label{chibar 2 deconf}
\overline{\chi}_{(2)}=\tilde{f}_{\sigma}+\delta^{(1)}_{\chi,\mathrm{th}}(T)+\delta^{(2)}_{\chi,\mathrm{th}}(\langle\!\langle\hat{\Phi}\rangle\!\rangle_{(2)},\langle\!\langle\hat{\Phi}\rangle\!\rangle_{(2)};T).
\ee
Because of $\overline{\chi}_{(2)} \simeq \delta^{(2)}_{\chi,\mathrm{th}}(\langle\!\langle\hat{\Phi}\rangle\!\rangle_{(2)},\langle\!\langle\hat{\Phi}\rangle\!\rangle_{(2)};T)$, we obtain
\be
\label{delta 1 high T}
1+\frac{\delta^{(1)}_{\chi,\mathrm{th}}(T)}{\tilde{f}_{\sigma}} \simeq0, \quad T>T_{\mathrm{c},N_{\mathrm{c}}}.
\ee
Then, the temperature dependence of $\delta^{(1)}_{\chi,\mathrm{th}}(T)/\tilde{f}_{\sigma}$ can be written as
\be
\label{delta 1 final}
1+\frac{\delta_{\chi,\mathrm{th}}^{(1)}(T)}{\tilde{f}_{\sigma}}=\theta\left(\frac{T_{\mathrm{c},N_{\mathrm{c}}}-T}{T_{\mathrm{c},N_{\mathrm{c}}}}\right)\left(1-\dfrac{T}{T_{0,N_{\mathrm{c}}}}\right),
\ee
where $\theta$ is the Heaviside step function. Substituting the above equation into Eq.~(\ref{Phi2 NLO}) we obtain
\be
\label{Phi2 NLO final}
\langle\!\langle\hat{\Phi}\rangle\!\rangle_{\text{NLO}} & \equiv & \langle\!\langle\hat{\Phi}\rangle\!\rangle_{(2)} \nm \\
& = & \theta\left(\frac{T-T_{\mathrm{c},N_{\mathrm{c}}}}{T_{\mathrm{c},N_{\mathrm{c}}}}\right)\left[\frac{1}{a_{N_{\mathrm{c}}}}\left(1-\frac{T_{0,N_{\mathrm{c}}}}{T}\right)\right]^{\frac{1}{N_{\mathrm{c}}}}.
\nm\\
\ee

By using the thermodynamic potential~(\ref{therm potential SPA}), we can check that the above equation indeed gives the minimum point of the thermodynamic potential $\Omega_{\chi}$ at $T>T_{\mathrm{c},N_{\mathrm{c}}}$. In the deconfinement phase, $\langle\!\langle\hat{\Phi}\rangle\!\rangle_{(2)}>0$ and 
\be
\label{d2VdPhi2 Nc at Phi2}
\left.\frac{\partial^{2}\Omega_{\chi}}{\partial\Phi^{2}}\right|_{\Phi=\langle\!\langle\hat{\Phi}\rangle\!\rangle_{(2)}} & = & \frac{(N_{\mathrm{c}}a_{N_{\mathrm{c}}})^{2}\tilde{m}_{\sigma}^{2}\tilde{f}_{\sigma}^{2}T^{2}(\langle\!\langle\hat{\Phi}\rangle\!\rangle_{(2)})^{2(N_{\mathrm{c}}-1)}}{T_{0,N_{\mathrm{c}}}^{2}} \nm \\
& > & 0.
\ee
Thus, $\langle\!\langle\hat{\Phi}\rangle\!\rangle_{(2)}$ is the minimum point at $T>T_{\mathrm{c},N_{\mathrm{c}}}$.

\subsection{The next-next-leading order of $\delta_{\chi,\mathrm{th}}^{(2)}/\tilde{f}_{\sigma}$ with respect to $\Phi^{N_{\mathrm{c}}}$}
\label{subsec: NNLO}

Next, we consider the next-next-leading order (NNLO) of $\delta_{\chi,\mathrm{th}}^{(2)}/\tilde{f}_{\sigma}$ with respect to $\Phi^{N_{\mathrm{c}}}$, $\delta_{\chi,\mathrm{NNLO}}^{(2)}/\tilde{f}_{\sigma}$. 

From Eqs.~(\ref{delta2 expa}) and (\ref{sol ODE ckk}), we have
\be
\label{expa delchi2 NNLO}
& & \frac{\delta^{(2)}_{\chi,\mathrm{th}}(\Phi,\Phi^{*};T)}{\tilde{f}_{\sigma}} =  \frac{\delta_{\chi,\mathrm{NNLO}}^{(2)}}{\tilde{f}_{\sigma}} \nonumber \\
& &{} = \frac{T}{T_{0,N_{\mathrm{c}}}}\left[1-\frac{a_{N_{\mathrm{c}}}}{2}\left(\varphi^{*}+\varphi\right)-\frac{b_{N_{c}}}{2}\left((\varphi^{*})^{2}+\varphi^{2}\right)\right],
\nm\\
\ee
where $b_{N_{\mathrm{c}}}\equiv-2\bar{c}_{02}/\bar{c}_{00}$. For pure gauge systems, the above equation reduces to
\be
\label{expa delchi2 NNLO pG}
\frac{\delta^{(2)}_{\chi,\mathrm{th}}(\Phi,\Phi;T)}{\tilde{f}_{\sigma}} 
& = & \frac{T}{T_{0,N_{\mathrm{c}}}}\left(1-a_{N_{\mathrm{c}}}\varphi-b_{N_{\mathrm{c}}}\varphi^{2}\right).
\ee
\begin{widetext}
In this case, the first-order derivative of $\Omega_{\chi}$ with respect to $\Phi$ becomes
\be
\label{dOmegadPhi NNLO}
\frac{\partial \Omega_{\chi}}{\partial \Phi} & = & \frac{\partial \Omega_{\chi}}{\partial \chi}\frac{\partial \chi}{\partial\varphi}\frac{\partial\varphi}{\partial\Phi} = {} - \frac{N_{\mathrm{c}}T\tilde{m}_{\sigma}^{2}\tilde{f}_{\sigma}^{2}}{T_{0,N_{\mathrm{c}}}}\left(\frac{\chi}{\tilde{f}_{\sigma}}\right)^{3}\ln\left(\frac{\chi}{\tilde{f}_{\sigma}}\right) \left(a_{N_{\mathrm{c}}}+2b_{N_{\mathrm{c}}}\Phi^{N_{\mathrm{c}}}\right) \Phi^{N_{\mathrm{c}}-1}.
\ee
The extreme points of $\Omega_{\chi}$ are given by $\partial\Omega_{\chi}/\partial\Phi=0$ and they are
\be
& & \langle\!\langle\hat{\Phi}\rangle\!\rangle_{(1)}=0 , \label{expec Phi 1 NNLO} \nm\\
& & (\langle\!\langle\hat{\Phi}\rangle\!\rangle_{(2)})^{N_{\mathrm{c}}}=-\frac{a_{N_{\mathrm{c}}}}{2b_{N_{\mathrm{c}}}}, \label{expec Phi 2 NNLO}	\nm\\
& &	\frac{\overline{\chi}_{(3)}}{\tilde{f}_{\sigma}}=1+\frac{\delta_{\chi,\mathrm{th}}^{(1)}}{\tilde{f}_{\sigma}}+\frac{T}{T_{0,N_{\mathrm{c}}}}\left[1-a_{N_{\mathrm{c}}}\varphi_{(3)}-b_{N_{\mathrm{c}}}\left(\varphi_{(3)}\right)^{2}\right]=0, \label{expec Phi 3 NNLO}	\nm\\
& &	\frac{\overline{\chi}_{(4)}}{\tilde{f}_{\sigma}}=1+\frac{\delta_{\chi,\mathrm{th}}^{(1)}}{\tilde{f}_{\sigma}}+\frac{T}{T_{0,N_{\mathrm{c}}}}\left[1-a_{N_{\mathrm{c}}}\varphi_{(4)}-b_{N_{\mathrm{c}}}\left(\varphi_{(4)}\right)^{2}\right]=1. \label{expec Phi 4 NNLO}
\ee
\end{widetext}	
Here, $\varphi_{(3)}\equiv(\langle\!\langle\hat{\Phi}\rangle\!\rangle_{(3)})^{N_{\mathrm{c}}}$ and $\varphi_{(4)}\equiv(\langle\!\langle\hat{\Phi}\rangle\!\rangle_{(4)})^{N_{\mathrm{c}}}$. We do not need to consider the first and second solutions $\langle\!\langle\hat{\Phi}\rangle\!\rangle_{(1)}$ and $\langle\!\langle\hat{\Phi}\rangle\!\rangle_{(2)}$ due to the triviality. Since 
\be
\label{Omega at Phi3,4}
\Omega_{\chi}(\langle\!\langle\hat{\Phi}\rangle\!\rangle_{(3)})=0>\Omega_{\chi}(\langle\!\langle\hat{\Phi}\rangle\!\rangle_{(4)})={}-\frac{\tilde{m}_{\sigma}^{2}\tilde{f}_{\sigma}^{2}}{16},
\ee
only the solution $\langle\!\langle\hat{\Phi}\rangle\!\rangle_{(4)}$ is physically interesting. The manifestation of the first-order deconfinement phase transition is totally the same as that in subsection~\ref{subsec: NLO}. The configuration of the dilaton corresponding to the minimal $\Omega_{\chi}$ reads
\be\label{chi4 expres}
\overline{\chi}_{(4)}=\tilde{f}_{\sigma}+\delta^{(1)}_{\chi,\mathrm{th}}(T)+\delta^{(2)}_{\chi,\mathrm{th}}(\langle\!\langle\hat{\Phi}\rangle\!\rangle_{(4)},\langle\!\langle\hat{\Phi}\rangle\!\rangle_{(4)};T).
\ee
In the deconfinement phase ($T>T_{\mathrm{c},N_{\mathrm{c}}}$), we have $\overline{\chi}_{(4)}=\delta^{(2)}_{\chi,\mathrm{th}}(\langle\!\langle\hat{\Phi}\rangle\!\rangle_{(4)},\langle\!\langle\hat{\Phi}\rangle\!\rangle_{(4)};T)$. Substituting Eq.~(\ref{delta 1 final}) into Eq.~(\ref{expec Phi 4 NNLO}), we have
\be
& & \left[a_{N_{\mathrm{c}}}+b_{N_{\mathrm{c}}}\varphi_{(4)}\right]\varphi_{(4)}=0, ~\mbox{for}~T<T_{\mathrm{c},N_{\mathrm{c}}}, \label{expec Phi 4 NNLO confined} \\	
& & b_{N_{\mathrm{c}}}\left(\varphi_{(4)}\right)^{2}+a_{N_{\mathrm{c}}}\varphi_{(4)}+\dfrac{T_{0,N_{\mathrm{c}}}}{T}-1=0, ~\mbox{for}~ T>T_{\mathrm{c},N_{\mathrm{c}}}. \nm\\
\label{expec Phi 4 NNLO deconfined}
\ee

In the confinement phase, Eq.~(\ref{expec Phi 4 NNLO confined}) yields two solutions $\langle\!\langle\hat{\Phi}\rangle\!\rangle_{(4)}=0$ and $(\langle\!\langle\hat{\Phi}\rangle\!\rangle_{(4)})^{N_{\mathrm{c}}}=-a_{N_{\mathrm{c}}}/b_{N_{\mathrm{c}}}$. It is obvious that the former is reasonable. In the deconfinement phase, we obtain the solutions from Eq.~(\ref{expec Phi 4 NNLO deconfined}) as
\be
\varphi_{(4,\pm)}=\frac{-a_{N_{\mathrm{c}}} \pm \sqrt{a_{N_{\mathrm{c}}}^{2}-4b_{N_{\mathrm{c}}}\left(\frac{T_{0,N_{\mathrm{c}}}}{T}-1\right)}}{2b_{N_{\mathrm{c}}}} , \label{expec Phi 4 NNLO deconfined +} 
\ee	
where $\varphi_{(4,\pm)}\equiv(\langle\!\langle\hat{\Phi}\rangle\!\rangle_{(4,\pm)})^{N_{\mathrm{c}}}$. The physical solution is the one that is a monotonic function of temperature $T$. If $b_{N_{\mathrm{c}}}>0$, $\langle\!\langle\hat{\Phi}\rangle\!\rangle_{(4,+)}$ increases while $\langle\!\langle\hat{\Phi}\rangle\!\rangle_{(4,-)}$ decreases with temperature. If $b_{N_{\mathrm{c}}}<0$, the situation is intact. For this reason, we choose 
\be
\label{expec Phi 4 NNLO deconfined + formu}
\langle\!\langle\hat{\Phi}\rangle\!\rangle_{(4,+)} = \left[\frac{-a_{N_{\mathrm{c}}}+\sqrt{a_{N_{\mathrm{c}}}^{2}-4b_{N_{\mathrm{c}}}\left(\frac{T_{0,N_{\mathrm{c}}}}{T}-1\right)}}{2b_{N_{\mathrm{c}}}}\right]^{\frac{1}{N_{\mathrm{c}}}}.
\nm\\
\ee

It is essential to check whether the solution $\langle\!\langle\hat{\Phi}\rangle\!\rangle_{(4)}=0$ is in the confinement phase and the solution $\langle\!\langle\hat{\Phi}\rangle\!\rangle_{(4,+)}$ is in the deconfinement phase and is exactly the point of the minimum value of $\Omega_{\chi}$. When the gluons are confined ($T<T_{\mathrm{c},N_{\mathrm{c}}}$), one obtains $\partial^{2}\Omega_{\chi}/\partial\Phi^{2}=0$ in $\Phi=\langle\!\langle\hat{\Phi}\rangle\!\rangle_{(4)}=0$. Therefore, similar to subsection~\ref{subsec: NLO}, we need to consider the first-order derivative of $\Omega_{\chi}$ at $\Phi=\langle\!\langle\hat{\Phi}\rangle\!\rangle_{(4)}+\delta\Phi=\delta\Phi$ with $0<\delta\Phi\ll 1$. Since
\be
\label{shift of chi NNLO}
& & \left.\frac{\chi(T)}{\tilde{f}_{\sigma}}\right|_{\Phi=\langle\!\langle\hat{\Phi}\rangle\!\rangle_{(4)}+\delta\Phi} \nm\\
& & = 1+\frac{\delta_{\chi,\mathrm{th}}^{(1)}(T)}{\tilde{f}_{\sigma}}+\frac{T}{T_{0,N_{\mathrm{c}}}}\left[1-a_{N_{\mathrm{c}}}(\delta\Phi)^{N_{\mathrm{c}}}-b_{N_{\mathrm{c}}}(\delta\Phi)^{2N_{\mathrm{c}}}\right] \nm \\
& & \approx 1+\frac{\delta_{\chi,\mathrm{th}}^{(1)}(T)}{\tilde{f}_{\sigma}}+\dfrac{T}{T_{0,N_{\mathrm{c}}}}\left[1-a_{N_{\mathrm{c}}}(\delta\Phi)^{N_{\mathrm{c}}}\right] \nm \\
& & = 1-\frac{a_{N_{\mathrm{c}}}T}{T_{0,N_{\mathrm{c}}}}(\delta\Phi)^{N_{\mathrm{c}}}<1,
\ee
the first-order derivative of $\Omega_{\chi}$ at $\Phi=\delta\Phi$ becomes
\be
\label{dOmegadPhi shift NNLO}
\left.\frac{\partial \Omega_{\chi}}{\partial \Phi}\right|_{\Phi=\delta\Phi} & = &{} - \frac{N_{\mathrm{c}}T\tilde{m}_{\sigma}^{2}\tilde{f}_{\sigma}^{2}}{T_{0,N_{\mathrm{c}}}}\left[\frac{\chi(T)}{\tilde{f}_{\sigma}}\right]^{3}\ln\left[\frac{\chi(T)}{\tilde{f}_{\sigma}}\right] \nm \\
& &{} \times\left[a_{N_{\mathrm{c}}}(\delta\Phi)^{N_{\mathrm{c}}}+2b_{N_{\mathrm{c}}}(\delta\Phi)^{2N_{\mathrm{c}}}\right](\delta\Phi)^{-1} \nonumber \\
& \approx &{} - \frac{N_{\mathrm{c}}T\tilde{m}_{\sigma}^{2}\tilde{f}_{\sigma}^{2}}{T_{0,N_{\mathrm{c}}}}\left[\frac{\chi(T)}{\tilde{f}_{\sigma}}\right]^{3}\ln\left[\frac{\chi(T)}{\tilde{f}_{\sigma}}\right] \nm \\
& & {} \times a_{N_{\mathrm{c}}}(\delta\Phi)^{N_{\mathrm{c}}-1} \nm\\
& > &0~.
\ee
Hence, the thermodynamic potential $\Omega_{\chi}$ monotonically increases with respect to $\Phi$ in the vicinity of $\langle\!\langle\hat{\Phi}\rangle\!\rangle_{(4)}=0$, indeed the point of the minimal value of $\Omega_{\chi}$ in the confinement phase. 

While glueballs are melted by the heat reservoir ($T>T_{\mathrm{c},N_{\mathrm{c}}}$), the second-order derivative of $\Omega_{\chi}$ at $\Phi=\langle\!\langle\hat{\Phi}\rangle\!\rangle_{(4,+)}$ is
\be
\label{d2OmegadPhi2 NNLO at Phi4+}
\left.\frac{\partial^{2}\Omega_{\chi}}{\partial\Phi^{2}}\right|_{\Phi=\langle\!\langle\hat{\Phi}\rangle\!\rangle_{(4,+)}} & = & \left(\frac{N_{\mathrm{c}}T\tilde{m}_{\sigma}\tilde{f}_{\sigma}}{T_{0,N_{\mathrm{c}}}}\right)^{2} \nm\\
& &{} \times \left[a_{N_{\mathrm{c}}}+2b_{N_{\mathrm{c}}}(\langle\!\langle\hat{\Phi}\rangle\!\rangle_{(4,+)})^{N_{\mathrm{c}}}\right]^{2} \nm \\
& & {} \times(\langle\!\langle\hat{\Phi}\rangle\!\rangle_{(4,+)})^{2(N_{\mathrm{c}}-1)}\nm\\
& > & 0,
\ee
indicating that $\langle\!\langle\hat{\Phi}\rangle\!\rangle_{(4,+)}$ is the point of minimum of $\Omega_{\chi}$ in the deconfinement phase.

Finally, the formula of the expectation value of the Polyakov loop is obtained as
\be
\label{Phi NNLO final}
\langle\!\langle\hat{\Phi}\rangle\!\rangle_{\text{NNLO}} & = & \theta\left(T-T_{\mathrm{c},N_{\mathrm{c}}} \right) \nm\\
& &{} \times \left[\frac{-a_{N_{\mathrm{c}}}+\sqrt{a_{N_{\mathrm{c}}}^{2}-4b_{N_{\mathrm{c}}}\left(\frac{T_{0,N_{\mathrm{c}}}}{T}-1\right)}}{2b_{N_{\mathrm{c}}}}\right]^{\frac{1}{N_{\mathrm{c}}}}.
\nm\\
\ee

\section{Constants in $\langle\!\langle\hat{\Phi}\rangle\!\rangle_{\text{NLO}}$ and $\langle\!\langle\hat{\Phi}\rangle\!\rangle_{\text{NNLO}}$}
\label{sec:three}

In this section, we will evaluate the three constants $a_{N_{\mathrm{c}}}$, $b_{N_{\mathrm{c}}}$ and $T_{0,N_{\mathrm{c}}}$ using LQCD results for the expectation value of the Polyakov loop in the pure-gauge sector~\cite{Kaczmarek:2002mc,Ratti:2005jh,Mykkanen:2012ri}. Specifically, we will focus on the center values of the LQCD data for the $\mathrm{SU(3)}$~\cite{Kaczmarek:2002mc,Ratti:2005jh}, $\mathrm{SU(4)}$ and $\mathrm{SU(5)}$~\cite{Mykkanen:2012ri} pure-gauge systems.

\begin{figure*}[htbp]	
	\centering
		\includegraphics[scale=0.255]{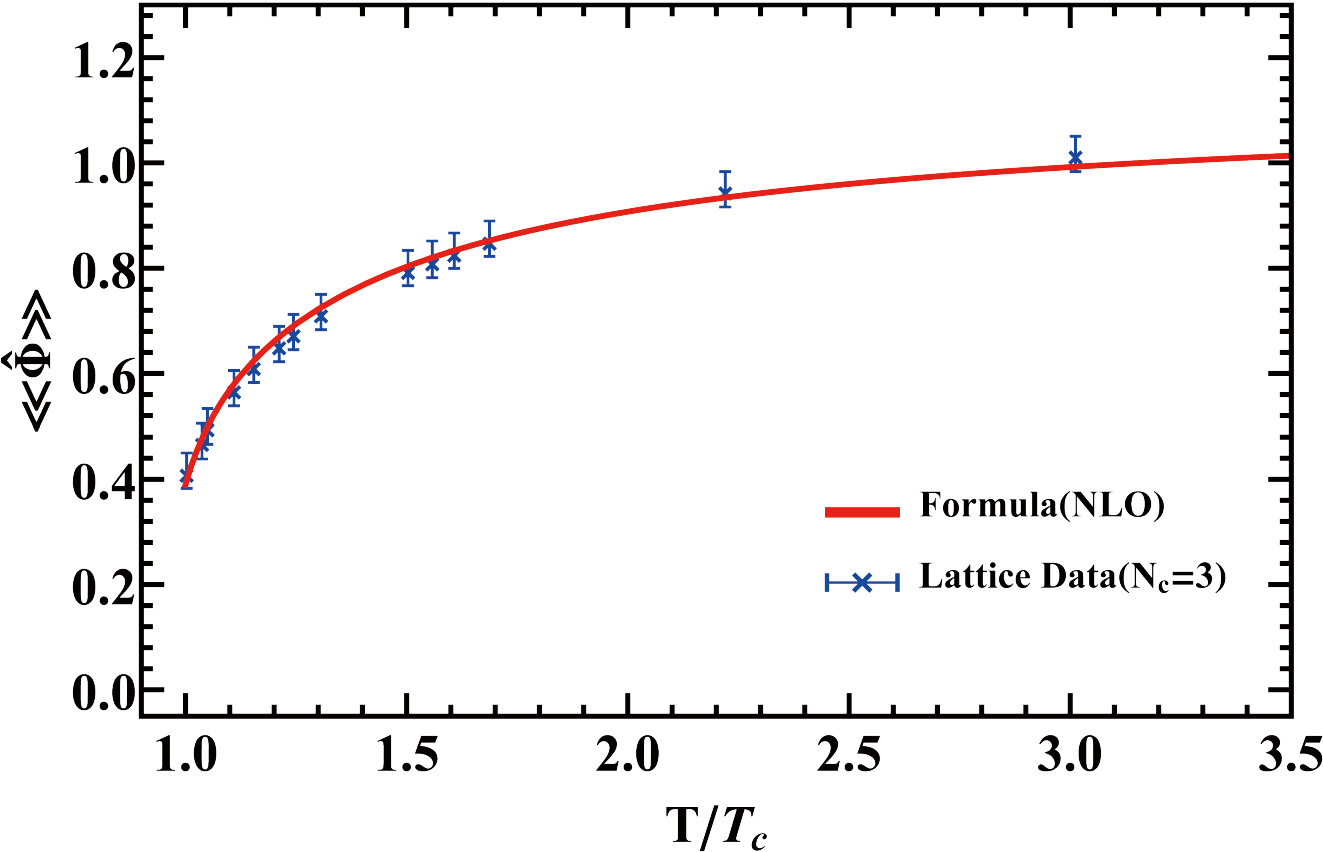}\quad	\includegraphics[scale=0.245]{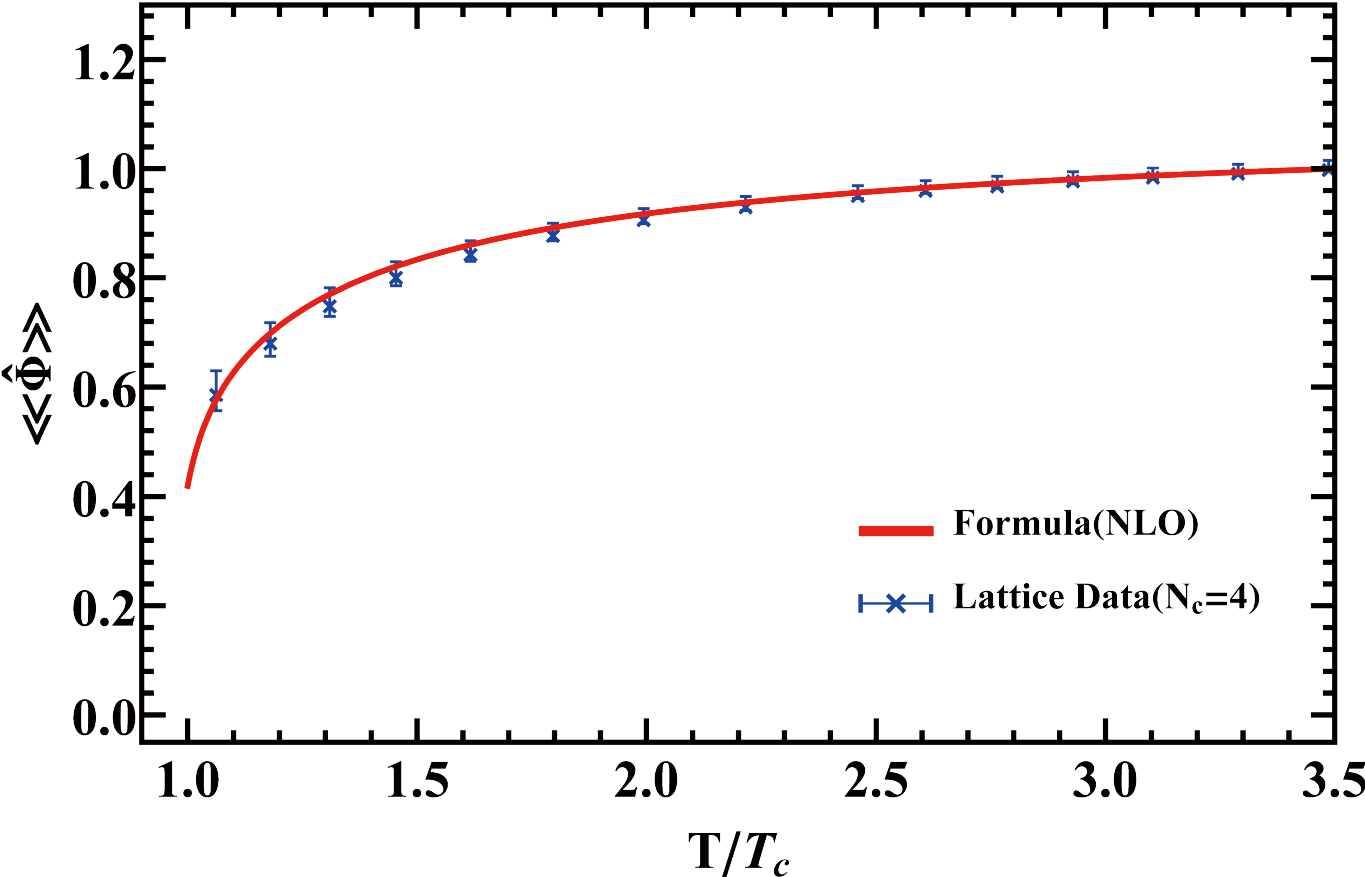}\quad
	\includegraphics[scale=0.245]{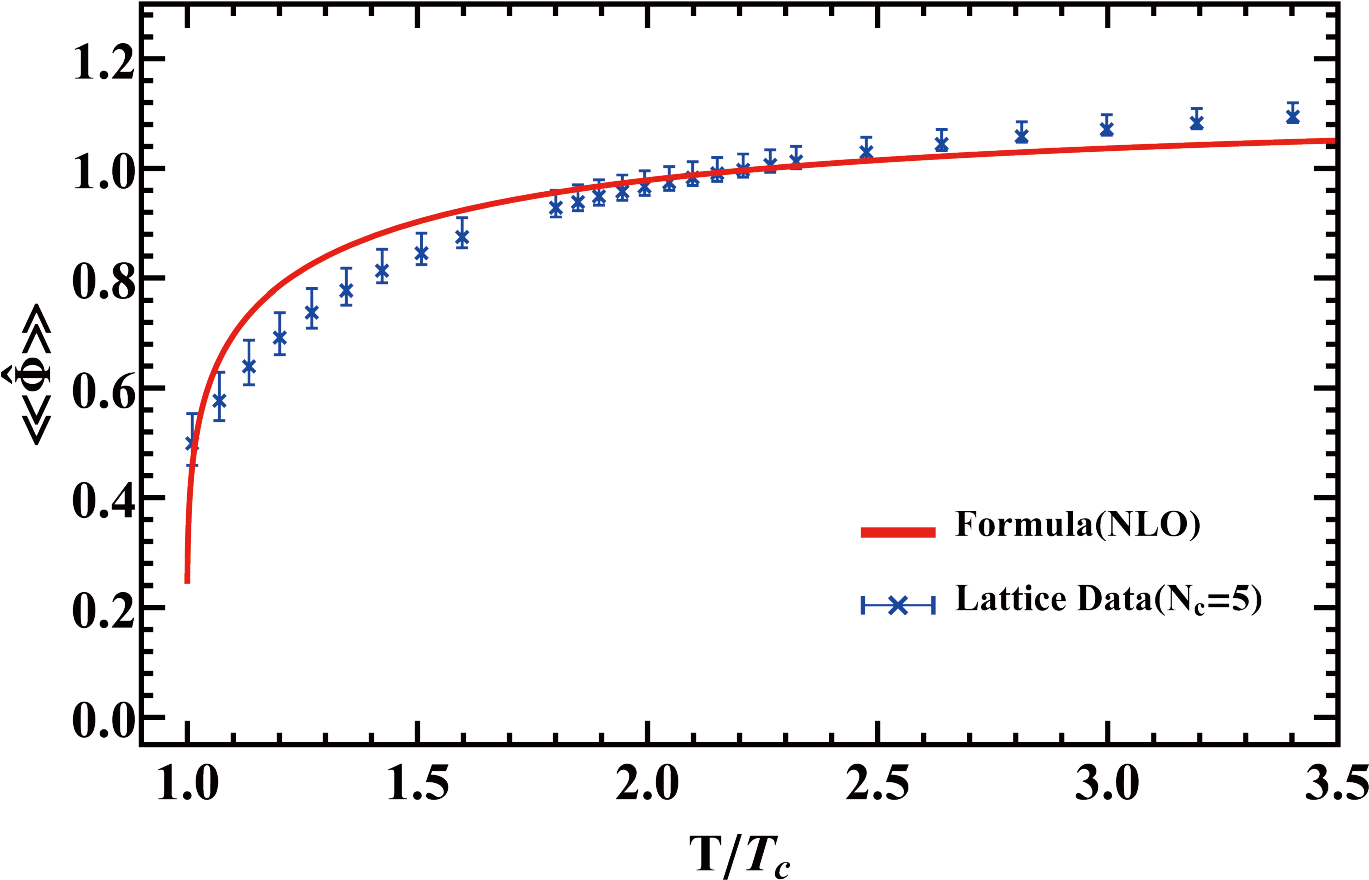}
	\caption{(color online). The curves of $\langle\!\langle\hat{\Phi}\rangle\!\rangle_{\text{NLO}}$ (red lines) compared with the LQCD data in the cases of $N_{\mathrm{c}}=3$~\cite{Kaczmarek:2002mc,Ratti:2005jh} and $N_{\mathrm{c}}=4,5$~\cite{Mykkanen:2012ri}.}
	\label{fig:Phi NLO}
\end{figure*}

We first fit the parameters in the NLO formula $\langle\!\langle\hat{\Phi}\rangle\!\rangle_{\text{NLO}}$, as given in Eq.~(\ref{Phi2 NLO final}). The results at the 95\% confidence level are presented in Table~\ref{tab:NLO para}.
\begin{table*}[htbp]
	\caption{The fixed parameters with the confidence intervals in $\langle\!\langle\hat{\Phi}\rangle\!\rangle_{\text{NLO}}$ for $\mathrm{SU(3)}$, $\mathrm{SU(4)}$ and $\mathrm{SU(5)}$ systems.}
    \label{tab:NLO para}
	\begin{ruledtabular}
		\begin{tabular}{cccc}
			Gauge group&
			$\mathrm{SU(3)}$&
			$\mathrm{SU(4)}$&
			$\mathrm{SU(5)}$\\
			\colrule
			$a_{N_{\mathrm{c}}}$ & $0.696735$ &  $0.720968$  & $0.557514$ \\
			\colrule
		95\% CI for $a_{N_{\mathrm{c}}}$ &$[0.670448,0.723021]$	& $[0.706921,0.735015]$&$[0.500769,0.614259]$ \\
        \colrule
        $T_{0,N_{\mathrm{c}}}/T_{\mathrm{c},N_{\mathrm{c}}}$ & $0.958802$ & $0.977624$ &  $0.999476$ \\
        \colrule
        95\% CI for $T_{0,N_{\mathrm{c}}}/T_{\mathrm{c},N_{\mathrm{c}}}$ & $[0.951533,0.966071]$ & $[0.968212,0.987036]$ & $[0.987320,1.011632]$
		\end{tabular}
	\end{ruledtabular}
\end{table*}
Note that we regard $T/T_{\mathrm{c},N_{\mathrm{c}}}$ as the argument in the formulas of $\langle\!\langle\hat{\Phi}\rangle\!\rangle_{\text{NLO}}$ and $\langle\!\langle\hat{\Phi}\rangle\!\rangle_{\text{NNLO}}$ when evaluating the parameters. Consequently, we only fix the ratio of the parameters $T_{0,N_{\mathrm{c}}}$ and $T_{\mathrm{c},N_{\mathrm{c}}}$. The critical temperature of $N_{\mathrm{c}}=3$ is $270~\text{MeV}$ and thus $T_{0,3}=0.958802\times 270~\text{MeV}\approx 258.9\text{MeV}$ in the NLO. Furthermore, we plot the curves of $\langle\!\langle\hat{\Phi}\rangle\!\rangle_{\text{NLO}}$ using the fixed parameters in Table~\ref{tab:NLO para} and compare our results with the LQCD data. As shown in Fig.~\ref{fig:Phi NLO}, our NLO results are in good agreement with the LQCD data for both the $\mathrm{SU(3)}$ and $\mathrm{SU(4)}$ cases. However, a deviation is observed in the $\mathrm{SU(5)}$ case.

\begin{figure*}[htbp]	
	\centering
		\includegraphics[scale=0.225]{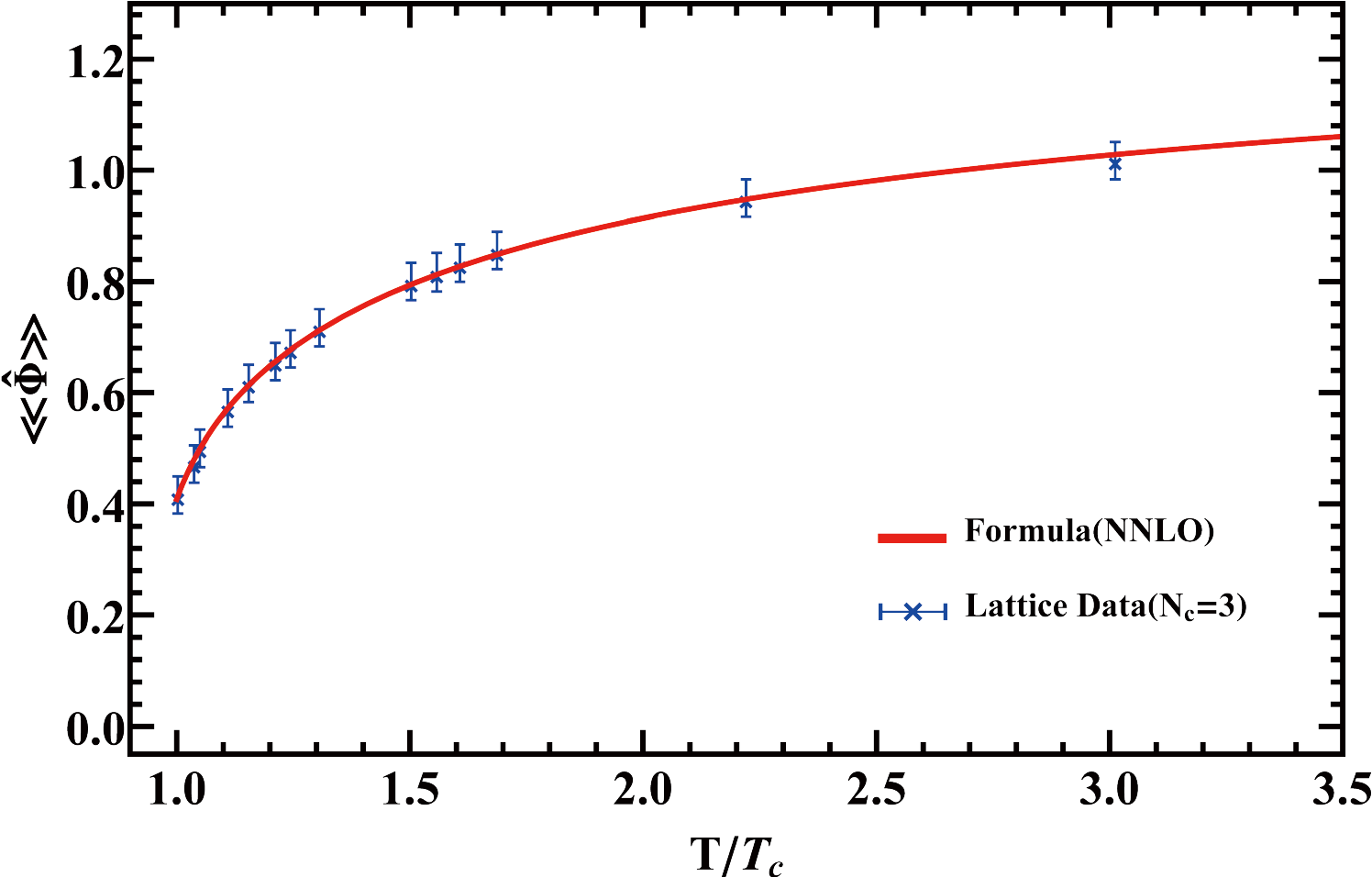}\quad	\includegraphics[scale=0.23]{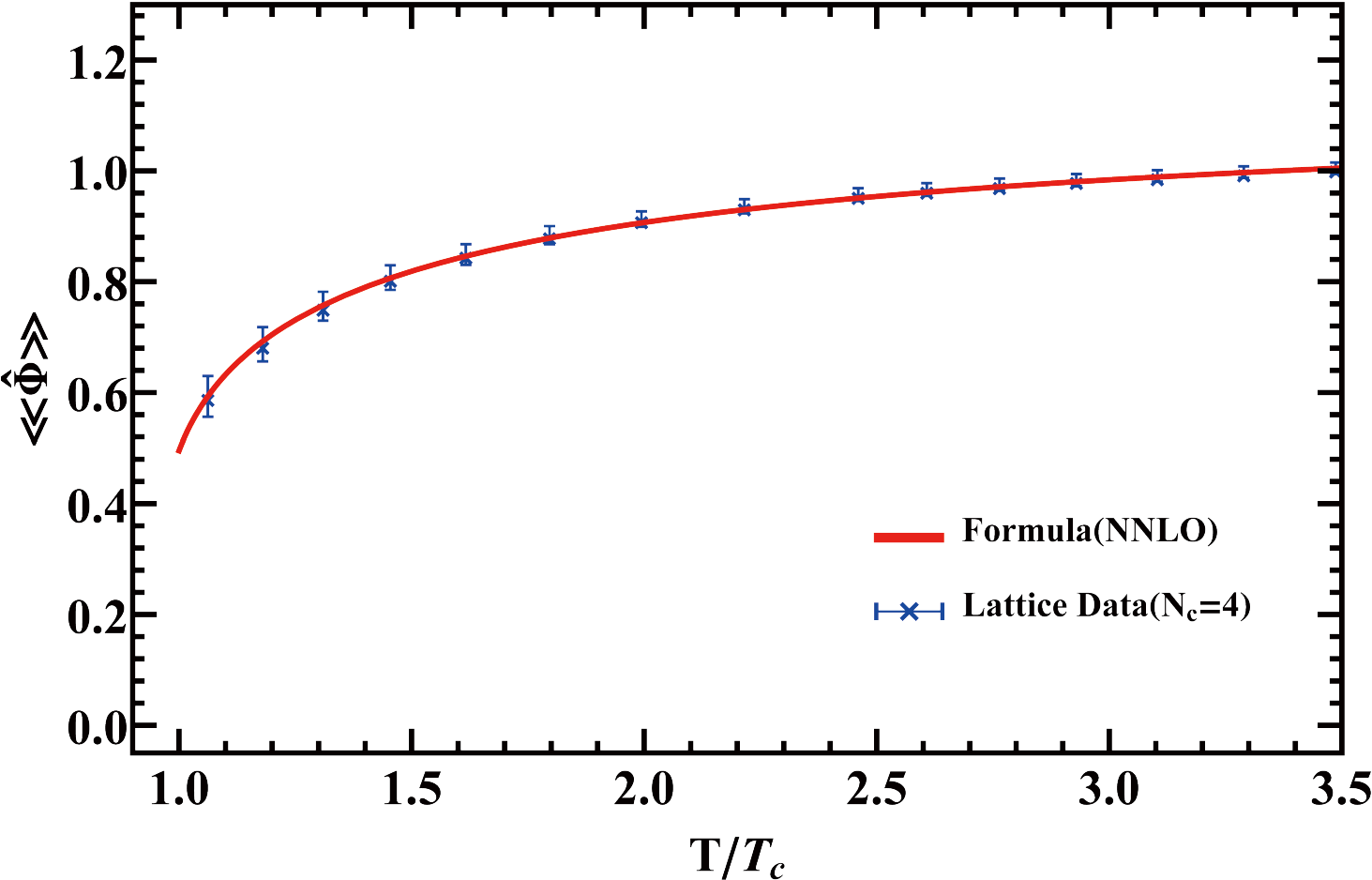}\quad
	\includegraphics[scale=0.23]{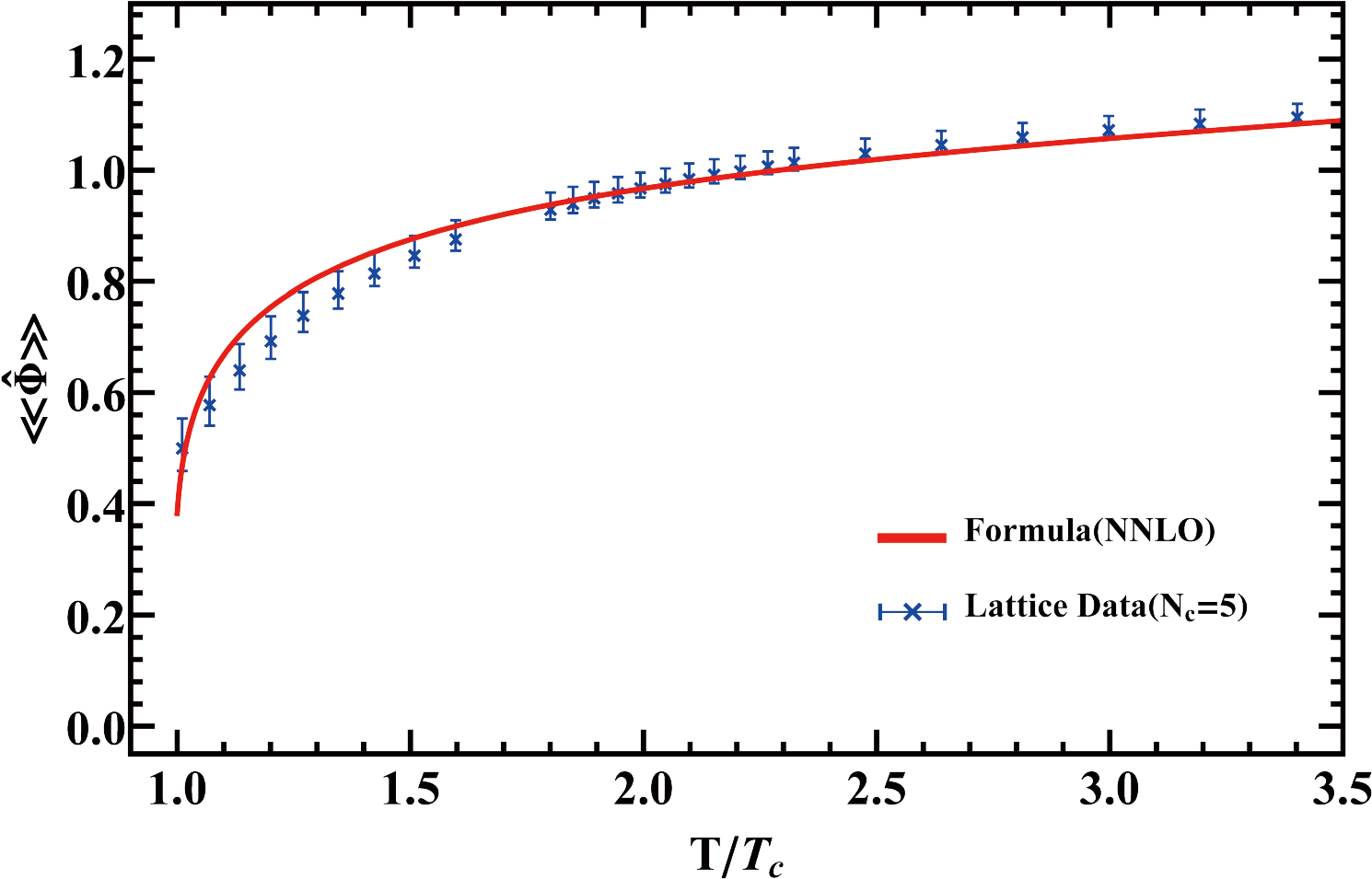}
	\caption{(color online). The curves of $\langle\!\langle\hat{\Phi}\rangle\!\rangle_{\text{NNLO}}$(red lines) compared with the LQCD data in the cases of $N_{\mathrm{c}}=3$~\cite{Kaczmarek:2002mc,Ratti:2005jh} and $N_{\mathrm{c}}=4,5$~\cite{Mykkanen:2012ri}.}
	\label{fig:Phi NNLO}
\end{figure*}

We now turn to the NNLO. We first fix the parameters for $\mathrm{SU(5)}$ case and obtain the best fit $b_{N_{\mathrm{c}}}={}-0.186231$. Notably, this value of $b_{N_{\mathrm{c}}}={}-0.186231$ is also suitable for both the $\mathrm{SU(4)}$ and $\mathrm{SU(5)}$ cases. Thus, we conclude that $b_{N_{\mathrm{c}}}$ is nearly independent of the number of colors, as our results fall within the error bars of the LQCD data. The values of other parameters at the 95\% confidence level are listed in Table~\ref{tab:NNLO para}, and the curves of $\langle\!\langle\hat{\Phi}\rangle\!\rangle_{\text{NNLO}}$ are displayed in Fig.~\ref{fig:Phi NNLO}. In this context, the value of $T_{0,3}$ is calculated to be $0.943952\times 270~\text{MeV}\approx254.9~\text{MeV}$, which is nearly the same as that obtained in the NLO case. Remarkably, we find that the deviation of our formula $\langle\!\langle\hat{\Phi}\rangle\!\rangle_{\text{NLO}}$ from the $\mathrm{SU(5)}$ LQCD data has been reduced to some extent. Additionally, the NNLO formula is also in accordance with the LQCD data for $N_{\mathrm{c}}=3,4$, as shown in Fig.~\ref{fig:Phi NNLO}. 

\begin{table*}[htbp]
	\caption{The fixed parameters with the confidence intervals in $\langle\!\langle\hat{\Phi}\rangle\!\rangle_{\text{NNLO}}$ for $\mathrm{SU(3)}$, $\mathrm{SU(4)}$ and $\mathrm{SU(5)}$ systems with $b_{N_{\mathrm{c}}}={}-0.186231$.}
    \label{tab:NNLO para}
	\begin{ruledtabular}
		\begin{tabular}{cccc}
			Gauge group&
			$\mathrm{SU(3)}$&
			$\mathrm{SU(4)}$&
			$\mathrm{SU(5)}$\\
			\colrule
			$a_{N_{\mathrm{c}}}$ & $0.834206$ &  $0.905033$  & $0.752401$ \\
			\colrule
		95\% CI for $a_{N_{\mathrm{c}}}$ & $[0.826099,0.842314]$	& $[0.898431,0.911634]$ & $[0.745672,0.759129]$ \\
        \colrule
        $T_{0,N_{\mathrm{c}}}/T_{\mathrm{c},N_{\mathrm{c}}}$ & $0.943952$ & $0.946053$ &  $0.993848$ \\
        \colrule
        95\% CI for $T_{0,N_{\mathrm{c}}}/T_{\mathrm{c},N_{\mathrm{c}}}$ & $[0.940958,0.946947]$ & $[0.939536,0.952571]$ & $[0.984730,1.002966]$
		\end{tabular}
	\end{ruledtabular}
\end{table*}

In addition, both formulae~(\ref{Phi2 NLO final}) and (\ref{Phi NNLO final}) tell us that, if the large-$N_{\mathrm{c}}$ limit is taken, both $\langle\!\langle\hat{\Phi}\rangle\!\rangle_{\text{NLO}}$ and $\langle\!\langle\hat{\Phi}\rangle\!\rangle_{\text{NNLO}}$ at $T>T_{\mathrm{c},N_{\mathrm{c}}}$ approach to unity. Furthermore, we may infer the behaviors of $\langle\!\langle\hat{\Phi}\rangle\!\rangle_{\text{NLO}}$ and $\langle\!\langle\hat{\Phi}\rangle\!\rangle_{\text{NNLO}}$ with increasing $N_{\mathrm{c}}$. In the interval $\langle\!\langle\hat{\Phi}\rangle\!\rangle_{\text{NLO}},\langle\!\langle\hat{\Phi}\rangle\!\rangle_{\text{NNLO}}\in (0,1)$, the terms in square brackets of Eqs.~(\ref{Phi2 NLO final}) and (\ref{Phi NNLO final}) are also smaller than unity. Hence, both $\langle\!\langle\hat{\Phi}\rangle\!\rangle_{\text{NLO}}$ and $\langle\!\langle\hat{\Phi}\rangle\!\rangle_{\text{NNLO}}$ increase with the color number $N_{\mathrm{c}}$. 

Imagine that we have many glueballs in the confinement phase. As the temperature increases, gluons carrying $N_{\mathrm{c}}$ colors will evaporate to the deconfinement phase, with larger $N_{\mathrm{c}}$ corresponding to more DoFs. Consequently, the discrepancy between the two phases becomes more pronounced, and the intensity of the first-order phase transition increases with $N_{\mathrm{c}}$. Therefore, the expectation value of the Polyakov loop, which serves as an order parameter of the deconfinement phase transition, becomes larger with the increasing of the color numbers. In the large-$N_{\mathrm{c}}$ limit, the order parameter jumps drastically from the confinement phase to the deconfinement phase, indicating the occurrence of the first-order phase transition.

\begin{figure*}[htbp]	
	\centering
	\includegraphics[scale=0.4]{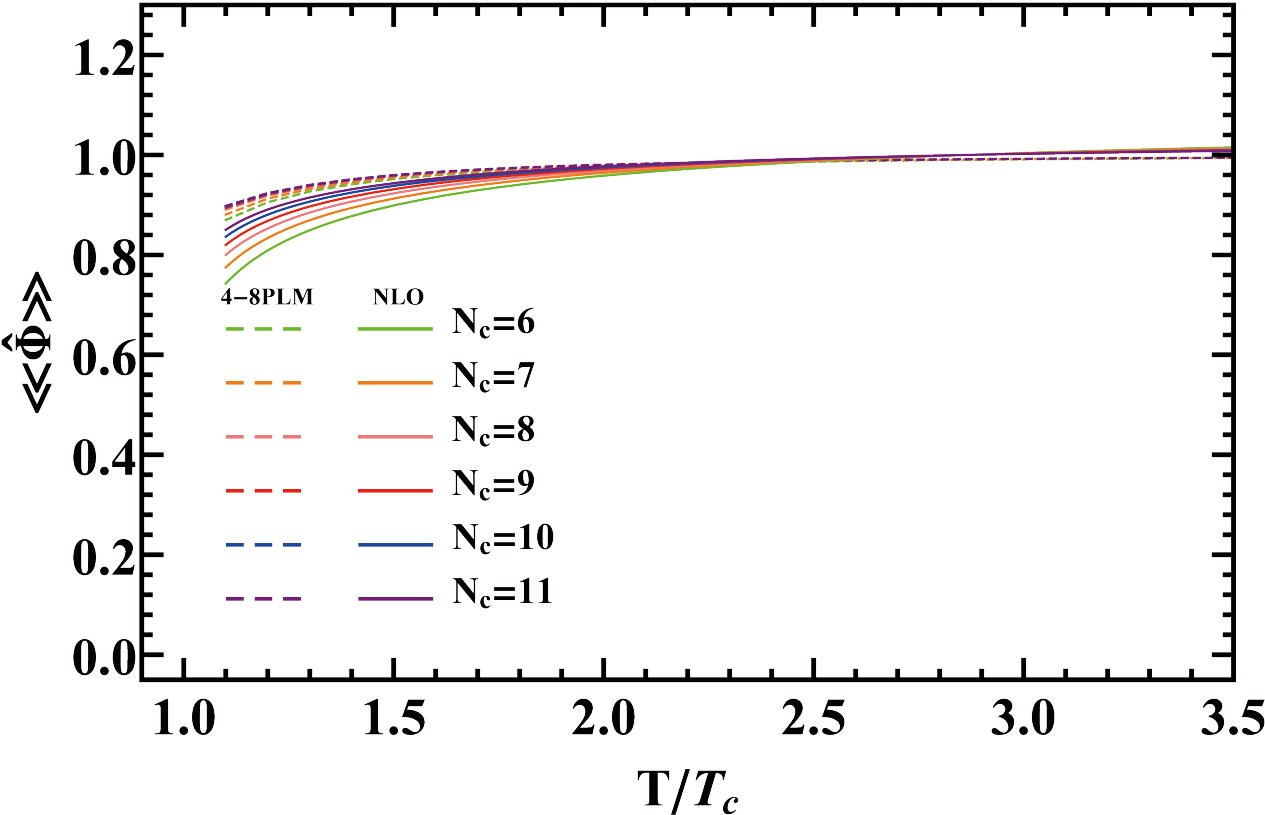}\quad	\includegraphics[scale=0.4]{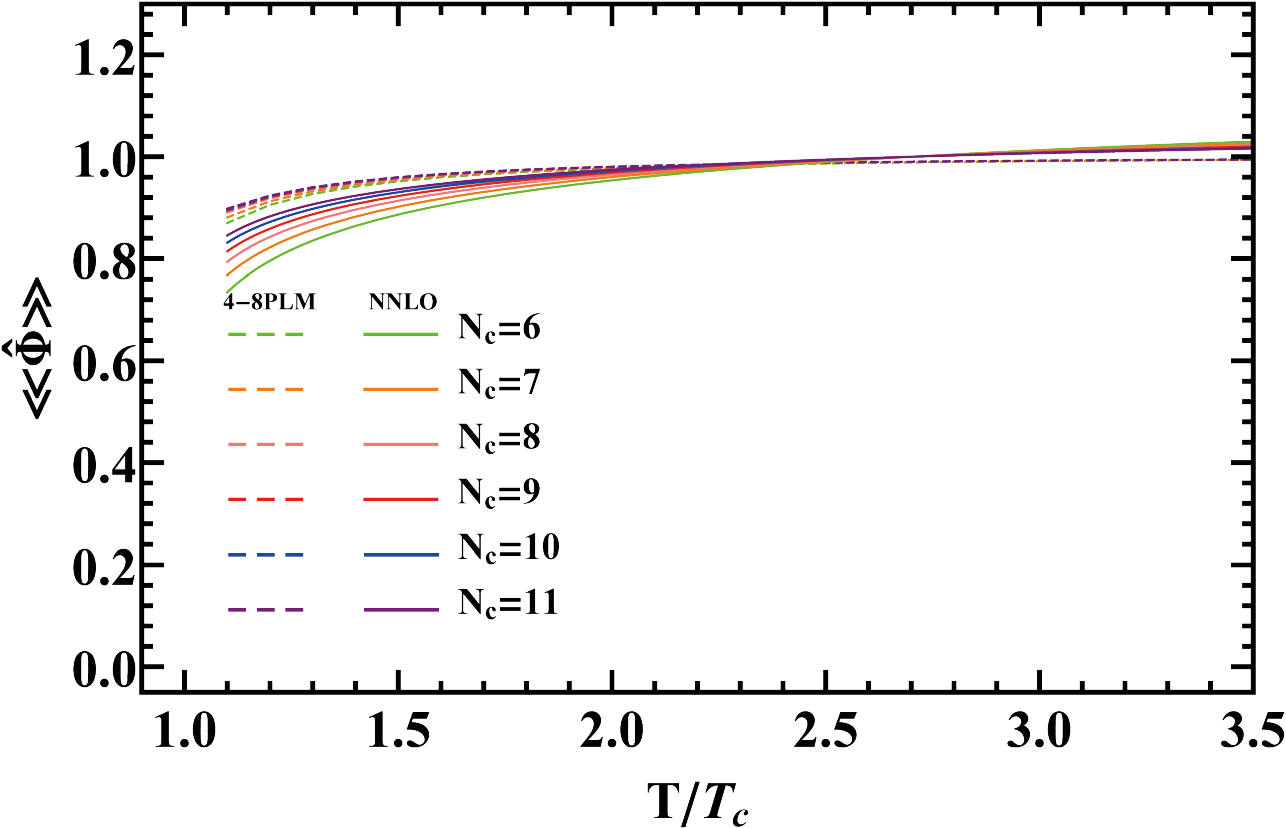}
	\caption{(color online). The curves of the expectation value of $\Phi$ as functions of $T$ in the large-$N_{\mathrm{c}}$ cases, i.e., $N_{\mathrm{c}}=6,7,\cdots,11$. The solid lines in the left panel and the right panel are the formulae $\langle\!\langle\hat{\Phi}\rangle\!\rangle_{\text{NLO}}$ with $(a_{N_{\mathrm{c}}})_{\text{ave}}=0.658406,~(T_{0,N_{\mathrm{c}}}/T_{c,N_{\mathrm{c}}})_{\text{ave}}=0.978634$ and $\langle\!\langle\hat{\Phi}\rangle\!\rangle_{\text{NNLO}}$ with $(a_{N_{\mathrm{c}}})_{\text{ave}}=0.830547,~b_{N_{\mathrm{c}}}\equiv-0.186231,~(T_{0,N_{\mathrm{c}}}/T_{c,N_{\mathrm{c}}})_{\text{ave}}=0.961284$ respectively. The dashed lines in both panels are the results of 4-8PLM model~\cite{Kang:2021epo}.}
	\label{fig:Phi largeN vs PLM}
\end{figure*}

As shown in Fig.~\ref{fig:Phi largeN vs PLM}, we qualitatively calculate $\langle\!\langle\hat{\Phi}\rangle\!\rangle_{\text{NLO}}$ and $\langle\!\langle\hat{\Phi}\rangle\!\rangle_{\text{NNLO}}$ for large-$N_{\mathrm{c}}$ cases where $N_{\mathrm{c}}=6,7,\cdots,11$, and we compare these results with those obtained using 4-8PLM Polyakov potentials from Ref.~\cite{Kang:2021epo}. For these large-$N_{\mathrm{c}}$ cases, the parameters in our formulas, i.e. $a_{N_{\mathrm{c}}}$ and $T_{0,N_{\mathrm{c}}}/T_{\mathrm{c},N_{\mathrm{c}}}$, are taken to be the arithmetic mean values listed in Table~\ref{tab:NLO para} for $\langle\!\langle\hat{\Phi}\rangle\!\rangle_{\text{NLO}}$ and in Table~\ref{tab:NNLO para} for $\langle\!\langle\hat{\Phi}\rangle\!\rangle_{\text{NNLO}}$. Note that $b_{N_{\mathrm{c}}}$ remains unchanged for these large-$N_{\mathrm{c}}$ values. We find that with the increasing of the number of colors, both $\langle\!\langle\hat{\Phi}\rangle\!\rangle_{\text{NLO}}$ and $\langle\!\langle\hat{\Phi}\rangle\!\rangle_{\text{NNLO}}$ curves are qualitatively coincident to that obtained using 4-8PLM Polyakov potentials. Specifically, when $T$ slightly exceeds the critical temperature, i.e., $T\sim 1.2 T_{\mathrm{c},N_{\mathrm{c}}}$, the expectation value of the Polyakov loop increases mildly with the growing $N_{\mathrm{c}}$ and the extent of this growth becomes smaller with larger color numbers, especially for $N_{\mathrm{c}}=10,11$, the expectation value changes very little.

\begin{figure}[htbp]
	\centering
	\includegraphics[scale=0.4]{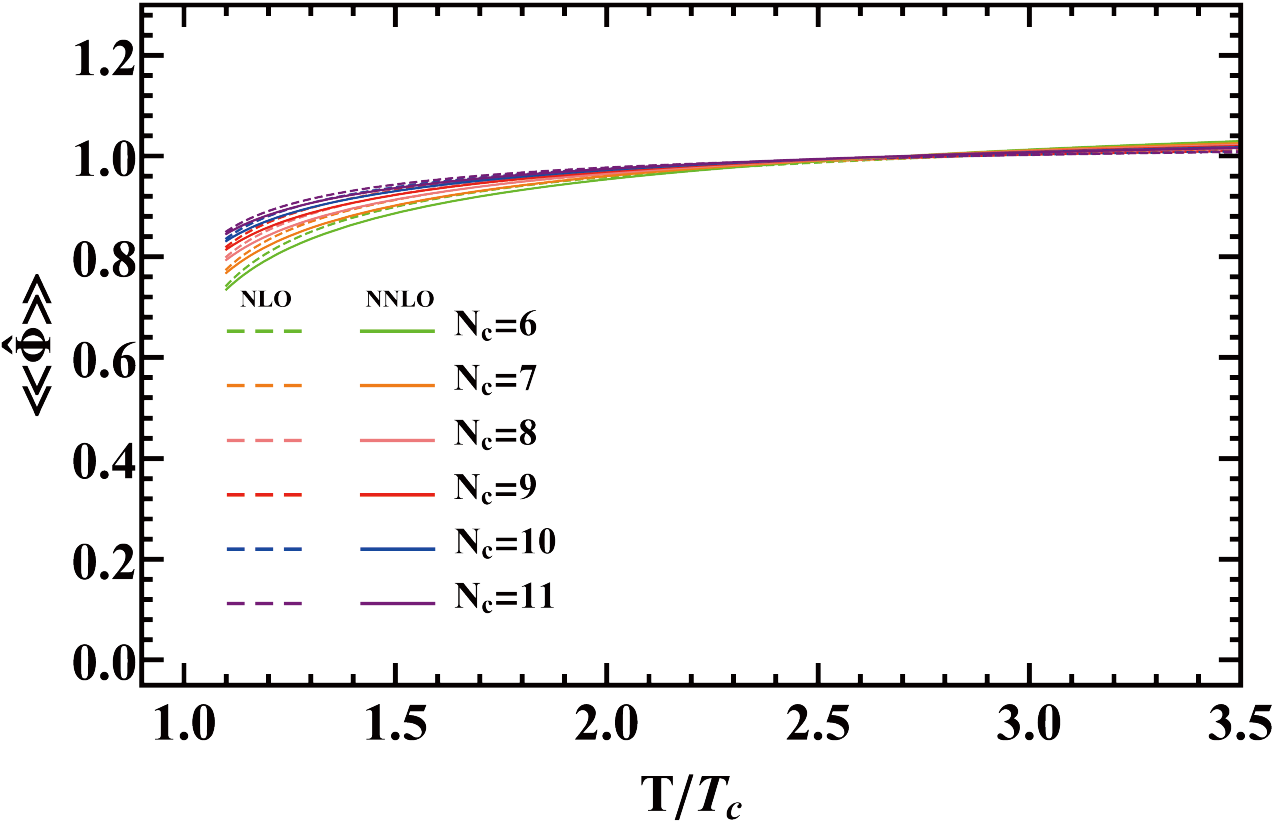}	
	\caption{\label{fig:Phi largeN NLO vs NNLO}(color online). The curves of $\langle\!\langle\hat{\Phi}\rangle\!\rangle_{\text{NLO}}$(dashed lines) with $(a_{N_{\mathrm{c}}})_{\text{ave}}=0.658406,~(T_{0,N_{\mathrm{c}}}/T_{c,N_{\mathrm{c}}})_{\text{ave}}=0.978634$ and $\langle\!\langle\hat{\Phi}\rangle\!\rangle_{\text{NNLO}}$(solid lines) with $(a_{N_{\mathrm{c}}})_{\text{ave}}=0.830547,~b_{N_{\mathrm{c}}}\equiv-0.186231,~(T_{0,N_{\mathrm{c}}}/T_{c,N_{\mathrm{c}}})_{\text{ave}}=0.961284$ in the large-$N_{\mathrm{c}}$ cases, i.e., $N_{\mathrm{c}}=6,7,\cdots,11$.}
\end{figure}

Before concluding this section, we compare the results of $\langle\!\langle\hat{\Phi}\rangle\!\rangle_{\text{NLO}}$ and $\langle\!\langle\hat{\Phi}\rangle\!\rangle_{\text{NNLO}}$ in the large-$N_{\mathrm{c}}$ limit. In Fig.~\ref{fig:Phi largeN NLO vs NNLO}, the curves of $\langle\!\langle\hat{\Phi}\rangle\!\rangle_{\text{NLO}}$ and $\langle\!\langle\hat{\Phi}\rangle\!\rangle_{\text{NNLO}}$ are nearly coincide. This is a rather natural result since in the large-$N_{\mathrm{c}}$ limit, both $\langle\!\langle\hat{\Phi}\rangle\!\rangle_{\text{NLO}}$ and $\langle\!\langle\hat{\Phi}\rangle\!\rangle_{\text{NNLO}}$ converge to unity.

\section{Summary and discussion}
\label{sec:Sum}

In this work, we devote ourselves to finding an underlying connection between the trace anomaly and the confinement phenomenon of QCD, and constructing a possible relationship between the dilaton field and the Polyakov loop.

To achieve our purpose, we first derived the scale transformation properties of temperature and chemical potentials within the framework of quantum statistical mechanics. We examined the scale covariance of the fundamental equation of quantum statistical mechanics, the Liouville equation, and concluded that the density operator $\hat{\rho}$ which describes a mixed quantum state at finite temperatures and/or densities, is a scale-invariant object for scale-symmetric systems. This conclusion leads to the scale transformation properties $T'=\lambda^{-1}T$ and $\mu'_{i}=\lambda^{-1}\mu_{i},i=1,\cdots,K$. With these properties, we established the scale invariance of the Polyakov loop $\Phi(\bm{x})$ and the scale covariance of the four laws of thermodynamics. 

An interesting and important conclusion is that for any scale-symmetric system, the EoS can be written down straightforwardly without the tedious calculation of the partition function based on a specific microscopic model. In other words, as long as one has the scale transformation properties of the fundamental thermodynamic quantities---such as pressure $P$, temperature $T$ and chemical potentials $\mu_{i},i=1,2,\cdots,K$---one can get the building blocks of a scale-symmetric EoS and construct the EoS directly. This procedure is similar to that of constructing a microscopic Lagrangian with specific symmetries once the transformation properties of relevant fields are known. We provided three typical examples of scale-symmetric EoSs. Two of them are the well-known Stefan\textendash Boltzmann law, i.e., $P\propto T^{4}$ and $P\propto \mu^{4}_{\mathrm{B}}$, which apply to thermodynamic systems dominated by massless particles at extremely high temperatures or high fermion densities~\cite{Fujimoto:2022ohj}. The third example is the pressure of a non-interacting gas of massless fermions at high temperatures but low chemical potentials~\cite{Kapusta_Gale_2023}. Furthermore, we derived a differential equation for scale-symmetric EoSs based on the scale transformation rules of pressure and temperature and found a corresponding solution. We confirmed that this differential equation is consistent with the traceless of energy-momentum tensor $\epsilon-3P=0$.

Considering the gluon field as a static and homogeneous background field, as adopted in the study of the deconfinement phase transition at high temperatures~\cite{Fukushima:2003fw,Ratti:2005jh}, we generally constructed a $\mathbb{Z}_{N_{\mathrm{c}}}$-symmetric power series of the Polyakov loop. We regarded this series is equal to the ratio of the thermal fluctuation $\delta_{\chi,\mathrm{th}}^{(2)}$ and the temperature-modified low energy constant $\tilde{f}_{\sigma}$. Subsequently, we derived the temperature dependence of coefficients in the series by using the scale transformation property of temperature that we obtained. In addition, we found that the thermal fluctuation of dilaton particles is proportional to temperature, i.e., $\delta^{(1)}_{\chi,\mathrm{th}}\propto T$. 

With the above general properties, we meticulously analyzed the minimum points of the thermodynamic potential $\Omega_{\chi}$ up to the NLO and the NNLO of the power series Eq.~(\ref{delta2 expa}). We derived the formulas of the expectation value of the Polyakov loop as functions of $T$, i.e., $\langle\!\langle\hat{\Phi}\rangle\!\rangle_{\text{NLO}}$ and $\langle\!\langle\hat{\Phi}\rangle\!\rangle_{\text{NNLO}}$ by regarding the first-order deconfinement phase transition in pure $\mathrm{SU(N_{\mathrm{c}})}$ gauge sector as a known condition.

Finally, we evaluated the constants $a_{N_{\mathrm{c}}}$, $b_{N_{\mathrm{c}}}$ and $T_{0,N_{\mathrm{c}}}/T_{\mathrm{c},N_{\mathrm{c}}}$ by performing a global fit to the LQCD data for $N_{\mathrm{c}}=3, 4, 5$. Our results using the formula $\langle\!\langle\hat{\Phi}\rangle\!\rangle_{\text{NLO}}$ are in good agreement with the LQCD data for both the $\mathrm{SU(3)}$ and $\mathrm{SU(4)}$ cases. However, a deviation is observed for $N_{\mathrm{c}}=5$, which can be partially corrected using the formula $\langle\!\langle\hat{\Phi}\rangle\!\rangle_{\text{NNLO}}$. Noticed that the dependence of the parameters in the formulae on the color number is not very drastic, particularly for $T_{0,N_{\mathrm{c}}}/T_{\mathrm{c},N_{\mathrm{c}}}$, we qualitatively estimated the behavior of the expectation value of $\Phi$ in the large-$N_{\mathrm{c}}$ cases, specifically for $N_{\mathrm{c}}=6,7,\cdots,11$, based on our formulae. It was found that both the formulae $\langle\!\langle\hat{\Phi}\rangle\!\rangle_{\text{NLO}}$ and $\langle\!\langle\hat{\Phi}\rangle\!\rangle_{\text{NNLO}}$ effectively describe the large-$N_{\mathrm{c}}$ behavior when compared the results from the 4-8PLM model~\cite{Kang:2021epo}. Our formulas predicted that if the large-$N_{\mathrm{c}}$ limit is taken within the temperature interval $T_{\mathrm{c},N_{\mathrm{c}}}<T\lesssim 2.5T_{\mathrm{c},N_{\mathrm{c}}}$, the expectation value of the Polyakov loop will approach unity. This indicates the presence of a strong first-order deconfinement phase transition.

The good agreement of our results with LQCD data, as well as the 4-8PLM results, further suggests that there are some relations between the trace anomaly and the confinement phenomenon. However, this work does not account for the dynamics of dilaton particles and the strongly coupled deconfined gluons. In addition, since our formulas are derived from the low-energy description of QCD trace anomaly, we anticipate that they may become less applicable at extremely high temperatures, for example $T\gg T_{\mathrm{c},N_{\mathrm{c}}}$. These issues, including the effect of quark, will be addressed in the future work.

\acknowledgments	

The work of Y.~L. M. is supported in part by the National Science Foundation of China (NSFC) under Grant No. 12347103, the National Key R\&D Program of China under Grant No. 2021YFC2202900 and Gusu Talent Innovation Program under Grant No. ZXL2024363.
	
\appendix

\section{The Liouville equation in classical and quantum field theories}

\label{app:Liouville}

To derive the scale transformation properties of temperature $T$, we first clarify some basic conceptions about the Liouville equation---the fundamental equation in statistical mechanics---in the framework of field theory.

Let us begin with a more intuitive yet accessible illustration of the fundamental notations used in classical Hamilton field theory. We consider a field defined in Minkowski spacetime, in one constant-time hypersurface. This system possesses an infinite number of DoFs in physical three-dimensional space $(\mathbb{R}^{3},\delta)$, with $\delta$ being the Euclidean metric. Specifically, we have mappings $\phi^{r}: (\mathbb{R}^{3},\delta) \to \mathbb{R}$ or equivalently $\forall \bm{x}\in (\mathbb{R}^{3},\delta) ,\phi^{r}: \bm{x} \mapsto \phi^{r}(\bm{x}), \phi^{r}(\bm{x})\in\mathbb{R}$ where $r=1,2,\cdots,D$ is the index of intrinsic DoFs such as spin, isospin, etc. In addition to the mappings of the field, we also have mappings of the canonical momenta $\pi_{r}: (\mathbb{R}^{3},\delta) \to \mathbb{R}$ or equivalently $\forall\bm{x}\in (\mathbb{R}^{3},\delta), \pi_{r}: \bm{x} \mapsto \pi_{r}(\bm{x}), \pi_{r}(\bm{x})\in\mathbb{R}$. Given the field is a distributed entity throughout the three-dimensional space, a particular state of the field is characterized by the complete set of mappings $\phi^{r}$ and $\pi_{r}$, rather than the values of $\phi^{r}(\bm{x}_{0})$ and $\pi_{r}(\bm{x}_{0})$ at a specific point $\bm{x}_{0}$ in the space. 

Different mappings of the field $\phi^{r}(\bm{x})$ and the associated canonical momenta $\pi_{r}(\bm{x})$ correspond to different states of the system. These mappings actually constitute a function space $\mathbb{M}_{(D)}$:
\be
\label{def M_D}
\mathbb{M}_{(D)} & \equiv & \mathbb{M}_{(\phi)}^{1}\times\mathbb{M}_{(\phi)}^{2}\times\cdots\times\mathbb{M}_{(\phi)}^{D} \nm \\
& & {} \times\mathbb{M}_{(\pi),1}\times\mathbb{M}_{(\pi),2}\times\cdots\times\mathbb{M}_{(\pi),D},
\ee
where $\displaystyle \mathbb{M}_{(\phi)}^{r}\equiv\left\{\phi^{r}|\phi^{r}:(\mathbb{R}^{3},\delta)\to\mathbb{R}\right\}$ and $\displaystyle\mathbb{M}_{(\pi),r}\equiv\{\pi_{r}|\pi_{r}:(\mathbb{R}^{3},\delta)\to\mathbb{R}\}$. For the same reason, since the field is a global object in the space, a dynamical variable, such as the Hamiltonian (total energy), momentum, charge density and so on, which is a real functional of the field and the canonical momentum, satisfies $F_{t}:\mathbb{M}_{(D)}\to \mathbb{R}$, that is
\be
F_{t} & : & (\phi^{1},\phi^{2},\cdots,\phi^{D},\pi_{1},\pi_{2},\cdots,\pi_{D}) \nonumber \\
& &{} \mapsto F_{t}[\phi^{1},\phi^{2},\cdots,\phi^{D},\pi_{1},\pi_{2},\cdots,\pi_{D}]\in \mathbb{R}, \nm\\
& \mbox {for} & \forall (\phi^{1},\phi^{2},\cdots,\phi^{D},\pi_{1},\pi_{2},\cdots,\pi_{D}) \in \mathbb{M}_{(D)} .\nm
\ee
Here, we use the following convention for convenience:
\be
\label{convention F_t}
F[\phi^{r},\pi_{r};t]\equiv F_{t}[\phi^{1},\phi^{2},\cdots,\phi^{D},\pi_{1},\pi_{2},\cdots,\pi_{D}].
\ee
It should be emphasized that the subscript of $F_{t}$ indicates that in general a dynamical variable is explicitly time-dependent. A typical example is the conserved charge of scale transformation
\be
\label{Q_scale}
Q_{{\ssp\mathrm{D}}}[\phi^{r},\pi_{r};t] & = & \int\limits_{\mathbb{R}^{3}}\mathrm{d}^{3}xJ_{{\ssp\mathrm{D}}}^{0} \nonumber \\
& = & \int\limits_{\mathbb{R}^{3}}\mathrm{d}^{3}x(x^{\nu}t^{0}_{~\nu}+\sum_{r}\pi_{r}\phi^{r}\Delta_{\phi^{r}}) \nonumber \\
& = & tH[\phi^{r},\pi_{r}] \nm\\
& &{} +\int\limits_{\mathbb{R}^{3}}\mathrm{d}^{3}x\sum_{r}\big[\bm{x}\cdot(\pi_{r}\bm{\nabla}\phi^{r})+\pi_{r}\phi^{r}\Delta_{\phi^{r}}\big], \nm\\
\ee 
where $t^{\mu}_{~\nu}$ is the energy-momentum tensor, $H$ is the Hamiltonian and $\Delta_{\phi^{r}}$ is the scale dimension of field $\phi^{r}$.
    
The phase space of field theory can be defined in analogy to that of classical mechanics:
\begin{widetext}
\be
\label{def phase space}
\mathbb{P}\equiv&\left\{(\cdots\phi^{1}(\bm{x})\cdots,\cdots,\cdots\phi^{D}(\bm{x})\cdots,\cdots\pi_{1}(\bm{x})\cdots,\cdots,\cdots\pi_{D}(\bm{x})\cdots)^{\mathrm{T}}|\phi^{r}\in\mathbb{M}^{r}_{(\phi)}, \pi_{r}\in\mathbb{M}_{(\pi),r}\right\}.
\ee
\end{widetext}	
Here, the symbol $\cdots\phi^{r}(\bm{x})\cdots$($\cdots\pi_{r}(\bm{x})\cdots$) means that the values of $\phi^{r}(\bm{x})$($\pi_{r}(\bm{x})$) are listed continuously and densely from left to right. An element in the phase space of field theory is termed phase point and is described by a $2D\times(+\infty)$-dimensional continuous real column matrix. Then, one can immediately realize that given an element of $\mathbb{M}_{(D)}$, i.e., $(\phi^{1},\cdots,\phi^{D},\pi_{1},\cdots,\pi_{D})$, the corresponding phase point $(\cdots\phi^{1}(\bm{x}),\cdots,\phi^{D}(\bm{x})\cdots,\cdots\pi_{1}(\bm{x}),\cdots, \cdots\pi_{D}(\bm{x})\cdots)^{\mathrm{T}}$ is fixed. In other words, a state of the field is locked to a phase point or each element of $\mathbb{M}_{(D)}$ is corresponding to a single phase point of the phase space, namely a single state of the field. 
    
With the above definitions and discussions, we are ready to investigate the time evolution of the state in an interval $I=[t_{\mathrm{i}},t_{\mathrm{f}}]$ or the motion of a phase point in the phase space with a certain initial condition. Here, $t_{\mathrm{i}}$ and $t_{\mathrm{f}}$ are initial time and final time respectively. Actually, a time evolution of the state corresponds to a curve in function space $\mathbb{M}_{(D)}$ that is parameterized by time $t\in I$. We denote the parametric equations as
\be
\phi^{k}=\phi^{k}_{(t)}, \; \pi_{k}=\pi_{k,(t)}, \; k=1,\cdots, D,
\label{para eq M_D}
\ee
with initial conditions
\be
\phi_{\mathrm{i}}^{k}=\phi^{k}_{(t_{\mathrm{i}})}, \; \pi_{k,\mathrm{i}}=\pi_{k,(t_{\mathrm{i}})},\; k=1,\cdots, D.
\label{ini cond M_D}
\ee
Equivalently, a time evolution of the state is corresponding to a curve in the phase space $\mathbb{P}$ and the phase point in this curve at time $t$ is
\be
\label{phase point at t}
& & (\cdots\phi^{1}_{(t)}(\bm{x})\cdots,\cdots,\cdots\phi^{D}_{(t)}(\bm{x})\cdots, \nm\\
& &{} \;\; \cdots\pi_{1,(t)}(\bm{x})\cdots,\cdots, \cdots\pi_{D,(t)}(\bm{x})\cdots)^{\mathrm{T}},
\ee
where $\phi^{r}_{(t)}(\bm{x})$ and $\pi_{r,(t)}(\bm{x})$ are the field and its canonical momentum defined in four-dimensional Minkowski spacetime, i.e., 
\be
\phi^{r}(x) & \equiv & \phi^{r}(\bm{x},t)\equiv\phi^{r}_{(t)}(\bm{x}), 
\label{phi in spacetime} \\
\pi_{r}(x) & \equiv & \pi_{r}(\bm{x},t)\equiv\pi_{r,(t)}(\bm{x}).
\label{pi in spacetime}
\ee
At the initial time $t_{\mathrm{i}}$, the phase point is located at
\be
\label{ini cond P}
& & (\cdots\phi^{1}_{\mathrm{i}}(\bm{x})\cdots,\cdots,\cdots\phi^{D}_{\mathrm{i}}(\bm{x})\cdots, \nm \\
& &{}\;\; \cdots\pi_{1,\mathrm{i}}(\bm{x})\cdots,\cdots, \cdots\pi_{D,\mathrm{i}}(\bm{x})\cdots)^{\mathrm{T}}.
\ee

The physical curve (or the so-called true trajectory of the phase point) in the phase space is given by the following canonical equations
\be
\label{cano eq}
\dot{\phi}^{r}(\bm{x},t) & = & \dfrac{\delta_{{\scriptscriptstyle\partial}}H[\phi^{r}_{(t)},\pi_{r,(t)}]}{\delta_{{\scriptscriptstyle\partial}}\pi_{r,(t)}(\bm{x})}=\dfrac{\delta_{{\scriptscriptstyle\partial}}H[\phi^{r}_{(t)},\pi_{r,(t)}]}{\delta_{{\scriptscriptstyle\partial}}\pi_{r}(\bm{x},t)} ,\\
\dot{\pi}_{r}(\bm{x},t) & = &{} -\dfrac{\delta_{{\scriptscriptstyle\partial}}H[\phi^{r}_{(t)},\pi_{r,(t)}]}{\delta_{{\scriptscriptstyle\partial}}\phi^{r}_{(t)}(\bm{x})}=-\dfrac{\delta_{{\scriptscriptstyle\partial}}H[\phi^{r}_{(t)},\pi_{r,(t)}]}{\delta_{{\scriptscriptstyle\partial}}\phi^{r}(\bm{x},t)}.\nm\\
\ee
Here, $\dot{\phi}^{r}(\bm{x},t)\equiv\dot{\phi}^{r}_{(t)}(\bm{x})\equiv\partial\phi^{r}(\bm{x},t)/\partial t\equiv\partial\phi^{r}_{(t)}(\bm{x})/\partial t$ and $\dot{\pi}_{r}(\bm{x},t)\equiv\dot{\pi}_{r,(t)}(\bm{x})\equiv\partial\pi_{r}(\bm{x},t)/\partial t\equiv\partial\pi_{r,(t)}(\bm{x})/\partial t$. The symbol $\delta_{{\scriptscriptstyle\partial}}$ represents the partial derivative of a multivariate functional $F_{t}$ in Eq.~(\ref{convention F_t}) which is defined as
\be
& & \int\limits_{\mathbb{R}^{3}}\mathrm{d}^{3}x\dfrac{\delta_{{\scriptscriptstyle\partial}}F[\phi^{r},\pi_{r};t]}{\delta_{{\scriptscriptstyle\partial}}\phi^{r}(\bm{x})}f^{r}(\bm{x}) \label{pd of F_t phi} \\
& & {} \equiv\lim_{\varepsilon^{r}\to 0}\Big\{F_{t}[\phi^{1},\cdots,\phi^{r}+\varepsilon^{r}f^{r},\cdots,\phi^{D},\pi_{1},\cdots,\pi_{D}]  \nonumber \\
& & {} \qquad\qquad \;\; -F_{t}[\phi^{1},\cdots,\phi^{r},\cdots,\phi^{D},\pi_{1},\cdots,\pi_{D}]\Big\}\Big/\varepsilon^{r}, \nm\\
& & \int\limits_{\mathbb{R}^{3}}\mathrm{d}^{3}x\dfrac{\delta_{{\scriptscriptstyle\partial}}F[\phi^{r},\pi_{r};t]}{\delta_{{\scriptscriptstyle\partial}}\pi_{r}(\bm{x})}g_{r}(\bm{x}) \label{pd of F_t pi} \\
& & {} \equiv\lim_{\eta_{r}\to 0}\Big\{F_{t}[\phi^{1},\cdots,\phi^{D},\pi_{1},\cdots,\pi_{r}+\eta_{r}g_{r},\cdots,\pi_{D}] \nonumber \\
& & {} \qquad\qquad \;\; -F_{t}[\phi^{1},\cdots,\phi^{D},\pi_{1},\cdots,\pi_{r},\cdots,\pi_{D}]\Big\}\Big/\eta_{r},\nm
\ee
where $f^{r}(\bm{x}), r=1,2,\cdots,D$ and $g_{r}(\bm{x}), r=1,2,\cdots,D$ are test functions which have arbitrary selectivity. Note the Einstein's convention of summation is not used for index $r$. The variation of $F[\phi^{r},\pi_{r};t]$ reads
\be
\label{vara F_t}
\delta F[\phi^{r},\pi_{r};t] & \equiv & F[\phi^{r}+\delta\phi^{r},\pi_{r}+\delta\pi_{r};t]-F[\phi^{r},\pi_{r};t] \nonumber \\
& = & \int\limits_{\mathbb{R}^{3}}\!\!\mathrm{d}^{3}x\!\!\sum_{r}\Bigg\{\frac{\delta_{\scriptscriptstyle\partial}F[\phi^{r},\pi_{r};t]}{\delta_{\scriptscriptstyle\partial}\phi^{r}(\bm{x})}\delta\phi^{r}(\bm{x}) \nm\\
& &{}\qquad\qquad\;\, +\frac{\delta_{\scriptscriptstyle\partial}F[\phi^{r},\pi_{r};t]}{\delta_{\scriptscriptstyle\partial}\pi_{r}(\bm{x})}\delta\pi_{r}(\bm{x})\Bigg\}.\nm\\
\ee

\begin{figure}[htbp]
	\centering
	\includegraphics[scale=0.3]{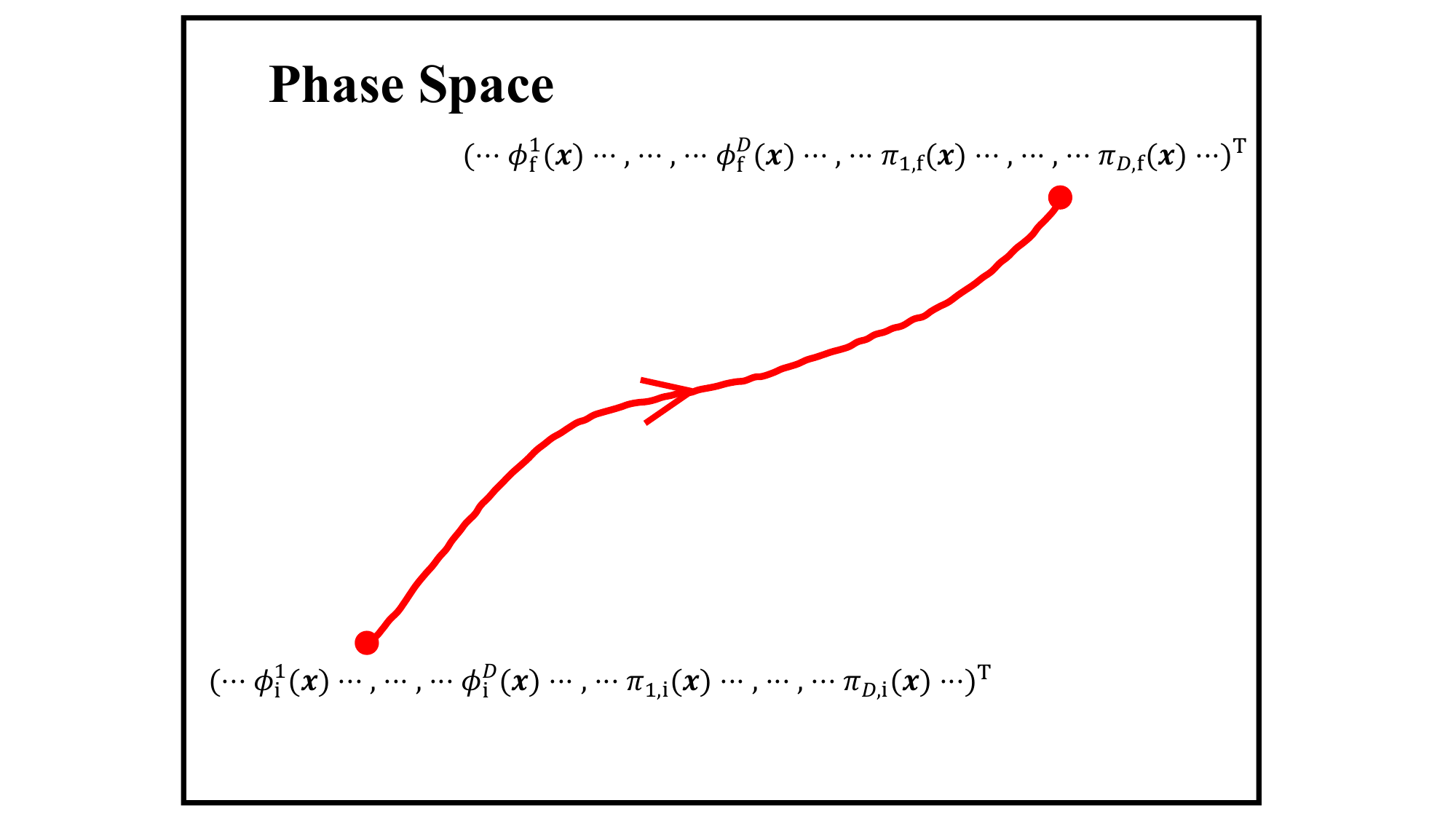}	
	\caption{The true trajectory of the phase point corresponding to a time evolution of the system from time $t_{\mathrm{i}}$ to $t_{\mathrm{f}}$.}
    \label{phase space sys}
\end{figure}

The trajectory of the phase point is sketched in Fig.~\ref{phase space sys}. Any dynamical variable $F[\phi^{r},\pi_{r};t]$ on the true trajectory---the trajectory corresponds to the minimal action of the system in classical field theory---is actually a function of time, i.e., $F(t)\equiv F[\phi^{r}_{(t)},\pi_{r,(t)};t]$. Equivalently, mappings (solutions) of the canonical equation set~(\ref{cano eq}) are substituted into $F[\phi^{r},\pi_{r};t]$. The equation of motion of $F(t)$ can be derived as 
\begin{widetext}
\be
\label{eom F}
\frac{\mathrm{d}F(t)}{\mathrm{d}t} & = & \lim_{\Delta t \to 0}\frac{F(t+\Delta t)-F(t)}{\Delta t}=\lim_{\Delta t\to 0}\frac{F\left[\phi^{r}_{(t+\Delta t)},\pi_{r,(t+\Delta t)};t+\Delta t\right]-F\left[\phi^{r}_{(t)},\pi_{r,(t)};t\right]}{\Delta t} \nonumber \\
& = & \lim_{\Delta t \to 0}\frac{1}{\Delta t}\left\{F\left[\phi^{r}_{(t)}+(\Delta t)\dot{\phi}^{r}_{(t)},\pi_{r,(t)}+(\Delta t)\dot{\pi}_{r,(t)};t+\Delta t \right]-F\left[\phi^{r}_{(t)},\pi_{r,(t)};t \right]\right\} \nonumber \\
& = & \lim_{\Delta t \to 0}\frac{1}{\Delta t}\Bigg\{\int\limits_{\mathbb{R}^{3}}\mathrm{d}^{3}x\sum_{r}\bigg(\frac{\delta_{{\scriptscriptstyle\partial}}F\left[\phi^{r}_{(t)},\pi_{r,(t)};t+\Delta t\right]}{\delta_{{\scriptscriptstyle\partial}}\phi^{r}_{(t)}(\bm{x})}(\Delta t)\dot{\phi}^{r}_{(t)}(\bm{x})+\frac{\delta_{{\scriptscriptstyle\partial}}F\left[\phi^{r}_{(t)},\pi_{r,(t)};t+\Delta t\right]}{\delta_{{\scriptscriptstyle\partial}}\pi_{r,(t)}(\bm{x})}(\Delta t)\dot{\pi}_{r,(t)}(\bm{x})\bigg) \nonumber \\
& &{}\qquad\qquad\; + F\left[\phi^{r}_{(t)},\pi_{r,(t)};t+\Delta t\right]-F\left[\phi^{r}_{(t)},\pi_{r,(t)};t\right]\Bigg\},
\ee
and the partial derivative of $F\left[\phi^{r}_{(t)},\pi_{r,(t)};t\right]$ with respect to time is defined as
\be
\label{pd of F t}
\frac{\partial F(t)}{\partial t}\equiv\frac{\partial F[\phi^{r}_{(t)},\pi_{r,(t)};t]}{\partial t}\equiv F'_{t}[\phi^{r}_{(t)},\pi_{r,(t)};t]\equiv\lim_{\Delta t\to 0}\frac{F[\phi^{r}_{(t)},\pi_{r,(t)};t+\Delta t]-F[\phi^{r}_{(t)},\pi_{r,(t)};t]}{\Delta t}.
\ee	
Then Eq.~(\ref{eom F}) becomes
\be
\label{eom F v2}
\frac{\mathrm{d}F(t)}{\mathrm{d}t} & = & \lim_{\Delta t\to 0}\frac{1}{\Delta t}\int\limits_{\mathbb{R}^{3}}\mathrm{d}^{3}x\sum_{r}\Bigg\{\frac{\delta_{{\scriptscriptstyle\partial}}}{\delta_{{\scriptscriptstyle\partial}}\phi^{r}_{(t)}(\bm{x})}\bigg(F\left[\phi^{r}_{(t)},\pi_{r,(t)};t\right]+\frac{\partial F\left[\phi^{r}_{(t)},\pi_{r,(t)};t\right]}{\partial t}(\Delta t)\bigg)(\Delta t)\dot{\phi}^{r}_{(t)}(\bm{x}) \nonumber \\ 
& & {}\qquad \qquad \qquad \qquad \quad + \frac{\delta_{{\scriptscriptstyle\partial}}}{\delta_{{\scriptscriptstyle\partial}}\pi_{r,(t)}(\bm{x})}\bigg(F\left[\phi^{r}_{(t)},\pi_{r,(t)};t\right]+\frac{\partial F\left[\phi^{r}_{(t)},\pi_{r,(t)};t\right]}{\partial t}(\Delta t)\bigg)(\Delta t)\dot{\pi}_{r,(t)}(\bm{x})\Bigg\}+\frac{\partial F(t)}{\partial t} \nonumber \\
& = & \int\limits_{\mathbb{R}^{3}}\mathrm{d}^{3}x\sum_{r}\Bigg\{\frac{\delta_{{\scriptscriptstyle\partial}}F\left[\phi^{r}_{(t)},\pi_{r,(t)};t\right]}{\delta_{{\scriptscriptstyle\partial}}\phi^{r}_{(t)}(\bm{x})}\dot{\phi}^{r}_{(t)}(\bm{x})+\frac{\delta_{{\scriptscriptstyle\partial}}F\left[\phi^{r}_{(t)},\pi_{r,(t)};t\right]}{\delta_{{\scriptscriptstyle\partial}}\pi_{r,(t)}(\bm{x})}\dot{\pi}_{r,(t)}(\bm{x})\Bigg\}+\frac{\partial F(t)}{\partial t}.
\ee
Recalling the canonical equation~(\ref{cano eq}), we finally obtain
\be
\label{eom F v3}
\frac{\mathrm{d}F(t)}{\mathrm{d}t} & = & \int\limits_{\mathbb{R}^{3}}\mathrm{d}^{3}x\sum_{r}\Bigg\{\frac{\delta_{{\scriptscriptstyle\partial}}F\left[\phi^{r}_{(t)},\pi_{r,(t)};t\right]}{\delta_{{\scriptscriptstyle\partial}}\phi^{r}_{(t)}(\bm{x})}\frac{\delta_{{\scriptscriptstyle\partial}}H\left[\phi^{r}_{(t)},\pi_{r,(t)}\right]}{\delta_{{\scriptscriptstyle\partial}}\pi_{r,(t)}(\bm{x})}-\frac{\delta_{{\scriptscriptstyle\partial}}F\left[\phi^{r}_{(t)},\pi_{r,(t)};t\right]}{\delta_{{\scriptscriptstyle\partial}}\pi_{r,(t)}(\bm{x})}\frac{\delta_{{\scriptscriptstyle\partial}}H\left[\phi^{r}_{(t)},\pi_{r,(t)}\right]}{\delta_{{\scriptscriptstyle\partial}}\phi^{r}_{(t)}(\bm{x})}\Bigg\}+\frac{\partial F(t)}{\partial t} \nonumber \\
& = & \Big\{F\left[\phi^{r}_{(t)},\pi_{r,(t)};t\right],H\left[\phi^{r}_{(t)},\pi_{r,(t)}\right]\Big\}_{\mathrm{PB}}+\frac{\partial F(t)}{\partial t},
\ee
where $\displaystyle \Big\{F[\phi^{r}_{(t)},\pi_{r,(t)};t],H[\phi^{r}_{(t)},\pi_{r,(t)}]\Big\}_{\mathrm{PB}}$ is the Poisson bracket at time $t$. Generally speaking, for two arbitrary dynamical variables $F[\phi^{r},\pi_{r};t]$ and $G[\phi^{r},\pi_{r};t]$, the Poisson bracket at time $t$ reads
\be
\label{PB}
\Big\{F\left[\phi^{r},\pi_{r};t],G[\phi^{r},\pi_{r};t\right]\Big\}_{\mathrm{PB}}\equiv\int\limits_{\mathbb{R}^{3}}\mathrm{d}^{3}x\sum_{r}\Bigg\{\frac{\delta_{{\scriptscriptstyle\partial}}F\left[\phi^{r},\pi_{r};t\right]}{\delta_{{\scriptscriptstyle\partial}}\phi^{r}(\bm{x})}\frac{\delta_{{\scriptscriptstyle\partial}}G\left[\phi^{r},\pi_{r};t\right]}{\delta_{{\scriptscriptstyle\partial}}\pi_{r}(\bm{x})}-\frac{\delta_{{\scriptscriptstyle\partial}}F\left[\phi^{r},\pi_{r};t\right]}{\delta_{{\scriptscriptstyle\partial}}\pi_{r}(\bm{x})}\frac{\delta_{{\scriptscriptstyle\partial}}G\left[\phi^{r},\pi_{r};t\right]}{\delta_{{\scriptscriptstyle\partial}}\phi^{r}(\bm{x})}\Bigg\}.
\ee
\end{widetext}

If the field $\phi^{r}_{(t)}(\bm{x})$ and the canonical momentum $\pi_{r,(t)}(\bm{x})$ are treated as special dynamical variables, i.e., 
\be
\phi^{r}_{(t)}(\bm{x})\equiv F^{r}_{(\phi)}[\phi^{r}_{(t)};\bm{x}]\equiv\int\limits_{\mathbb{R}^{3}}\mathrm{d}^{3}y\phi^{r}_{(t)}(\bm{y})\delta^{(3)}(\bm{y}-\bm{x}),\nm\\
\pi_{r,(t)}(\bm{x})\equiv F_{(\pi),r}[\pi_{r,(t)};\bm{x}]\equiv\int\limits_{\mathbb{R}^{3}}\mathrm{d}^{3}y\pi_{r,(t)}(\bm{y})\delta^{(3)}(\bm{y}-\bm{x}) , 
\nm\\
\label{pi as dynam}
\ee
one can check that Eq.~(\ref{eom F v3}) comprises the canonical equation set (\ref{cano eq}). In this case, the functionals $F^{r}_{(\phi)}[\phi^{r}_{(t)};\bm{x}]$ and $F_{(\pi),r}[\pi_{r,(t)};\bm{x}]$ are called local functionals at space point $\bm{x}$.

One can be conscious of the explicit time-independence of $F^{r}_{(\phi)}[\phi^{r}_{(t)};\bm{x}]$ and $F_{(\pi),r}[\pi_{r,(t)};\bm{x}]$. Actually, using definition~(\ref{pd of F t}) we have
\be
\frac{\partial F^{r}_{(\phi)}[\phi^{r}_{(t)};\bm{x}]}{\partial t} & = & \lim_{\Delta t\to 0}\frac{F^{r}_{(\phi)}[\phi^{r}_{(t)};\bm{x}]-F^{r}_{(\phi)}[\phi^{r}_{(t)};\bm{x}]}{\Delta t}=0,\nm\\
\frac{\partial F_{(\pi),r}[\pi_{r,(t)};\bm{x}]}{\partial t} & = & \lim_{\Delta t\to 0}\dfrac{F_{(\pi),r}[\pi_{r,(t)};\bm{x}]-F_{(\pi),r}[\pi_{r,(t)};\bm{x}]}{\Delta t} \nm\\
& = & 0.\label{pd of F t}
\ee
Note that the partial derivative with respect to time in in the left hands of Eq.~(\ref{pd of F t}) should not be confused with $\dot{\phi}^{r}_{(t)}(\bm{x})=\lim_{\Delta t \to 0}[\phi^{r}_{(t+\Delta t)}(\bm{x})-\phi^{r}_{(t)}(\bm{x})]/\Delta t$ and $\dot{\pi}_{r,(t)}(\bm{x})=\lim_{\Delta t\to 0}[\pi_{r,(t+\Delta t)}(\bm{x})-\pi_{r,(t)}(\bm{x})]/\Delta t$. Then, we have
\be
\frac{\mathrm{d}F^{r}_{(\phi)}[\phi^{r}_{(t)};\bm{x}]}{\mathrm{d}t} & = & \Big\{F^{r}_{(\phi)}[\phi^{r}_{(t)};\bm{x}],H[\phi^{r}_{(t)},\pi_{r,(t)}]\Big\}_{\mathrm{PB}} \label{dF_phi dt} \\
\frac{\mathrm{d}F_{(\pi),r}[\pi_{r,(t)};\bm{x}]}{\mathrm{d}t} & = & \Big\{F_{(\pi),r}[\pi_{r,(t)};\bm{x}],H[\phi^{r}_{(t)},\pi_{r,(t)}]\Big\}_{\mathrm{PB}}. 
\nm\\
\label{dF_pi dt}
\ee	
Since
\be
\frac{\mathrm{d}F^{r}_{(\phi)}[\phi^{r}_{(t)};\bm{x}]}{\mathrm{d}t} 
 = \dot{\phi}^{r}_{(t)}(\bm{x}), \nm\\
\frac{\mathrm{d}F_{(\pi),r}[\pi_{r,(t)};\bm{x}]}{\mathrm{d}t} 
 =  \dot{\pi}_{r,(t)}(\bm{x}), 
\ee
Eqs.~(\ref{dF_phi dt}) and (\ref{dF_pi dt}) become
\be
\dot{\phi}^{r}_{(t)}(\bm{x}) & = & \Big\{\phi^{r}_{(t)}(\bm{x}),H[\phi^{r}_{(t)},\pi_{r,(t)}]\Big\}_{\mathrm{PB}} \label{cano eq phi PB} \\
\dot{\pi}_{r,(t)}(\bm{x}) & = & \Big\{\pi_{r,(t)}(\bm{x}),H[\phi^{r}_{(t)},\pi_{r,(t)}]\Big\}_{\mathrm{PB}}. \label{cano eq pi PB}
\ee
\begin{widetext}
These are actually the Poisson bracket representation of the canonical equation set due to:
\be
\Big\{\phi^{r}_{(t)}(\bm{x}),H\left[\phi^{r}_{(t)},\pi_{r,(t)}\right]\Big\}_{\mathrm{PB}} & = & \int\limits_{\mathbb{R}^{3}}\mathrm{d}^{3}y\sum_{r'}\Bigg\{\frac{\delta_{{\scriptscriptstyle\partial}}\phi^{r}_{(t)}(\bm{x})}{\delta_{{\scriptscriptstyle\partial}}\phi^{r'}_{(t)}(\bm{y})}\frac{\delta_{{\scriptscriptstyle\partial}}H\left[\phi^{r}_{(t)},\pi_{r,(t)}\right]}{\delta_{{\scriptscriptstyle\partial}}\pi_{r',(t)}(\bm{y})}-\frac{\delta_{{\scriptscriptstyle\partial}}\phi^{r}_{(t)}(\bm{x})}{\delta_{{\scriptscriptstyle\partial}}\pi_{r',(t)}(\bm{y})}\frac{\delta_{{\scriptscriptstyle\partial}}H\left[\phi^{r}_{(t)},\pi_{r,(t)}\right]}{\delta_{{\scriptscriptstyle\partial}}\phi^{r'}_{(t)}(\bm{y})}\Bigg\} \nonumber \\
& = & 
\frac{\delta_{{\scriptscriptstyle\partial}}H\left[\phi^{r}_{(t)},\pi_{r,(t)}\right]}{\delta_{{\scriptscriptstyle\partial}}\pi_{r,(t)}(\bm{x})},\label{PB of phi H}\\
\Big\{\pi_{r,(t)}(\bm{x}),H[\phi^{r}_{(t)},\pi_{r,(t)}]\Big\}_{\mathrm{PB}} & = & \int\limits_{\mathbb{R}^{3}}\mathrm{d}^{3}y\sum_{r'}\Bigg\{\frac{\delta_{{\scriptscriptstyle\partial}}\pi_{r,(t)}(\bm{x})}{\delta_{{\scriptscriptstyle\partial}}\phi^{r'}_{(t)}(\bm{y})}\frac{\delta_{{\scriptscriptstyle\partial}}H[\phi^{r}_{(t)},\pi_{r,(t)}]}{\delta_{{\scriptscriptstyle\partial}}\pi_{r',(t)}(\bm{y})}-\frac{\delta_{{\scriptscriptstyle\partial}}\pi_{r,(t)}(\bm{x})}{\delta_{{\scriptscriptstyle\partial}}\pi_{r',(t)}(\bm{y})}\frac{\delta_{{\scriptscriptstyle\partial}}H[\phi^{r}_{(t)},\pi_{r,(t)}]}{\delta_{{\scriptscriptstyle\partial}}\phi^{r'}_{(t)}(\bm{y})}\Bigg\} \nonumber \\
& = &{}-\frac{\delta_{{\scriptscriptstyle\partial}}H[\phi^{r}_{(t)},\pi_{r,(t)}]}{\delta_{{\scriptscriptstyle\partial}}\phi^{r}_{(t)}(\bm{x})}.\label{PB of pi H}
\ee
\end{widetext}	

\begin{figure}[htbp]
	\centering
	\includegraphics[scale=0.3]{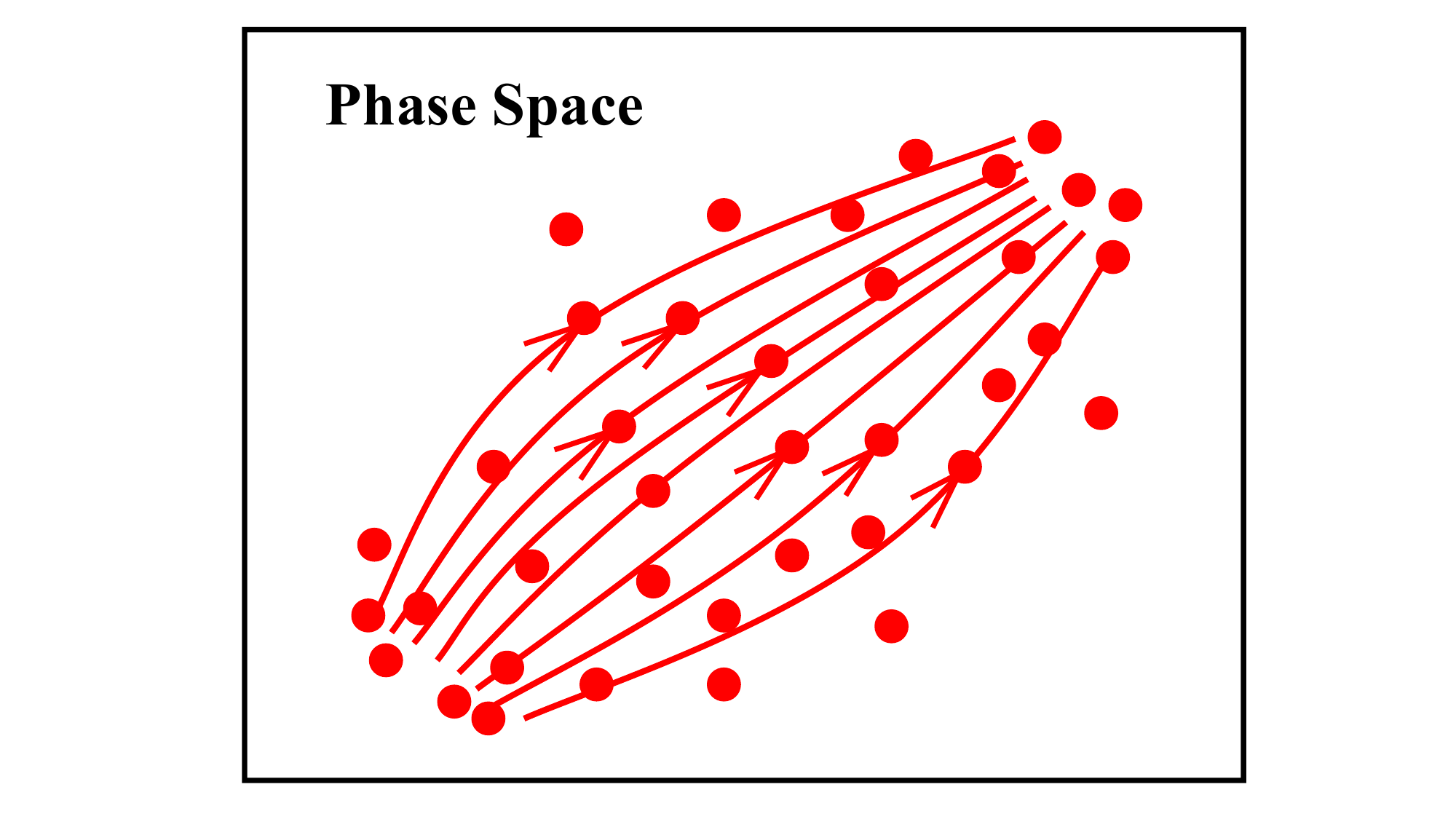}	
	\caption{A fluid corresponding to a statistical ensemble is sketched here. The red dots are the phase points which comprise the fluid and the red curves are streamlines of the fluid.}
    \label{phase space ens}
\end{figure}

The above discussion concentrated on the system in terms of of Hamiltonian formulism in classical field theory. Now, we turn to the statistical ensemble which is actually a set composed of vast quantities of systems which have the same Hamiltonian but different initial conditions. The phase points corresponding to these systems form a fluid in the phase space---termed phase fluid---sketched in Fig.~\ref{phase space ens}. The red dots in Fig.~\ref{phase space ens} represent the phase points which comprise the fluid. In the classical statistical mechanics, every phase point in the fluid moves according to the Hamiltonian equations of motion, i.e., Eq.~(\ref{cano eq}). For this reason, the velocity distribution of the phase fluid in the phase space is actually given by the Hamiltonian equations of motion. Specifically, Eq.~(\ref{cano eq}) describes the motion of a phase point and hence gives its trajectory. A certain initial condition corresponds to a certain trajectory. If all different initial conditions are involved, there are different trajectories in the phase space and all these trajectories are exactly the streamlines of the phase fluid. With these concepts and pictures, we reassess Eq.~(\ref{cano eq}). We find that Eq.~(\ref{cano eq}) gives the integral curves parameterized by the variable $t$ for the "vector field" in the phase space of which the components are
\be
\label{phase velocity field}
& & \left(\cdots\frac{\delta_{{\scriptscriptstyle\partial}}H[\phi^{r},\pi_{r}]}{\delta_{{\scriptscriptstyle\partial}}\pi_{1}(\bm{x})}\cdots,\cdots,\cdots\frac{\delta_{{\scriptscriptstyle\partial}}H[\phi^{r},\pi_{r}]}{\delta_{{\scriptscriptstyle\partial}}\pi_{D}(\bm{x})}\cdots, \right. \nonumber \\
& &{} \; \left.\cdots-\frac{\delta_{{\scriptscriptstyle\partial}}H[\phi^{r},\pi_{r}]}{\delta_{{\scriptscriptstyle\partial}}\phi^{1}(\bm{x})}\cdots,\cdots,\cdots-\frac{\delta_{{\scriptscriptstyle\partial}}H[\phi^{r},\pi_{r}]}{\delta_{{\scriptscriptstyle\partial}}\phi^{D}(\bm{x})}\cdots\right)^{\mathrm{T}}.\nm\\
\ee  
Thus, Eq.~(\ref{cano eq}) can be rewritten in terms of the velocity distribution of the phase fluid, namely
\be
\label{cano eq as vel distribution}
\dot{\phi}^{r}(\bm{x}) & \equiv & \dot{\phi}^{r}[\phi^{r},\pi_{r};\bm{x}] = \frac{\delta_{{\scriptscriptstyle\partial}}H[\phi^{r},\pi_{r}]}{\delta_{{\scriptscriptstyle\partial}}\pi_{r}(\bm{x})},\nm\\
\dot{\pi}_{r}(\bm{x}) & \equiv & \dot{\pi}_{r}[\phi^{r},\pi_{r};\bm{x}] = {} - \frac{\delta_{\scriptscriptstyle\partial}H[\phi^{r},\pi_{r}]}{\delta_{{\scriptscriptstyle\partial}}\phi^{r}(\bm{x})}.
\ee

In the phase space, the number density of phase points at time $t$ denoted by $\varrho$ is a functional of the field and its canonical momentum, i.e., $\varrho[\phi^{r},\pi_{r};t]$. In analogy with the equation of continuity in hydrodynamics
\be
\label{EOC hydro}
\frac{\partial\rho_{\mathrm{m}}}{\partial t}+\nabla\cdot(\rho_{\mathrm{m}}\bm{v})=0,
\ee
where $\rho_{\mathrm{m}}$ is the mass density of a fluid and $\bm{v}=\bm{v}(\bm{x},t)$ is the velocity distribution of the fluid. The equation of continuity of the phase fluid can be written as
\begin{widetext}
\be
\label{EOC phase fluid}
\frac{\partial\varrho\left[\phi^{r},\pi_{r};t\right]}{\partial t}+\int\limits_{\mathbb{R}^{3}}\mathrm{d}^{3}x\sum_{r}\Bigg\{\frac{\delta_{{\scriptscriptstyle\partial}}\big(\varrho\left[\phi^{r},\pi_{r};t\right]\dot{\phi}^{r}(\bm{x})\big)}{\delta_{{\scriptscriptstyle\partial}}\phi^{r}(\bm{x})}+\frac{\delta_{{\scriptscriptstyle\partial}}\big(\varrho\left[\phi^{r},\pi_{r};t\right]\dot{\pi}_{r}(\bm{x})\big)}{\delta_{{\scriptscriptstyle\partial}}\pi_{r}(\bm{x})} \Bigg\}=0.
\ee
Then, by using Eq.~(\ref{cano eq as vel distribution}), we have
\be
\label{deri LEQ}
& & \frac{\partial\varrho\left[\phi^{r},\pi_{r};t\right]}{\partial t}+\int\limits_{\mathbb{R}^{3}}\mathrm{d}^{3}x\sum_{r}\Bigg\{\frac{\delta_{{\scriptscriptstyle\partial}}\varrho\left[\phi^{r},\pi_{r};t\right]}{\delta_{{\scriptscriptstyle\partial}}\phi^{r}(\bm{x})}\frac{\delta_{{\scriptscriptstyle\partial}}H\left[\phi^{r},\pi_{r}\right]}{\delta_{{\scriptscriptstyle\partial}}\pi_{r}(\bm{x})}-\frac{\delta_{{\scriptscriptstyle\partial}}\varrho\left[\phi^{r},\pi_{r};t\right]}{\delta_{{\scriptscriptstyle\partial}}\pi_{r}(\bm{x})}\frac{\delta_{{\scriptscriptstyle\partial}}H\left[\phi^{r},\pi_{r}\right]}{\delta_{{\scriptscriptstyle\partial}}\phi^{r}(\bm{x})} \nm\\
& &\qquad\qquad\qquad\qquad\qquad\qquad {} +  \varrho\left[\phi^{r},\pi_{r};t\right]\Big[\frac{\delta^{2}_{{\scriptscriptstyle\partial}}H\left[\phi^{r},\pi_{r}\right]}{\delta_{{\scriptscriptstyle\partial}}\phi^{r}(\bm{x})\delta_{{\scriptscriptstyle\partial}}\pi_{r}(\bm{x})}-\frac{\delta^{2}_{{\scriptscriptstyle\partial}}H\left[\phi^{r},\pi_{r}\right]}{\delta_{{\scriptscriptstyle\partial}}\pi_{r}(\bm{x})\delta_{{\scriptscriptstyle\partial}}\phi^{r}(\bm{x})}\Big]\Bigg\}=0.
\ee
\end{widetext}
The last term in the brace of Eq.~(\ref{deri LEQ}) vanishes due to 
\be
\label{partial H partial phi pi}
\frac{\delta^{2}_{{\scriptscriptstyle\partial}}H[\phi^{r},\pi_{r}]}{\delta_{{\scriptscriptstyle\partial}}\phi^{r}(\bm{x})\delta_{{\scriptscriptstyle\partial}}\pi_{r}(\bm{x})}=\frac{\delta^{2}_{{\scriptscriptstyle\partial}}H[\phi^{r},\pi_{r}]}{\delta_{{\scriptscriptstyle\partial}}\pi_{r}(\bm{x})\delta_{{\scriptscriptstyle\partial}}\phi^{r}(\bm{x})}.
\ee
From the definition of the Poisson bracket Eq.~(\ref{PB}), we finally obtain
\be
\label{Liouville eq varrho}
\frac{\partial\varrho[\phi^{r},\pi_{r};t]}{\partial t}+\Big\{\varrho[\phi^{r},\pi_{r};t],H[\phi^{r},\pi_{r}]\Big\}_{\mathrm{PB}}=0.
\ee
This is the Liouville equation in classical field theory and it almost has the same mathematical form as the one in classical Hamiltonian mechanics. However, one should note that in classical field theory, the number density of phase points $\varrho$ is a functional of the field and its canonical momentum, instead of a function in classical Hamiltonian mechanics. 

The Liouville equation can be equivalently expressed as the total derivative of $\varrho$ with respect to time. Concretly, imagine that one enters into the phase space and is following a certain phase point of the phase fluid, then the number density of phase points around this person is actually a function of time, i.e., $\varrho(t)\equiv\varrho[\phi^{r}_{(t)},\pi_{r,(t)};t]$. In this case, the number density of phase points $\varrho(t)$ is a kind of dynamical variable which we have introduced in Eq.~(\ref{convention F_t}). Hence, we have 
\be
\label{Liouville eq d_varrho d_t}
\frac{\mathrm{d}\varrho(t)}{\mathrm{d}t} & = & \frac{\partial \varrho[\phi^{r}_{(t)},\pi_{r,(t)};t]}{\partial t} \nm\\
& &{} +\Big\{\varrho[\phi^{r}_{(t)},\pi_{r,(t)};t],H[\phi^{r}_{(t)},\pi_{r,(t)}]\Big\}_{\mathrm{PB}} \nonumber \\
& = & 0.
\ee
This version of Liouville equation indicates that in the vicinity of a certain moving phase point of the phase fluid, the number density of phase points is conserved.

Furthermore, since a statistical ensemble involves enormous number of systems, or equivalently, the phase fluid contains enormous number of phase points, a distribution probability of a phase point at time $t$ can be defined by
\be\label{probability rho}
\rho[\phi^{r},\pi_{r};t]=\frac{\varrho[\phi^{r},\pi_{r};t]}{\mathcal N},
\ee
where $\mathcal N$ is a constant representing the total number of phase points of the phase fluid, or equivalently, the total number of systems in the statistical ensemble. Then, the Liouville equation (\ref{Liouville eq varrho}) can be rewritten as
\be
\label{Liouville eq rho}
\frac{\partial\rho[\phi^{r},\pi_{r};t]}{\partial t}+\Big\{\rho[\phi^{r},\pi_{r};t],H[\phi^{r},\pi_{r}]\Big\}_{\mathrm{PB}}=0,
\ee
which gives the distribution of probability of the statistical ensemble at time $t$. Furthermore, Eq.~(\ref{Liouville eq d_varrho d_t}) becomes
\be
\label{Liouville eq d_rho d_t}
\frac{\mathrm{d}\rho(t)}{\mathrm{d}t}=0,
\ee
which indicates that, around a moving phase point of the phase fluid, the distribution probability is time-independent. Eqs.~(\ref{Liouville eq rho}) and (\ref{Liouville eq d_rho d_t}) are two equivalent versions of the Liouville equation and both are the fundamental equations for classical equilibrium and nonequilibrium systems.

At the end of this appendix, we give the Liouville equation in quantum field theories. This can be achieved straightforwardly by utilizing the replacement $\{~~,~~\}_{\mathrm{PB}} \to [~~,~~]/i$ of canonical quantization formalism. From Eqs.~(\ref{Liouville eq rho}) and (\ref{Liouville eq d_rho d_t}), we obtain
\be
\label{quantum Liouville eq rho}
\frac{\partial \rho[\hat{\phi}^{r},\hat{\pi}_{r};t]}{\partial t}+\frac{1}{i}\Big[\rho[\hat{\phi}^{r},\hat{\pi}_{r};t],H[\hat{\phi}^{r},\hat{\pi}_{r}]\Big]=0,
\ee
and
\be
\label{quantum Liouville eq d_rho d_t}
\frac{\mathrm{d}\hat{\rho}(t)}{\mathrm{d}t}=0,
\ee
where $\rho[\hat{\phi}^{r},\hat{\pi}_{r};t]$ or $\hat{\rho}(t)$ is exactly the density operator in quantum statistical mechanics. Both (\ref{quantum Liouville eq rho}) and (\ref{quantum Liouville eq d_rho d_t}) are equivalent and termed quantum Liouville equation. Here, one should note that the standard canonical quantization gives equations in Heisenberg picture rather than Schr\"odinger picture. It should also be underlined that Eq.~(\ref{quantum Liouville eq d_rho d_t}) is more fundamental than (\ref{quantum Liouville eq rho}) in quantum theory since in quantum theory the density operator $\hat{\rho}(t)$ may not have a classical counterpart.

\section{Quantum scale transformation in Heisenberg picture}

\label{app:ScaleHeisenberg}

In this appendix, we discuss the quantum scale transformation in the Heisenberg picture. We start our discussion from the equation of continuity obtained from a continuous symmetry in the Lagrangian of classical field theory
\be
\label{eoc from sym}
\partial_{\mu}J^{\mu}=0,
\ee
the corresponding conserved charge
\be
\label{cons charge}
Q(t)=\int\limits_{\mathbb{R}^{3}}\mathrm{d}^{3}xJ^{0}(\bm{x},t).
\ee
In the Hamiltonian representation of classical field theory, the conserved charge is a dynamical variable which is a functional of the field and its canonical momentum. In general, the conserved charge is distinctly time-dependent, i.e., $Q(t)\equiv Q[\phi^{r}_{(t)},\pi_{r,(t)};t]$. Combining Eq.~(\ref{eoc from sym}) and Eq.~(\ref{eom F v3}), we have
\be
\label{eoc from sym Hamiltonian}
\frac{\mathrm{d}Q(t)}{\mathrm{d}t} & = & \Big\{Q[\phi^{r}_{(t)},\pi_{r,(t)};t],H\left[\phi^{r}_{(t)},\pi_{r,(t)}\right]\Big\}_{\mathrm{PB}} \nm\\
& &{} +\frac{\partial \left[\phi^{r}_{(t)},\pi_{r,(t)};t\right]}{\partial t} \nm\\
& = & 0.
\ee
This equation can be regarded as the Hamiltonian representation of the Noether's theorem in classical field theory. Then, one can immediately obtain the following quantum version of the charge conservation equation:
\be
\label{quantum eoc from sym}
\frac{\mathrm{d}Q(t)}{\mathrm{d}t} & = & \frac{1}{i}\Big[Q\left[\hat{\phi}^{r}_{(t)},\hat{\pi}_{r,(t)};t\right],H\left[\hat{\phi}^{r}_{(t)},\hat{\pi}_{r,(t)}\right]\Big] \nm\\
& &{} +\frac{\partial Q\left[\hat{\phi}^{r}_{(t)},\hat{\pi}_{r,(t)};t\right]}{\partial t} \nm\\
& = & 0.
\ee
It should be stressed that only when the conserved charge $Q$ has no distinct time-dependence, it commutes with Hamiltonian $H[\hat{\phi}^{r}_{(t)},\hat{\pi}_{r,(t)}]$. Eq.~(\ref{quantum eoc from sym}) plays a crucial role in the following discussions.

We now consider the scale transformation of spacetime coordinates
\be
\label{scale trans coor}
x^{\prime \mu}=e^{a}x^{\mu}\equiv \lambda x^{\mu}~, \quad a\in\mathbb{R}~,\quad \lambda>0,
\ee
and the corresponding transformation of the field
\be\label{scale trans field}
\phi'^{r}(\bm{x}',t')=\lambda^{-\Delta_{\phi^{r}}}\phi^{r}(\bm{x},t)=\exp(-a\Delta_{\phi^{r}})\phi^{r}(\bm{x},t),
\ee
where $\Delta_{\phi^{r}}$ is the scaling dimension of $\phi^{r}(\bm{x},t)$. For any scale-symmetric system, the scale transformation of the energy-momentum tensor $t^{\mu}_{\,\,\,\nu}(\bm{x},t)$ reads
\begin{widetext}
\be
\label{scale trans EMT}
t'^{\mu}_{\,\,\,\,\,\nu}(\bm{x}',t') & = & \sum_{r}\frac{\partial\mathcal{L}\big(\phi'^{r}(\bm{x}',t'),\partial'_{\mu}\phi'^{r}(\bm{x}',t')\big)}{\partial\big(\partial'_{\mu}\phi'^{r}(\bm{x}',t')\big)}\partial'_{\nu}\phi'^{r}(\bm{x}',t')-\delta^{\mu}_{\nu}\mathcal{L}\big(\phi'^{r}(\bm{x}',t'),\partial'_{\mu}\phi'^{r}(\bm{x}',t')\big) \nm \\
& = & \lambda^{-4}\left[\sum_{r}\frac{\partial\mathcal{L}\big(\phi^{r}(\bm{x},t),\partial_{\mu}\phi^{r}(\bm{x},t)\big)}{\partial\big(\partial_{\mu}\phi^{r}(\bm{x},t)\big)}\partial_{\nu}\phi^{r}(\bm{x},t)-\delta^{\mu}_{\nu}\mathcal{L}\big(\phi^{r}(\bm{x},t),\partial_{\mu}\phi^{r}(\bm{x},t)\big)\right] \nm \\
& = & \lambda^{-4}t^{\mu}_{\,\,\,\,\nu}(\bm{x},t).
\ee
\end{widetext}
Hence, the classical scale transformation of the Hamiltonian $H(t)= H[\phi^{r}_{(t)},\pi_{r,(t)}]$ becomes
\be
\label{scale trans Hamiltonian}
H'(t') 
& = & \int\limits_{\mathbb{R}^{3}}\mathrm{d}^{3}x't'^{0}_{\,\,\,\,\,0}(\bm{x}',t') = \int\limits_{\mathbb{R}^{3}}\mathrm{d}^{3}x\lambda^{3}\cdot\lambda^{-4}t^{0}_{\,\,\,\,0}(\bm{x},t) \nm \\
& = & \lambda^{-1}H(t).
\ee

As for the quantum scale transformation of the Hamiltonian operator $\hat{H}(t)$, we can proceed from its eigenvalue equation 
\be
\label{eigen eq H}
\hat{H}(t)\left|E,t\right\rangle=E\left|E,t\right\rangle~.
\ee
Here, although the Hamiltonian operator is conserved and the corresponding eigenvectors are of time-independence, we retain the time variable $t$ in the eigenvalue equation for the sake of clarity. 

Now, let us make the first postulate that the eigenvalue equation of the Hamiltonian operator is covariant under the quantum scale transformation, namely 
\be
\label{eigen eq H prime}
\hat{H}'(t')\left|E',t'\right\rangle'=E'\left|E',t'\right\rangle'.
\ee
Here, $\big\{\hat{H}'(t'),\left|E',t'\right\rangle',E'\big\}$ and $\big\{\hat{H}(t),\left|E,t\right\rangle,E\big\}$ are defined at the same spacetime point and describe the same system. The second postulate we want to make is that $\hat{H}'(t')$ and $E'$ have the same scaling properties as their classical counterpart,i.e., 
\be
\label{scale trans H op, E}
\hat{H}'(t')=\lambda^{-1}\hat{H}(t), \quad E'=\lambda^{-1}E.
\ee
Combining Eq.~(\ref{eigen eq H}) and Eq.~(\ref{scale trans H op, E}), we immediately obtain
\be
\label{scale trans eigen eq H}
\hat{H}'(t')\left|E,t\right\rangle=E'\left|E,t\right\rangle. 
\ee
And, comparing Eq.~(\ref{eigen eq H prime}) with Eq.~(\ref{scale trans eigen eq H}), we obtain 
\be
\label{prime eignvec}
\left|E',t'\right\rangle'=\eta_{{\ssp\mathrm{D}}}(a)\left|E,t\right\rangle,
\ee
where $\eta_{{\ssp\mathrm{D}}}(a)$ is a constant factor related to the scale transformation, i.e., $\eta_{{\ssp\mathrm{D}}}(0)=1$. 

We have introduced $\big\{\hat{H}'(t'),\left|E',t'\right\rangle',E'\big\}$ and $\big\{\hat{H}(t),\left|E,t\right\rangle,E\big\}$ at the same spacetime point. One may ask what are the scale transformation properties of a quantum state $\left|\Psi\right\rangle$, the expectation value of energy $\big\langle\hat{H}(t)\big\rangle_{{\ssp\Psi}}\equiv\left\langle\Psi\right|\hat{H}(t)\left|\Psi\right\rangle$ and the squared quantum fluctuation of the energy
\be
\label{squared QF}
\big\langle\big[\delta_{{\ssp\Psi}}\hat{H}(t)\big]^{2}\big\rangle_{{\ssp\Psi}} & = & \left\langle\Psi\right|\big[\delta_{{\ssp\Psi}}\hat{H}(t)\big]^{2}\left|\Psi\right\rangle \nm \\
& \equiv & \left\langle\Psi\right|\big[\hat{H}(t)-\big\langle\hat{H}(t)\big\rangle_{{\ssp\Psi}}\hat{I}\big]^{2}\left|\Psi\right\rangle \nm \\
& = & \big\langle\big[\hat{H}(t)\big]^{2}\big\rangle_{{\ssp\Psi}}-\big(\big\langle\hat{H}(t)\big\rangle_{{\ssp\Psi}}\big)^{2}.
\ee

To answer these questions, we should concern Eq.~(\ref{prime eignvec}) and give a clarification about its physical meaning. Since the eigenvectors are basis vectors of the Hilbert space, Eq.~(\ref{prime eignvec}) indicates that the Hilbert space with basis vectors $\big\{|E,t\rangle\big\}$ in the coordinates $\big\{x^{\mu}\big\}$ is totally equivalent to the Hilbert space with basis vectors $\big\{|E',t'\rangle'\big\}$ in the coordinates $\big\{x'^{\mu}\big\}$ or in other words, two different coordinates at the same spacetime point actually share one Hilbert space. In addition, the probability distribution of energy for a given state $|\Psi\rangle$ should be invariant under the scale transformation. For these reasons, it is reasonable to stipulate that the state $|\Psi\rangle$ is invariant under scale transformation. Therefore, the expectation value of energy transforms as
\be
\label{scale trans expc(H)}
\big\langle\hat{H}'(t')\big\rangle_{{\ssp\Psi}} & \equiv & \left\langle\Psi\right|\hat{H}'(t')\left|\Psi\right\rangle = \lambda^{-1}\left\langle\Psi\right|\hat{H}(t)\left|\Psi\right\rangle \nm \\
& = & \lambda^{-1}\big\langle\hat{H}(t)\big\rangle_{{\ssp\Psi}},
\ee  
which is the same as the classical scale transformation of energy Eq.~(\ref{scale trans Hamiltonian}). Moreover, the squared quantum fluctuation of the energy transforms as
\be
\label{scale trans qufluc(H)^2}
\big\langle\big[\delta_{{\ssp\Psi}}\hat{H}'(t')\big]^{2}\big\rangle_{{\ssp\Psi}} & = & \big\langle\big[\hat{H}'(t')\big]^{2}\big\rangle_{{\ssp\Psi}}-\big(\big\langle\hat{H}'(t')\big\rangle_{{\ssp\Psi}}\big)^{2} \nm \\
& = & \left\langle\Psi\right|\lambda^{-2}\big[\hat{H}(t)\big]^{2}\left|\Psi\right\rangle-\lambda^{-2}\big(\big\langle\hat{H}(t)\big\rangle_{{\ssp\Psi}}\big)^{2} \nm \\
& = & \lambda^{-2}\big\langle\big[\delta_{{\ssp\Psi}}\hat{H}(t)\big]^{2}\big\rangle_{{\ssp\Psi}}.
\ee

In addition, the scale transformation does not change the orthonormal relation and the projection operators of subspace of the Hilbert space (completeness relation is the projection operator of whole Hilbert space). We argue this issue using the one-particle states of a real-scalar fields as an example. The one-particle states $|E(\bm{p}),t\rangle$ are basis vectors of a subspace that is crucial in quantum field theory. The orthonormal relation and the projection operator are
\be
& & \langle E(\bm{p}),t | E(\bm{q}),t \rangle =(2\pi)^{3}\big(2E(\bm{p})\big)\delta^{(3)}(\bm{p}-\bm{q}), \label{ortho}\\
& & \int\limits_{\mathbb{R}^{3}}\frac{\mathrm{d}^{3}p}{(2\pi)^{3}}\dfrac{1}{2E(\bm{p})}|E(\bm{p}),t\rangle \langle E(\bm{p}),t |=\hat{I}_{\mathrm{1p}}, \label{pro}
\ee
respectively. Here, the subscript of $\hat{I}_{\mathrm{1p}}$ means ``one particle" and $E(\bm{p})=\sqrt{|\bm{p}|^{2}+\big[M(|\bm{p}|/\Lambda)\big]^{2}}$ is the dispersion relation where $M(|\bm{p}|/\Lambda)$ has the dimension of mass and $\Lambda$ is an energy scale. The function $\big[M(|\bm{p}|/\Lambda)\big]^{2}$ is actually a Lorentz scalar since the square of four-momentum $p^{\mu}=(E(\bm{p}),\bm{p})$ is
\be\label{p squared}
\eta_{\mu\nu}p^{\mu}p^{\nu}=\big[M(|\bm{p}|/\Lambda)\big]^{2}.
\ee
For a given energy scale $\Lambda$, if the three-momentum carried by one particle is rather small, i.e., $|\bm{p}|/\Lambda\ll 1$, we can take a low-momentum expansion 
\be
\label{LME}
\big[M(|\bm{p}|/\Lambda)\big]^{2}=\big[M(0)\big]^{2}+\sum_{k=1}^{+\infty}\frac{\Big\{\big[M(0)\big]^{2}\Big\}^{(k)}}{k!}\left(\frac{|\bm{p}|}{\Lambda}\right)^{k},\nm\\
\ee
where $\Big\{\big[M(0)\big]^{2}\Big\}^{(k)}$ is the $k\text{th}$ derivative evaluated at the point $|\bm{p}|/\Lambda=0$. 

Before deriving the scale transformation property for one-particle states $| E(\bm{p}),t \rangle$, it is necessary to recall the $\mathrm{SO(1,3)}$ transformation for these states, i.e., $|E'(\bm{p}'),t'\rangle'=\eta_{{\ssp\mathrm{R,B}}}(\bm{\theta},\bm{\eta})|E(\bm{p}),t\rangle$. Here, the subscript of $\eta_{{\ssp\mathrm{R,B}}}$ represents the rotation and the boost. Explicitly, $\bm{\theta}=\theta\bm{n}_{{\ssp\mathrm{R}}}$ with $\bm{n}_{{\ssp\mathrm{R}}}$ being the unit vector along the rotation axis and $\theta$ standing for the anticlockwise rotation for an observer facing the unit vector. In addition, $\bm{\eta}=\eta\bm{n}_{{\ssp\mathrm{B}}}$ where the magnitude of the rapidity $\eta\geqslant 0$ and $\bm{n}_{{\ssp\mathrm{B}}}=\bm{v}/|\bm{v}|$ with the velocity $\bm{v}$ of one inertial reference frame relative to another one. To derive the $\mathrm{SO(1,3)}$ transformation of $| E(\bm{p}),t \rangle$, we should firstly prove $\mathrm{d}^{3}p/E(\bm{p})$ and $E(\bm{p})\delta^{(3)}(\bm{p}-\bm{q})$ are two $\mathrm{SO(1,3)}$-invariant objects. For a given vector $V^{\mu}=(V^{0},\bm{p})$ with the dimension of momentum, we have 
\be
\label{V2 minus M2}
& & \eta_{\mu\nu}V^{\mu}V^{\nu}-\big[M(|\bm{p}|/\Lambda)\big]^{2} \nm\\
& & \;\;\; {}= \eta_{\mu\nu}V'^{\mu}V'^{\nu}-\big[M'(|\bm{p}'|/\Lambda')\big]^{2},
\ee
therefore
\be
\label{delta V02 minus E2}
\delta\left[\left(V^{0}\right)^{2}-\left(E(\bm{p})\right)^{2}\right]=\delta\left[\left(V'^{0}\right)^{2}-\left(E'(\bm{p}')\right)^{2}\right].
\ee
Using the property of $\delta$-function and the $\mathrm{SO(1,3)}$ invariant volume element $\mathrm{d}^{3}p\mathrm{d}V^{0}$, we have
\be
& & \frac{\mathrm{d}^{3}p}{E(\bm{p})}\!\!\int\limits_{-\infty}^{+\infty}\!\!\!\mathrm{d}V^{0}\!\!\left[\delta\left(V^{0}-E(\bm{p})\right)+\delta\left(V^{0}+E(\bm{p})\right)\right] \nm\\
& & =\!\!\frac{\mathrm{d}^{3}p'}{E'(\bm{p}')}\!\!\int\limits_{-\infty}^{+\infty}\!\!\!\mathrm{d}V'^{0}\!\!\left[\delta\!\left(V'^{0}-E'(\bm{p}')\right)+\delta\!\left(V'^{0}+E'(\bm{p}')\right)\right]. \nm
\ee
Thus we obtain 
\be
\mathrm{d}^{3}p/E(\bm{p})=\mathrm{d}^{3}p'/E'(\bm{p}').
\label{eq:SO13Phase}
\ee 

Next, we can suppose that, under $\mathrm{SO(1,3)}$ transformation
\be
\label{inv X delta3}
X(\bm{p})\delta^{(3)}(\bm{p}-\bm{q})=X'(\bm{p}')\delta^{(3)}(\bm{p}'-\bm{q}').
\ee
Since $\mathrm{d}^{3}p/E(\bm{p})=\mathrm{d}^{3}p'/E'(\bm{p}')$, we have
\be
\label{integral inv X delta3}
& & \int\limits_{\mathbb{R}^{3}}\dfrac{\mathrm{d}^{3}p}{E(\bm{p})}X(\bm{p})\delta^{(3)}(\bm{p}-\bm{q}) \nm\\
& & \qquad\qquad {} =\int\limits_{\mathbb{R}^{3}}\dfrac{\mathrm{d}^{3}p'}{E'(\bm{p}')}X'(\bm{p}')\delta^{(3)}(\bm{p}'-\bm{q}'),
\ee
namely
\be\label{X/E}
\dfrac{X(\bm{q})}{E(\bm{q})}=\dfrac{X'(\bm{q}')}{E'(\bm{q}')}.
\ee
Hence, substituting Eq.~(\ref{X/E}) into Eq.~(\ref{inv X delta3}), we have
\be\label{E delta3}
E(\bm{p})\delta^{(3)}(\bm{p}-\bm{q})=E'(\bm{p}')\delta^{(3)}(\bm{p}'-\bm{q}').
\ee

Combining Eq.~(\ref{eq:SO13Phase}) and Eq.~(\ref{E delta3}) and recalling Eqs.~(\ref{ortho}) and (\ref{pro}), we have the following two equations:
\be\label{SO1,3 trans ortho}
& & '\!\langle E'(\bm{p}'), t'|E'(\bm{q}'),t'\rangle' \nm \\
& & =|\eta_{{\ssp\mathrm{R,B}}}(\bm{\theta},\bm{\eta})|^{2}(2\pi)^{3}\big(2E'(\bm{p}')\big)\delta^{(3)}(\bm{p}'-\bm{q}'),
\ee
and
\be\label{SO1,3 trans pro}
& & \int\limits_{\mathbb{R}^{3}}\frac{\mathrm{d}^{3}p'}{(2\pi)^{3}}\frac{1}{2E'(\bm{p}')}|E'(\bm{p}'),t'\rangle'\,'\!\langle E'(\bm{p}'),t'| \nm \\
& &\;\;\;\;\;\;\;\;\;\;\;\;\;\;\; {} = |\eta_{{\ssp\mathrm{R,B}}}(\bm{\theta},\bm{\eta})|^{2}\hat{I}_{{\ssp\mathrm{1p}}}.
\ee
Since the orthonormal relation and projection operators at the same spacetime point have the same mathematical form, comparing Eqs.~(\ref{SO1,3 trans ortho}) and (\ref{SO1,3 trans pro}) with Eqs.~(\ref{ortho}) and (\ref{pro}), we obtain $|\eta_{{\ssp\mathrm{R,B}}}(\bm{\theta},\bm{\eta})|=1$, i.e., $\eta_{{\ssp\mathrm{R,B}}}(\bm{\theta},\bm{\eta})$ is a phase factor.

To discuss the $\mathrm{SO(1,3)}$ invariance of the cumulative distribution function (CDF) of the three-momentum of a particle, we first define the following normalized one-particle states
\be\label{N 1P states}
|E(\bm{p}),t\rangle_{{\ssp\mathrm{N}}}\equiv\frac{1}{\sqrt{(2\pi)^{3}2E(\bm{p})}}|E(\bm{p}),t\rangle.
\ee
The corresponding orthonormal relation and projection operator read
\be
\label{N ortho}
{}_{{\ssp\mathrm{N}}}\!\langle E(\bm{p}),t|E(\bm{q}),t\rangle_{{\ssp\mathrm{N}}} =
\delta^{(3)}(\bm{p}-\bm{q}),
\ee
and
\be\label{N pro}
\hat{I}_{{\ssp\mathrm{1p}}} & = & \int\limits_{\mathbb{R}^{3}}\frac{\mathrm{d}^{3}p}{(2\pi)^{3}}\frac{1}{2E(\bm{p})}(2\pi)^{3}2E(\bm{p})|E(\bm{p}),t\rangle_{{\ssp\mathrm{N}}}{}_{{\ssp\mathrm{N}}}\!\langle E(\bm{p}),t| \nm \\
& = & \int\limits_{\mathbb{R}^{3}}\mathrm{d}^{3}p|E(\bm{p}),t\rangle_{{\ssp\mathrm{N}}}{}_{{\ssp\mathrm{N}}}\!\langle E(\bm{p}),t|.
\ee
If the system stands on a certain quantum state $\left|\Psi\right\rangle$, the CDF $\mathscr{F}_{{\ssp\mathrm{M}}}\left(\tilde{p}^{1},\tilde{p}^{2},\tilde{p}^{3},t\right)$ is
\be
\label{1P CDF_M}
\mathscr{F}_{{\ssp\mathrm{M}}}\left(\tilde{p}^{1},\tilde{p}^{2},\tilde{p}^{3},t\right)\equiv\int\limits_{-\infty}^{\tilde{p}^{1}}\int\limits_{-\infty}^{\tilde{p}^{2}}\int\limits_{-\infty}^{\tilde{p}^{3}}\!\!\!\mathrm{d}p^{1}\mathrm{d}p^{2}\mathrm{d}p^{3}|{}_{{\ssp\mathrm{N}}}\!\langle E(\bm{p}),t|\Psi\rangle|^{2},\nm\\
\ee
which gives the probability that at time $t$, the components of the three-momentum of one-particle are less than or equal to $\tilde{p}^{1}$, $\tilde{p}^{2}$ and $\tilde{p}^{3}$ respectively. In addition, the density of probability $|{}_{{\ssp\mathrm{N}}}\!\langle E(\bm{p}),t|\Psi\rangle|^{2}$ is not invariant under $\mathrm{SO(1,3)}$ transformation. Then, we can check the $\mathrm{SO(1,3)}$ invariance of this CDF immediately as follows:
\be
\label{check SO1 3 inv CDF}
& & \mathscr{F}'_{{\ssp\mathrm{M}}}\left(\tilde{p}'^{1},\tilde{p}'^{2},\tilde{p}'^{3},t'\right) \nm \\
& & \;\; =\int\limits_{-\infty}^{\tilde{p}'^{1}}\int\limits_{-\infty}^{\tilde{p}'^{2}}\int\limits_{-\infty}^{\tilde{p}'^{3}}\!\!\!\mathrm{d}p'^{1}\mathrm{d}p'^{2}\mathrm{d}p'^{3}|{}_{{\ssp\mathrm{N}}}\!\!{}'\!\langle E'(\bm{p}'),t'|\Psi\rangle|^{2} \nm \\
& & \;\; =\int\limits_{-\infty}^{\tilde{p}'^{1}}\int\limits_{-\infty}^{\tilde{p}'^{2}}\int\limits_{-\infty}^{\tilde{p}'^{3}}\!\!\!\mathrm{d}p'^{1}\mathrm{d}p'^{2}\mathrm{d}p'^{3}\frac{E(\bm{p})}{E'(\bm{p}')}|{}_{{\ssp\mathrm{N}}}\!\langle E(\bm{p}),t|\Psi\rangle|^{2} \nm \\
& & \;\; =\int\limits_{-\infty}^{\tilde{p}^{1}}\int\limits_{-\infty}^{\tilde{p}^{2}}\int\limits_{-\infty}^{\tilde{p}^{3}}\!\!\!\mathrm{d}p^{1}\mathrm{d}p^{2}\mathrm{d}p^{3}|{}_{{\ssp\mathrm{N}}}\!\langle E(\bm{p}),t|\Psi\rangle|^{2} \nm \\
& & \;\; =\mathscr{F}_{{\ssp\mathrm{M}}}\left(\tilde{p}^{1},\tilde{p}^{2},\tilde{p}^{3},t\right)~.
\ee

Now, let us come back to the scale transformation of one-particle states $|E(\bm{p}),t\rangle$. It is easy to derive the scale transformation of three-momentum utilizing Eq.~(\ref{scale trans EMT}) as
\be
\label{scale trans momentum}
P'^{k}(t') &\equiv & P^{k}[\phi'^{r}_{(t')},\pi'_{r,(t')}]=\int\limits_{\mathbb{R}^{3}}\mathrm{d}^{3}x't'^{0k}(\bm{x}',t') \nm \\
& = & \lambda^{-1}\int\limits_{\mathbb{R}^{3}}\mathrm{d}^{3}x\eta^{k\mu}t^{0}_{\,\,\,\mu}(\bm{x},t) \nm \\
& = & \lambda^{-1}P^{k}(t), ~k=1,2,3. 
\ee
By using the scale transformation of energy $E(\bm{p})$ and three-momentum $\bm{p}$, combining Eq.~(\ref{prime eignvec}) with Eqs.~(\ref{ortho}) and (\ref{pro}), we have
\be
\label{scale trans ortho}
& & {}'\!\langle E'(\bm{p}'),t'|E'(\bm{q}'),t'\rangle' \nm \\
& & \;\; =\lambda^{-2}|\eta_{{\ssp\mathrm{D}}}(a)|^{2}(2\pi)^{3}\big(2E'(\bm{p}')\big)\delta^{(3)}(\bm{p}'-\bm{q}'),
\ee
and
\be
\label{scale trans pro}
& & \int\limits_{\mathbb{R}^{3}}\frac{\mathrm{d}^{3}p'}{(2\pi)^{3}}\frac{1}{2E'(\bm{p}')}|E'(\bm{p}'),t'\rangle'{}'\!\langle E'(\bm{p}'),t'| \nm\\
& &\;\;\;\;\; =\lambda^{-2}|\eta_{{\ssp\mathrm{D}}}(a)|^{2}\hat{I}_{\mathrm{1p}}.
\ee
Since the scale transformation cannot change the mathematical form of the orthonormal and complete relation, we obtain $|\eta_{{\ssp\mathrm{D}}}(a)|=\lambda$ and $\eta_{{\ssp\mathrm{D}}}(a)$ is not a phase factor. 

Finally, we would like to check the scale invariance of the CDF $\mathscr{F}_{{\ssp\mathrm{M}}}\left(\tilde{p}^{1},\tilde{p}^{2},\tilde{p}^{3},t\right)$ as follows:
\be
\label{check scale inv CDF}
& & \mathscr{F}'_{{\ssp\mathrm{M}}}\left(\tilde{p}'^{1},\tilde{p}'^{2},\tilde{p}'^{3},t'\right) \nm \\
& & =\int\limits_{-\infty}^{\tilde{p}'^{1}}\int\limits_{-\infty}^{\tilde{p}'^{2}}\int\limits_{-\infty}^{\tilde{p}'^{3}}\!\!\!\mathrm{d}p'^{1}\mathrm{d}p'^{2}\mathrm{d}p'^{3}|{}_{{\ssp\mathrm{N}}}\!\!{}'\!\langle E'(\bm{p}'),t'|\Psi\rangle|^{2} \nm \\
& & =\int\limits_{-\infty}^{\tilde{p}'^{1}}\int\limits_{-\infty}^{\tilde{p}'^{2}}\int\limits_{-\infty}^{\tilde{p}'^{3}}\!\!\!\mathrm{d}p'^{1}\mathrm{d}p'^{2}\mathrm{d}p'^{3}\left|\frac{[\eta_{{\ssp\mathrm{D}}}(a)]^{*}}{\sqrt{(2\pi)^{3}2E'(\bm{p}')}}\langle E(\bm{p}),t|\Psi\rangle\right|^{2} \nm \\
& & =\int\limits_{-\infty}^{\tilde{p}^{1}}\int\limits_{-\infty}^{\tilde{p}^{2}}\int\limits_{-\infty}^{\tilde{p}^{3}}\!\!\!\mathrm{d}p^{1}\mathrm{d}p^{2}\mathrm{d}p^{3}\frac{\lambda^{-3}|\eta_{{\ssp\mathrm{D}}}(a)|^{2}}{(2\pi)^{3}2\lambda^{-1}E(\bm{p})}|\langle E(\bm{p}),t|\Psi\rangle|^{2} \nm \\
& & =\mathscr{F}_{{\ssp\mathrm{M}}}\left(\tilde{p}^{1},\tilde{p}^{2},\tilde{p}^{3},t\right).
\ee

The above discussion actually shows one paradigm of the quantum scale transformation for the Hamiltonian and its spectrum of eigenvalues and the corresponding eigenvectors. In this paradigm, we focus on the relations of the quantities, such as $\hat{H}(t)$, $E$, $|E,t\rangle$, $|\Psi\rangle$ and the CDF of one-particle's three- momentum, at the same spacetime point which is parameterized by two different coordinates $\left\{x'^{\mu}=\lambda x^{\mu}\right\}$ and $\left\{x^{\mu}\right\}$. This paradigm has the feature that the basis vectors of a Hilbert space are not ``rotated" by an operator.

We next construct another paradigm that the basis vectors of a Hilbert space are ``rotated" by a unitary operator.

We now try to find a reversible transformation $\hat{D}(t,a)$---which will be seen unitary later---such that
\be
\label{D(t,a) for H}
\hat{H}'(t)=\hat{D}(t,a)\hat{H}(t)[\hat{D}(t,a)]^{-1}.
\ee
Consider the infinitesimal scale transformation $x'^{\mu}=(1+a)x^{\mu},|a|\ll 1$, the transformation of the Hamiltonian~(\ref{scale trans H op, E}) becomes $\hat{H}'(t)=(1-a)\hat{H}\big((1-a)t\big)$. Then, the transformation of Eq.~(\ref{D(t,a) for H}) becomes
\be\label{infini D(t,a) for H}
\big[\hat{I}+a\delta\hat{D}(t)\big]\hat{H}(t)\big[\hat{I}-a\delta\hat{D}(t)\big]=(1-a)\hat{H}\big((1-a)t\big),\nm\\
\ee
which can be simplified as $\Big[\delta\hat{D}(t),\hat{H}(t)\Big]=-\hat{H}(t)$ by considering the energy conservation. Using the equation of motion of the conserved charge $\hat{Q}_{\ssp\mathrm{D}}(t)\equiv Q_{\ssp\mathrm{D}}[\hat{\phi}^{r}_{(t)},\hat{\pi}_{r,(t)};t]$ defined by Eq.~(\ref{Q_scale}), i.e., 
\be
\label{EOM Q dil}
\dfrac{\mathrm{d}\hat{Q}_{\ssp\mathrm{D}}(t)}{\mathrm{d}t}=\frac{1}{i}\Big[\hat{Q}_{\ssp\mathrm{D}}(t),\hat{H}(t)\Big]+\frac{\partial\hat{Q}_{\ssp\mathrm{D}}(t)}{\partial t}=0~,
\ee
and considering $\partial\hat{Q}_{\ssp\mathrm{D}}(t)/\partial t=\hat{H}(t)$, we have 
\be
\label{commutator Q dil H}
\frac{1}{i}\Big[\hat{Q}_{\ssp\mathrm{D}}(t),\hat{H}(t)\Big]=-\hat{H}(t).
\ee
Consequently, $\delta\hat{D}(t)=\frac{1}{i}\hat{Q}_{\ssp\mathrm{D}}(t)$. Here, it should be noted that, to guarantee the hermitian of charge operator $\big[\hat{Q}_{\ssp\mathrm{D}}(t)\big]^{\dagger}=\hat{Q}_{\ssp\mathrm{D}}(t)$,  we used the following expression:
\be
\label{Hermitian Q dil}
\hat{Q}_{\ssp\mathrm{D}}(t) & = & t\hat{H}(t) \nm\\
& &{} +\int\limits_{\mathbb{R}^{3}}\mathrm{d}^{3}x\sum_{r}\bigg[\frac{\bm{x}}{2}\cdot\big(\hat{\pi}_{r}\bm{\nabla}\hat{\phi}^{r}+(\hat{\pi}_{r}\bm{\nabla}\hat{\phi}^{r})^{\dagger}\big) \nm \\
& & \qquad\qquad\qquad\;\; {} +\frac{\Delta_{\phi^{r}}}{2}\big(\hat{\pi}_{r}\hat{\phi}^{r}+(\hat{\pi}_{r}\hat{\phi}^{r})^{\dagger}\big)\bigg].
\nm\\
\ee
We then finally find that, for a finite scale transformation, $\hat{D}(t,a)=\exp\left[-i a\hat{Q}_{\ssp\mathrm{D}}(t)\right]$, that is, $\hat{D}(t,a)$ is a unitary operator. 

Furthermore, the eigenvalue equation $\hat{H}(t)|E,t\rangle=E|E,t\rangle$ transforms as
\be
\hat{D}(t,a)\hat{H}(t)[\hat{D}(t,a)]^{\dagger}\hat{D}(t,a)|E,t\rangle=E\hat{D}(t,a)|E,t\rangle.
\nm\\
\ee
that is,
\be\label{D(t,a) for EVE H}
\hat{H}'(t)\hat{D}(t,a)|E,t\rangle=E \hat{D}(t,a)|E,t\rangle.
\ee
Then we make replacements $t\to t'$ and $E \to E'$ in $\hat{H}'(t)\hat{D}(t,a)|E,t\rangle=E \hat{D}(t,a)|E,t\rangle$ and obtain
\be\label{rep t E in D(t,a) for EVE H}
\hat{H}'(t')\hat{D}(t',a)|E',t'\rangle=E' \hat{D}(t',a)|E',t'\rangle.
\ee
Here, we assume that the domains of $\hat{H}'(t')$ and $|E,t\rangle$ with respect to variables $E$, $t$ and $t'$ are $\mathbb{R}$ and hence the replacements make sense.
Comparing Eq.~(\ref{eigen eq H prime}) with the above equation, we have
\be
\label{D(t,a) for EVs primed E t}
\hat{D}(t',a)|E',t'\rangle=\zeta_{{\ssp\mathrm{D}}}(a,E',t')|E',t'\rangle',
\ee
where $\zeta_{{\ssp\mathrm{D}}}(a,E',t')$ is a factor related to the scale transformation group. We can make replacements $t'\to t$ and $E'\to E$ in this equation and get
\be
\label{D(t,a) for EVs}
\hat{D}(t,a)|E,t\rangle=\zeta_{{\ssp\mathrm{D}}}(a,E,t)|E,t\rangle'.
\ee
By using Eq.~(\ref{D(t,a) for EVs primed E t}), we can obatin the following inner product
\be
\label{in pro E tilde E}
&\langle \widetilde{E}',t'|E',t'\rangle=[\zeta_{{\ssp\mathrm{D}}}(a,\widetilde{E}',t')]^{*}\zeta_{{\ssp\mathrm{D}}}(a,E',t'){}'\!\langle \widetilde{E}',t'|E',t'\rangle'.\nm\\
\ee
Since $\langle \widetilde{E}',t'|E',t'\rangle=\delta_{\widetilde{E}'E'}$, we have
\be
\label{prime in pro E tilde E}
'\!\langle \widetilde{E}',t'|E',t'\rangle' & = & \dfrac{\delta_{\widetilde{E}'E'}}{[\zeta_{{\ssp\mathrm{D}}}(a,\widetilde{E}',t')]^{*}\zeta_{{\ssp\mathrm{D}}}(a,E',t')} \nm \\
& = & \dfrac{\delta_{\widetilde{E}'E'}}{|\zeta_{{\ssp\mathrm{D}}}(a,E',t')|^{2}}.
\ee
Because the orthonormal relations at the same spacetime point have the same mathematical form, i.e., ${}'\!\langle \widetilde{E}',t'|E',t'\rangle'=\delta_{\widetilde{E}'E'}$, we obtain
\be\label{|zeta D|}
|\zeta_{{\ssp\mathrm{D}}}(a,E',t')|=1,
\ee
which means that $\zeta_{{\ssp\mathrm{D}}}(a,E',t')$ is actually a phase factor.

As for the completeness relation, from Eq.~(\ref{D(t,a) for EVs primed E t}) we have
\be
& & \sum_{E'}\hat{D}(t',a)|E',t'\rangle\langle E',t'|[\hat{D}(t',a)]^{\dagger} \nm\\
& &{} =\sum_{E'}\zeta_{{\ssp\mathrm{D}}}(a,E',t')|E',t'\rangle'{}'\!\langle E',t'|[\zeta_{{\ssp\mathrm{D}}}(a,E',t')]^{*} ,\nm\\
\ee
where $\sum\limits_{E'}$ denotes a certain integral when the energy spectrum is continuous.
Consequently, 
\be
\hat{D}(t',a)[\hat{D}(t',a)]^{\dagger}=\sum_{E'}|\zeta_{{\ssp\mathrm{D}}}(a,E',t')|^{2}|E',t'\rangle'{}'\!\langle E',t'|,\nm\\
\ee
that is,
\be
\label{comp rel E vec}
\sum_{E'}|E',t'\rangle'{}'\!\langle E',t'|=\hat{I}~,
\ee
From the above discussion, we choose $\zeta_{\ssp\mathrm{D}} = 1$.

Finally, according to the above discussion, one can easily conclude that the conserved charges $Q_{i}(t),i=1,2,\cdots,K$ corresponding to the global unitary symmetry in quantum field theory are scale invariant.

\section{Scale properties of the laws of thermodynamics}
\label{app:4laws}

In this appendix, we argue that the four laws of thermodynamics are scale covariant. 

Using the relations between the thermodynamic quantities and the partition function, we have
\be
P^\prime & = & \frac{\partial\Big\{T'\ln\big[Z'(t',T',V',\mu'_{1},\cdots,\mu'_{K})\big]\Big\}}{\partial V'} \nm \\
& = & \frac{\partial\Big\{\lambda^{-1}T\ln\big[Z(t,T,V,\mu_{1},\cdots,\mu_{K})\big]\Big\}}{\partial V}\frac{\partial V}{\partial V'} \nm \\
& = & \lambda^{-4}P, \label{scale trans pressure}\\
N^\prime_{i} & = & \frac{\partial\Big\{T'\ln\big[Z'(t',T',V',\mu'_{1},\cdots,\mu'_{K})\big]\Big\}}{\partial\mu'_{i}} \nm \\
& = & \sum_{j=1}^{K}\frac{\partial\Big\{\lambda^{-1}T\ln\big[Z(t,T,V,\mu_{1},\cdots,\mu_{K})\big]\Big\}}{\partial\mu_{j}}\frac{\partial\mu_{j}}{\partial\mu'_{i}} \nm \\
& = & N_{i}~, \quad i=1,\cdots,K~,\label{scale trans particle number} \\
S^\prime & = & \frac{\partial\Big\{T'\ln\big[Z'(t',T',V',\mu'_{1},\cdots,\mu'_{K})\big]\Big\}}{\partial T'} \nm \\
& = & \frac{\partial\Big\{\lambda^{-1}T\ln\big[Z(t,T,V,\mu_{1},\cdots,\mu_{K})\big]\Big\}}{\partial T}\frac{\partial T}{\partial T'} \nm \\
& = & S~.\label{scale trans entropy}
\ee
Consequently, one can obtain
\be
U^\prime & = &{} -P'V'+T'S'+\sum_{i=1}^{K}\mu'_{i}N'_{i} = \lambda^{-1}U,\label{scale trans internal energy}\\
H^\prime & = & U'+P'V'=\lambda^{-1}H, \label{scale trans enthalpy}\\
F^\prime & = & U'-T'S'=
\lambda^{-1}F,
\label{scale trans Helmholtz free-energy}\\
G^\prime & = & \sum_{i=1}^{K}\mu'_{i}N'_{i} 
=\lambda^{-1}G,
\label{scale trans Gibbs free-energy}\\
\varPsi^\prime & = & F'-G'= 
\lambda^{-1}\varPsi.
\label{scale trans thermodynamic potential}
\ee
Here, $P, N_{i}, S, U, H, G$ and $\varPsi$ are pressure, particle numbers, entropy, internal energy, enthalpy, Helmholtz free energy, Gibbs free energy and thermodynamic potential, respectively.  

In addition to the above thermodynamic quantities, we need to obtain the scale transformation property of the heat $\delta Q$ that is absorbed by the system in a thermodynamic process. Since the scale transformation itself is independent of any thermodynamic process, we consider a reversible process. According to the second law of thermodynamics, $\delta Q=T\mathrm{d}S$, we obtain 
\be
\delta Q'=T'\mathrm{d}S'=\lambda^{-1}T\mathrm{d}S=\lambda^{-1}\delta Q.
\ee
Here, it should be emphasized that in the above derivation, we regard the second law of thermodynamics for reversible processes as scale covariance. In general, the second law reads $\mathrm{d}S\geq\delta Q/T$ with the greater-than sign for irreversible processes and the equal sign for reversible processes. Nevertheless, there is no inconsistency and the scale covariance of the second law in general is obvious:
\be
\label{scale covar 2nd law}
\mathrm{d}S'\geq\frac{\delta Q'}{T'}~~ \Rightarrow ~~\mathrm{d}S\geq\frac{\lambda^{-1}\delta Q}{\lambda^{-1}T}~~\Rightarrow~~\mathrm{d}S\geq\frac{\delta Q}{T}.
\ee

As for the covariances of other three laws of thermodynamics, let us check one by one as follows: 
\begin{itemize}
\item The zeroth law: For any three thermodynamic systems $\mathcal{A}$, $\mathcal{B}$ and $\mathcal{C}$ in the coordinates $\{x^{\mu}\}$, the system $\mathcal{A}$ equilibrates with the system $\mathcal{C}$ ($T_{\mathcal{A}}=T_{\mathcal{C}}$) and it also equilibrates with the system $\mathcal{B}$ ($T_{\mathcal{A}}=T_{\mathcal{B}}$), then $\mathcal{B}$ and $\mathcal{C}$ must equilibrate with each other ($T_{\mathcal{B}}=T_{\mathcal{A}}=T_{\mathcal{C}}$). We perform the scale transformation $T'=\lambda^{-1}T$, and the coordinates $\{x'^{\mu}=\lambda x^{\mu}\}$. $T'_{\mathcal{A}}=T'_{\mathcal{C}}$ means that $\mathcal{A}$ equilibrates with $\mathcal{C}$ and $T'_{\mathcal{A}}=T'_{\mathcal{B}}$ indicates that $\mathcal{A}$ equilibrates with $\mathcal{B}$, then equation $T'_{\mathcal{B}}=T'_{\mathcal{A}}=T'_{\mathcal{C}}$ tells us that $\mathcal{B}$ must equilibrate with $\mathcal{C}$. For this reason, the zeroth law is scale covariant.

\item The first law: The differential form of the first law of thermodynamics reads
\be
\label{DF of first law}
\mathrm{d}U=\delta Q-P\mathrm{d}V+\sum_{i=1}^{K}\mu_{i}\mathrm{d}N_{i}.
\ee
Performing the scale transformation
\be
\label{scale trans DF of first law}
\lambda^{-1}\mathrm{d}U=\lambda^{-1}\delta Q-\lambda^{-1}P\mathrm{d}V+\lambda^{-1}\sum_{i=1}^{K}\mu_{i}\mathrm{d}N_{i},
\ee
we have
\be
\label{scale trans DF of first law 2}
\mathrm{d}U'=\delta Q'-P'\mathrm{d}V'+\sum_{i=1}^{K}\mu'_{i}\mathrm{d}N'_{i},
\ee
which is obviously scale covariant.
	
\item The third law: For any thermodynamic system, it is impossible to refrigerate it to arrive at zero temperature with finite procedures. In the coordinate $\{x^{\mu}\}$, the third law is valid. In the coordinate $\{x'^{\mu}=\lambda x^{\mu}\}$, $T'=\lambda^{-1}T,~\lambda>0$ tells us that with finite procedures, we cannot arrive at $T'=0$. Hence, the third law is scale covariant as well. It is rather interesting that $T'=\lambda^{-1}T,~\lambda>0$ also tells us that we cannot find a well-defined coordinate $\{x'^{\mu}\}$ in which we could arrive at $T'=0$. In other words, if we expect to obtaining $T'=0$ using scale transformation then $\lambda$ is of divergence and $x'^{\mu}=\lambda x^{\mu}$ are divergent and the coordinate $\{x'^{\mu}\}$ is not well-defined in this case.
\end{itemize}

\bibliography{ref.bib}	
\end{document}